\pdfoutput=1

\documentclass{aa}  

\usepackage{xcolor}
\usepackage[colorlinks]{hyperref}
\definecolor{linkcolor}{rgb}{0.0,0.3,0.5}
\hypersetup{colorlinks=true, linkcolor=linkcolor, filecolor=linkcolor, citecolor=linkcolor, urlcolor=linkcolor}
\usepackage{orcidlink}
\usepackage{graphicx}
\usepackage{txfonts}

\newcommand{\sn}{\ifmmode {\rm S/N}\else${\rm S/N}$\fi}
\newcommand{\Pm}{\ifmmode P_{\rm m}\else$P_{\rm m}$\fi}
\newcommand{\Pf}{\ifmmode P_{\rm F}\else$P_{\rm F}$\fi}
\newcommand{\Po}{\ifmmode P_{\rm 1O}\else$P_{\rm 1O}$\fi}
\newcommand{\Poo}{\ifmmode P_{\rm 2O}\else$P_{\rm 2O}$\fi}
\newcommand{\Px}{\ifmmode P_{\rm x}\else$P_{\rm x}$\fi}
\newcommand{\Psh}{\ifmmode P_{\rm sh}\else$P_{\rm sh}$\fi}
\newcommand{\fF}{\ifmmode \nu_{\rm F}\else$\nu_{\rm F}$\fi}
\newcommand{\fo}{\ifmmode \nu_{\rm 1O}\else$\nu_{\rm 1O}$\fi}
\newcommand{\fx}{\ifmmode \nu_{\rm x}\else$\nu_{\rm x}$\fi}
\newcommand{\fsh}{\ifmmode \nu_{\rm sh}\else$\nu_{\rm sh}$\fi}
\renewcommand{\fm}{\ifmmode \nu_{\rm m}\else$\nu_{\rm m}$\fi}
\newcommand{\Af}{\ifmmode A_{\rm F}\else$A_{\rm F}$\fi}
\newcommand{\Ao}{\ifmmode A_{\rm 1O}\else$A_{\rm 1O}$\fi}
\newcommand{\Aoo}{\ifmmode A_{\rm 2O}\else$A_{\rm 2O}$\fi}
\newcommand{\Ax}{\ifmmode A_{\rm x}\else$A_{\rm x}$\fi}
\newcommand{\Cepa}{\ifmmode \mathrm{Cep}_{0.61}\else Cep$_{0.61}$\fi}
\newcommand{\Cepb}{\ifmmode \mathrm{Cep}_{0.68}\else Cep$_{0.68}$\fi}
\newcommand{\oc}{\ensuremath{O\!-\!C}\xspace}

\makeatletter
\renewcommand*\aa@numarticle{}
\renewcommand*\aa@textidlineempty{\aa@headfont \aa@headings}
\renewcommand*\aa@pageof{page \thepage{} of \pageref*{LastPage}}
\renewcommand*\aa@journalname{Accepted in Astronomy \& Astrophysics}
\renewcommand*\aa@manuscriptname{\hspace{\stretch{1}}\copyright ESO \the\year}
\makeatother
\usepackage{letltxmacro}
\LetLtxMacro{\originaleqref}{\eqref}
\renewcommand{\eqref}{Eq.~\originaleqref}

\defcitealias{Rathour2024A&A}{Paper~I}
\defcitealias{Rathour2025A&A}{Paper~II}

\newcommand{\PaperI}{\citetalias{Rathour2024A&A}}
\newcommand{\PaperII}{\citetalias{Rathour2025A&A}}

\begin{document} 

\title{Non-evolutionary effects on period change in Magellanic Cepheids.}
\subtitle{III. Empirical connection with additional low-amplitude variability}

\titlerunning{Non-evolutionary effects on period change in Magellanic Cepheids III}
\authorrunning{Rathour et al.}
\author{Rajeev Singh Rathour\inst{1,2}\orcidlink{0000-0002-7448-4285}\corrauth{rajeev.rathour@oca.eu}
\and Rados{\l}aw Smolec\inst{2}\orcidlink{0000-0001-7217-4884}\email{smolec@camk.edu.pl}
\and Vincent Hocd\'e\inst{1}\orcidlink{0000-0002-3643-0366}\email{vincent.hocde@oca.eu}
\and Nicolas Nardetto\inst{1}\orcidlink{0000-0002-7399-0231}\email{nicolas.nardetto@oca.eu}
\and Oliwia Zi\'o{\l}kowska\inst{2}\orcidlink{0000-0002-0696-2839}\email{oliwiakz@camk.edu.pl}
\and Igor Soszy\'nski\inst{3}\orcidlink{0000-0002-7777-0842}\email{soszynsk@astrouw.edu.pl}
\and Andrzej Udalski\inst{3}\orcidlink{0000-0001-5207-5619}\email{udalski@astrouw.edu.pl}}

\institute{Universit\'e Côte d'Azur, Observatoire de la C\^ote d'Azur, CNRS, Laboratoire Lagrange, France\\
email : \texttt{rajeev.rathour@oca.eu}
\and
Nicolaus Copernicus Astronomical Centre, Polish Academy of Sciences, Bartycka 18, 00-716 Warszawa, Poland
\and
Astronomical Observatory, University of Warsaw, Aleje Ujazdowskie 4, Warsaw, 00-478, Poland
} 

\date{Received 21 September 2026; accepted 10 November 2026}
\abstract
{The subject of the \oc (Observed minus Calculated) method in Classical Cepheid is generally associated with the measurement of secular period change, also known as evolutionary period change. While this offers a direct way to track the evolution of a star, the technique also helps uncover period variations on shorter timescales (non-evolutionary period changes). In our previous works, we compiled a large sample of Magellanic Classical Cepheids that manifest irregularities in their \oc diagrams, finding irregular period changes in $\sim$33\% of the sample. The study, however, deliberately excluded Cepheids showing additional low-amplitude variability, which provides the context for the present work.
}
{We aim to establish whether the first-overtone Cepheids hosting additional low-amplitude variability differ in their period-change behaviour from the general first-overtone population. Including this population brings the census of irregular period changes in Magellanic Cepheids a step closer to completeness. 
}
{We analyse a sample of 544 Magellanic Cepheids with additional low-amplitude variability, using Optical Gravitational Lensing Experiment (OGLE) photometry. Specifically, these are Cepheids with (i) a period ratio \Px/\Po $\sim$ 0.60--0.65 or (ii) \Po/\Px $\sim$ 0.68. This sample was investigated for period changes using \oc diagrams, and the detected period changes were classified and characterised with several statistical methods, following the methodology of our previous studies. We then combine these results with our previous first-overtone sample to update the incidence rates of the period-change classes.
}
{Based on their \oc diagrams, we classified these Cepheids into three categories: class 1 (constant period), class 2 (linear period change), and class 3 (irregular period change). We find irregular period changes in 339 first-overtone Cepheids (195 in the LMC and 144 in the SMC), corresponding to an incidence rate of $\sim$75.5\%. This is only modestly higher than the $\sim$68.1\% found for the far larger single-mode first-overtone sample of our previous study. Including our candidates shifts the combined first-overtone incidence only marginally, to $\sim$69.7\%. 
}
{For the first time, we provide an empirical connection between additional low-amplitude variability and irregular period changes in first-overtone Cepheids. In our additional-variability sample, irregular period changes are more prominent, showing a statistically significant increase over purely single-mode Cepheids. At a fixed pulsation period, the irregularity strength anticorrelates with the dominant pulsation amplitude but not with the additional-signal amplitude, suggesting that the additional periodicity marks a weakly saturated pulsation state rather than driving the irregularity. We further detect cycle-to-cycle (C2C) period variation in 37\% of the class 3 sample, with the associated C2C jitter strength ($\sigma_\eta$) roughly seven times larger than in non-Blazhko RRc stars, and scaling with pulsation period at a shallower slope. In the future, extending the study of period changes to other Cepheid types, in particular Type-II Cepheids, will further probe the origin of these irregularities.
}

\maketitle
\nolinenumbers
\section{Introduction}
\label{sec: Introduction}

Classical Cepheids, or Type-I Cepheids (hereafter Cepheids), are a cornerstone of stellar evolution and pulsation studies. This class of stars is particularly notable for its role as a distance indicator \citep{Freedman2001ApJ,Riess2021ApJ} owing to its tight Period--Luminosity ($P$--$L$) relation \citep{leavitt19081777,Leavitt1912}, also known as the Leavitt Law. Cepheids are $\sim$3--13 M$_{\odot}$ stars that are predominantly in the core-helium-burning phase and are characterised by their distinctive light curves, which exhibit high-amplitude variability.

Classically, Cepheids were known to pulsate in single radial modes such as fundamental (F) mode, first overtone (1O) mode or second overtone (2O) mode. Eventually, with more observations, multi-mode Cepheids were discovered ranging from double-mode Cepheids in combinations (F+1O, 1O+2O, 1O+3O) to even triple-mode detections (F+1O+2O, 1O+2O+3O) \citep[e.g.,][]{soszynski2008optical,soszynski2010optical,Soszynski2015AcA....65..297S,soszynski2016multi,soszynski2017AcA....67..297S,Soszynski2019AcA....69...87S}. Even though the physics of excitation of multi-mode radial mode pulsation is a bone of contention \citep[e.g.,][]{Kollath2002A&A...385..932K,Smolec2008AcA....58..233S}, the presence of multiple modes in Cepheids has aided in constraining physical parameters and in turn evolution/pulsation theory \citep[e.g.,][]{MoskalikDziembowski2005A&A,Beaulieu2006ApJ}.

In the past two decades, Cepheids have been recognised as exhibiting greater complexity than previously assumed, mainly the presence of secondary low-amplitude ($\sim$mmag) periodicities that cannot be attributed to radial modes. One important group of such variability in Cepheids manifests an additional periodicity (\Px) alongside the radial first-overtone period (\Po), with a period ratio \Px/\Po $\in$ 0.60-0.65 \citep{soszynski2008optical,soszynski2010optical}, also referred to as the 0.62-mode in other studies \citep{Suveges2018a,Suveges2018b}. We refer to them in this work as the \Cepa\ variability group. These additional periodicities arrange themselves in three distinct sequences on the Petersen diagram \citep{petersen1973masses}. Many studies across Galactic and Magellanic fields have been dedicated to detecting these additional signatures using Optical Gravitational Lensing Experiment (OGLE) photometry \citep{Moskalik2008CoAst.157..343M,Moskalik2009MNRAS.394.1649M,soszynski2010optical,Pietrukowicz2013AcA....63..379P,Smolec2016MNRAS.458.3561S,Rathour2021MNRAS.505.5412R, Smolec2023MNRAS.519.4010S} and, more recently, TESS data \citep{Plachy2021ApJS..253...11P,Plachy2025A&A...703A.155P}. The model attempting to explain the origin of these sequences, introduced by \citet{Dziembowski2016CoKon.105...23D}, states that these signals are harmonics of non-radial modes of degree $\ell = 7$ (top sequence), 8 (middle sequence), and 9 (bottom sequence). According to this model, geometric cancellation reduces the observed amplitude of the non-radial mode itself, thereby obscuring its direct detection. Analogues of such periodicities are also present in RR Lyrae pulsators, RRc stars in particular \citep{Olech2009A&A...494L..17O,Moskalik2015MNRAS.447.2348M,Smolec2017MNRASRRL.467.2349S,Jurcsik2015ApJS..219...25J,Netzel2015MNRAS.447.1173N,Netzel2019MNRAS.487.5584N}.

In addition to $\Cepa$ stars, first-overtone classical Cepheids also display another distinct type of additional variability, with a period $\Px$ longer than that of the dominant first overtone, and a period ratio $\Po/\Px$ $\sim$ 0.68 (hereafter referred to as the $\Cepb$ group). Several studies have reported their occurrences \citep{Suveges2018a,Suveges2018b,Smolec2023MNRAS.519.4010S}. The signal typically has an amplitude at the millimagnitude level and often appears coherent, with no apparent variation in phase or amplitude. Similar to the $\Cepa$ group, RR Lyrae stars also manifest the presence of $\Cepb$-group variability \citep{Netzel2015Cep68,Netzel2019MNRAS.487.5584N}. Interestingly, a Kepler RR Lyrae star (KIC~9453114) exhibits the simultaneous presence of signals belonging to both 0.61 and 0.68-type variability \citep{Moskalik2015MNRAS.447.2348M}. The incidence rate of the $\Cepb$ group is estimated to be 1.3\% in the LMC and 0.3\% in the SMC \citep{Smolec2023MNRAS.519.4010S}. Despite these observational constraints, the physical origin of the additional periodicities remains unclear.

Apart from secondary periodicities, Cepheids also undergo long-term changes in their pulsation periods. This occurs as a result of secular evolution during their passage through the instability strip, leading to gradual period increases (during the first and third crossings) or decreases (during the second crossing). Such a phenomenon is recorded by measuring the period-change rate of Cepheids, as done by several studies \citep[e.g.,][]{Berdnikov2000ASPC..203..244B,Turner2006PASP..118..410T,PietrukowiczLMC2001AcA....51..247P,PietrukowiczSMC2002AcA....52..177P,Poleski2008AcA....58..313P,Rodriguez-Segovia2022MNRAS.509.2885R,Csornyei2022MNRAS.511.2125C}. The measured rates have been only broadly consistent with theoretical predictions \citep{Turner2006PASP..118..410T,Neilson2012ApJ...760L..18N,Anderson2014A&A...564A.100A,Anderson2016A&A...591A...8A,Miller2020ApJ...896..128M,Espinoza2022MNRAS.tmp.2545E}, with scope for improvement through the inclusion of additional physics (e.g., core overshooting, rotation, or mass loss). However, even secular period changes have at times been found to exhibit deviations from this expected behaviour (e.g.\ S~Vul, Y~Oph; \citealp{Csornyei2022MNRAS.511.2125C}), including sudden period shifts, cyclic modulations, and stochastic variations \citep[e.g.,][]{Parenago1956PZ.....11..236P,Berdnikov2000NewA....4..625B,Poleski2008AcA....58..313P,Csornyei2022MNRAS.511.2125C,Rathour2025A&A}. These irregularities suggest additional physical processes influencing Cepheid pulsations beyond standard evolutionary effects. 

Departures from the above-mentioned secular period changes usually occur on shorter timescales ($\sim$10$^{4}$ days) and are referred to as non-linear period changes. A fraction of these non-secular period variations (and consequently the \oc diagram) are cyclic in nature and are attributed to binarity. Such phenomena manifest due to the light-travel time effect \citep[LTTE;][]{Irwin1952ApJ...116..211I,Irwin1959AJ.....64..149I}, which has been explored for Cepheids \citep{Rathour2024A&A} and RR Lyrae stars \citep[e.g.,][]{Hajdu2015MNRAS.449L.113H,Hajdu2021ApJ...915...50H} to identify binary candidates. The remaining fraction of non-linear period changes is substantially more complex and is most often manifested as irregular behaviour in the \oc diagram. Despite decades of observational signatures, the physical mechanism responsible for this puzzling behaviour has not yet been established. A number of possible explanations have been put forward, including binary interactions \citep{Szabados1992ASPC...32..255S}, discrete mixing events in the core \citep{Sweigart1979A&A....71...66S}, magnetic fields \citep{Stothers1980PASP...92..475S}, and episodic mass loss \citep{Laskarides1974Ap&SS..27..485L,Koopmann1994ApJ...423..380K}. However, none of these explanations can account for the observed timescales.

In our previous works, we investigated period changes in the context of the LTTE, indicative of binarity (\citealt{Rathour2024A&A}, hereafter \PaperI), and the irregular behaviour of period changes of yet unknown origin (\citealt{Rathour2025A&A}, hereafter \PaperII). However, in these studies, in order to ensure a clean sample of purely single radial-mode Cepheids, we intentionally excluded stars that showed additional low-amplitude signals. This sample cut was motivated by the concern that additional low-amplitude periodicities could influence the stability of the radial mode. This resulted in hundreds of Cepheids being rejected from the analysis, which may have biased the calculation of the incidence rate of irregular period changes across the metallicity environment and pulsation mode. In this study, our goal is to investigate whether the presence of additional low-amplitude periodicities affects or correlates with irregular period changes in Magellanic Cloud Cepheids. Therefore, we focus explicitly on such candidates, using OGLE photometry to examine two groups of additional periodicity: $\Cepa$ and $\Cepb$. The motivation for focusing on these specific types of low-amplitude periodicities is because a considerable sample of Cepheids exhibiting these periodicities has been identified in the OGLE survey \citep{Smolec2023MNRAS.519.4010S}.

Sect.~\ref{sec: Methodology} introduces our OGLE \citep{udalski2015ogle} sample and the methodology used for both frequency analysis and \oc diagram construction. Thereafter, in Sect.~\ref{sec: Results}, we compile our findings, probing the connection between the additional low-amplitude periodicities of Cepheids and their \oc curve shapes, followed by a discussion in Sect.~\ref{sec: Discussion} and our concluding remarks in Sect.~\ref{sec: Conclusions}.

\section{Methodology}
\label{sec: Methodology}

\subsection{Photometry data}
\label{subsec: Photometry data}

Our data are similar to those presented in \PaperI\ and \PaperII, extended by the most recent observations from the ongoing OGLE-V survey phase. The photometry consists of homogeneous OGLE\footnote{\url{http://www.astrouw.edu.pl/ogle/}} survey photometry spanning several phases: (i) OGLE-III \citep[2001–2009;][]{Udalski2008AcA....58...69U,soszynski2008optical,soszynski2010optical}, which also contains OGLE-II \citep[1997–2000;][]{Udalski1997AcA....47..319U,Udalski1999AcA....49..223U,Udalski1999bAcA....49..437U}; and (ii) OGLE-IV \citep[2010–February 2025;][]{udalski2015ogle,Soszynski2015AcA....65..297S,Soszynski2017aAcA....67..103S,Soszynski2019AcA....69...87S}, which also contains high-cadence, non-public post-2022 data from the extended OGLE-IV survey; and (iii) the ongoing OGLE-V phase, which began in February 2025 and continues the high-cadence monitoring of the Magellanic Clouds. During the OGLE-IV extended and OGLE-V observations, the survey's primary focus was on detecting short-duration microlensing events, which resulted in an unprecedented observational cadence over approximately three years (see high-cadence survey details in \citealp{Mroz2024ApJ...976L..19M}). The extended OGLE-IV and ongoing OGLE-V survey covers only the central fields of the Magellanic Clouds, comprising five fields in the LMC and four in the SMC. Overall, this provides more than two decades of photometric coverage of the Small Magellanic Cloud (SMC) and Large Magellanic Cloud (LMC) Cepheids. Our analysis was restricted to the \textit{I}-band data, which provide substantially denser temporal coverage than the \textit{V}-band observations. 

We note that blending by nearby sources is not expected to significantly affect our \oc analysis, as a constant or slowly varying blend alters the light-curve amplitude but introduces no significant time-variable phase shift, while a variable blend of nearly matching period is highly improbable and would likely have been flagged in our frequency analysis.

\subsection{Sample and data preparation}
\label{subsec: Sample and data preparation}

As mentioned earlier, in this study we investigate period changes in Cepheids with two types of low-amplitude periodicity, the $\Cepa$ and $\Cepb$ groups. The largest reported sample of Magellanic first-overtone Cepheids exhibiting these periodicities was compiled by \citet{Smolec2023MNRAS.519.4010S}. We note that fundamental-mode Cepheids do not show this form of variability but instead exhibit another kind of low-amplitude modulation \citep[e.g.,][]{Smolec2017MNRAS.468.4299S,Smolec2023MNRAS.519.4010S}. We start with the sample of \citet{Smolec2023MNRAS.519.4010S} for the $\Cepa$ group, consisting of 291 LMC and 225 SMC Cepheids, for an overall sample of 516 candidates. For the $\Cepb$ group, we start with a sample of 28 Cepheids, comprising 23 from the LMC and 5 from the SMC. Overall, this constitutes a starting total sample of 544 Cepheids.

After obtaining the data, the initial step is merging the photometry from all OGLE survey phases. Following an instrument upgrade between the OGLE-III and OGLE-IV phases \citep{udalski2015ogle}, combining the data required caution \citep[see Sect.~2.1 of][]{Hajdu2015MNRAS.449L.113H}. We refer to Sect.~3.1 of \PaperI\ or Sect.~2.3 of \PaperII, where we describe in detail the procedure to combine the data, which involves transforming each individual OGLE-phase dataset to an arbitrary intensity scale, followed by an appropriate Fourier-order fit to correct for the shift in the mean intensity, and ending with recovering the dataset on a magnitude scale. Such a method is also employed in similar work on RR Lyrae stars using OGLE photometry \citep{Hajdu2021ApJ...915...50H}.

Similar to our previous works, we have also imposed sample cuts on the Cepheids under investigation (see Table~\ref{tab:data sample}). To place all candidates on the same footing, we rejected candidates that were missing either of the OGLE survey phases (Step 1 in Table~\ref{tab:data sample}). From the $\Cepa$ LMC sample, 59 Cepheids lacked OGLE-III data and 1 lacked OGLE-IV data. In the SMC sample, 7 lacked OGLE-III data and 1 lacked OGLE-IV data. This resulted in the rejection of 68 candidates overall, consisting of 60 from the LMC and 8 from the SMC. Similarly, in the $\Cepb$ sample, we rejected 6 in the LMC and 1 in the SMC, leading to a total of 7 rejected candidates, all of which were missing OGLE-III survey-phase photometry. Additionally, 20 Cepheids in total were rejected during the subsequent filtering procedure (Step 2 in Table~\ref{tab:data sample}) for various reasons: (i) consecutive seasons with sparse sampling, leading to a gap in the \oc diagram (fewer data points in a season yield an \oc measurement with a high error, so these points were omitted); (ii) pulsation periods close to integer values, leading to a sampling bias that in certain cases affects the resulting Fourier template; and (iii) a systematic magnitude shift across individual OGLE seasons.

\subsection{Analysis}
\label{subsec: Analysis}

In this study, we deal with the detection and characterisation of irregular period changes in Cepheids with low-amplitude periodicities; hence, the two backbone procedures of our work are frequency and \oc analysis. We briefly describe both. The frequency analysis was conducted on our Cepheid sample in a similar way to previous studies on comparable targets and databases \citep[e.g.,][]{Netzel2015MNRAS.447.1173N,Smolec2016MNRAS.458.3561S,Smolec2017MNRAS.468.4299S,Rathour2021MNRAS.505.5412R,Smolec2023MNRAS.519.4010S}. 

As an initial step, we compute the Discrete Fourier Transform of the time series to identify the dominant frequency. The subsequent stage of the analysis involves applying the standard consecutive pre-whitening method \citep[e.g.,][]{Moskalik2009MNRAS.394.1649M}. For Cepheids with comparable cadences in both the standard OGLE-IV and extended datasets, the data are merged and analysed collectively. When the extended OGLE-IV data differ significantly in cadence or coverage, we conduct manual frequency analyses on the OGLE-IV and extended datasets separately, following the same methodology as above. This distinction is necessary due to their differing observational characteristics, particularly variations in sampling cadence, which influence detection limits \citep[see Fig.~5 in][]{Smolec2017MNRAS.468.4299S} and affect frequency identification. Additionally, the signals under investigation may not be strictly coherent (especially $\Cepa$-group candidates); small intrinsic variations over long timescales can affect their stability. Analysing separate datasets allows for more effective pre-whitening.

To detect non-evolutionary period changes in Cepheids with low-amplitude periodicities, we compute their \oc diagrams. The \oc diagram is a powerful tool for studying timing variations in both eclipsing binaries and pulsating stars \citep[for the canonical methodology, see][]{Zhou1999PBeiO..33...17Z,Sterken2005ASPC..335....3S}. An alternative approach to analysing period variations is the Hertzsprung method \citep{Hertzsprung1919AN....210...17H}, which focuses on phase shifts rather than the precise timing of extrema. Instead of measuring individual maxima or minima, this technique tracks how the observed light curve shifts in phase relative to a constructed template. The phase shift over time provides a direct measure of period variations. A modified version is adopted, as described in Sect.~3.2 of \PaperI\ or Sect.~2.5 of \PaperII, and used in other similar studies \citep{Hajdu2015MNRAS.449L.113H,Hajdu2021ApJ...915...50H,Rodriguez-Segovia2022MNRAS.509.2885R}.

\begin{table*}
\caption{Cepheid sample distribution.}
\label{tab:data sample}
\centering\footnotesize
\begin{tabular}{lc|cc|ccc}
\hline
\hline
\textbf{Type} & \textbf{Initial OGLE sample} & \multicolumn{2}{c|}{\textbf{Sample Cleaning Steps}} & \multicolumn{3}{c}{\textbf{Final Sample}} \\
\hline
& from \cite{Smolec2023MNRAS.519.4010S} & \textbf{Step 1} & \textbf{Step 2} & \textbf{Class 1} & \textbf{Class 2} & \textbf{Class 3} \\
\hline
\hline
\textbf{\Cepa\ LMC 1O}  & 291  & 60 & 9 & 27 & 7 & 188 \\
\textbf{\Cepa\ SMC 1O}  & 225  & 8  & 9 & 52 & 14 & 142 \\
\hline
\textbf{\Cepb\ LMC 1O}  & 23   & 6  & 2 & 3 & 5 & 7 \\ 
\textbf{\Cepb\ SMC 1O}  & 5    & 1  & 0 & 1 & 1 & 2 \\ 
\hline
\hline
\textbf{Total}          & 544  & 75 & 20 & 83 & 27 & 339 \\ 
\hline
\end{tabular}
\tablefoot{Step 1: Candidates missing either of the OGLE survey phase data; Step 2: Rejected during final filtering. Classes are defined by \oc morphology: class 1 for constant period (flat \oc); class 2 for linear period change (parabolic \oc); class 3 for irregular period change (see Sect.~\ref{subsec: Classification}).}
\end{table*}

\section{Results}
\label{sec: Results}

\subsection{Justification: Effect of low-amplitude variability}
\label{subsec: Justification: Effect of low-amplitude variability}

Before computing the \oc diagrams and investigating their connection with the low-amplitude periodicities, it is important to quantify the contribution of these periodicities to the overall behaviour of the period changes depicted in the \oc shapes. We note that in \PaperI, for only a few targets, we performed a similar investigation of Cepheids that not only showed a binary signature in the \oc diagram but also contained low-amplitude signals in the frequency spectrum. We analysed such Cepheids within the binary sample and checked whether there was any significant effect on the binary parameters extracted from the \oc diagram when the low-amplitude signals were cleaned, compared with when they were not. The results indicated a negligible effect, as already demonstrated for several Cepheids (see \PaperI, Sect.~4.4). In \PaperII, we rejected from the outset all Cepheids reported to show any kind of low-amplitude variability, in order to investigate only purely single-radial-mode Cepheids. However, in this study, as we work specifically with the low-amplitude-periodicity sample showing non-linear period changes, it is all the more important to test whether these periodicities have an effect on the \oc diagram.

\begin{figure*}
\begin{center}
{\includegraphics[height=3.5cm,width=0.24\linewidth]{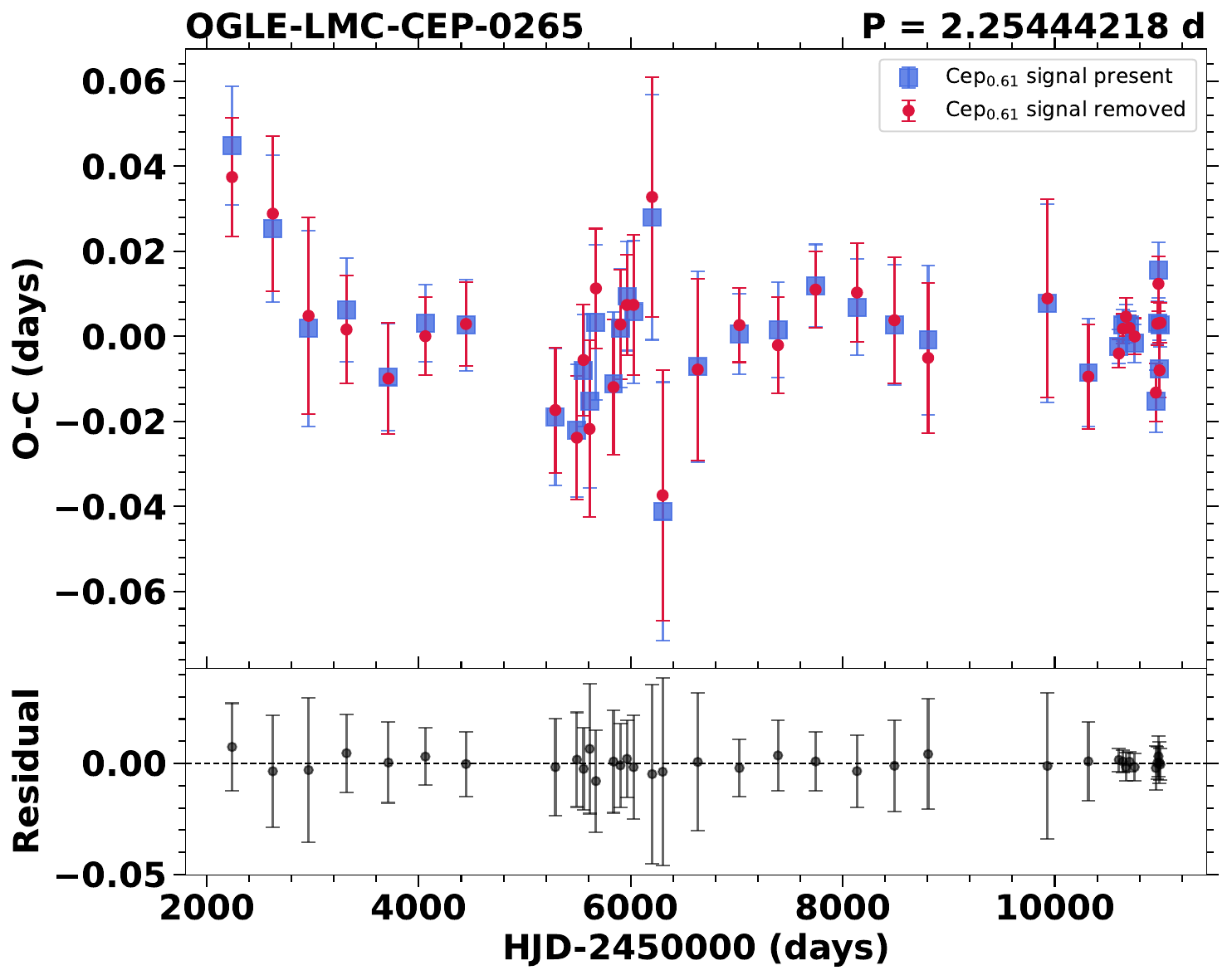}}
{\includegraphics[height=3.5cm,width=0.24\linewidth]{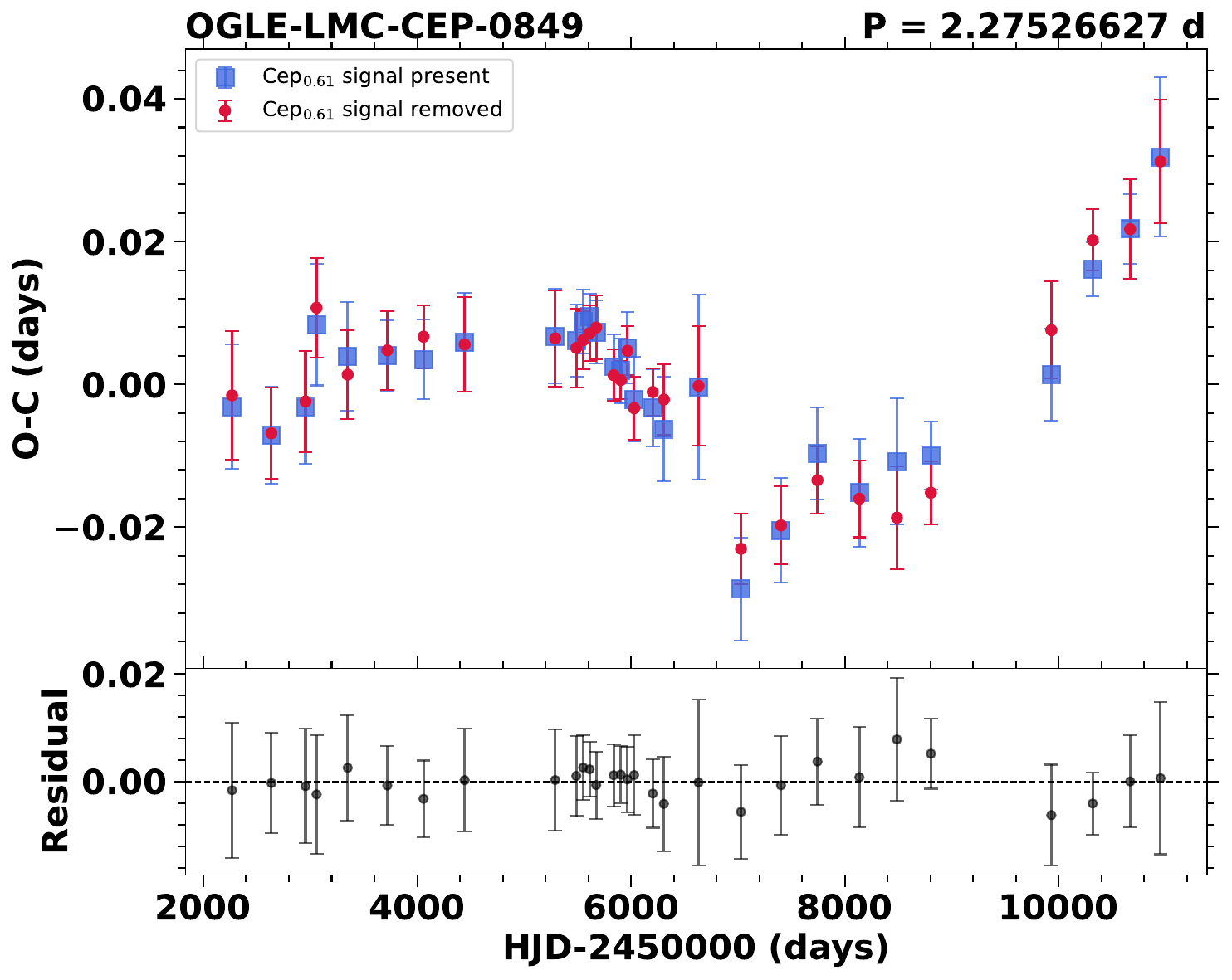}}
{\includegraphics[height=3.5cm,width=0.24\linewidth]{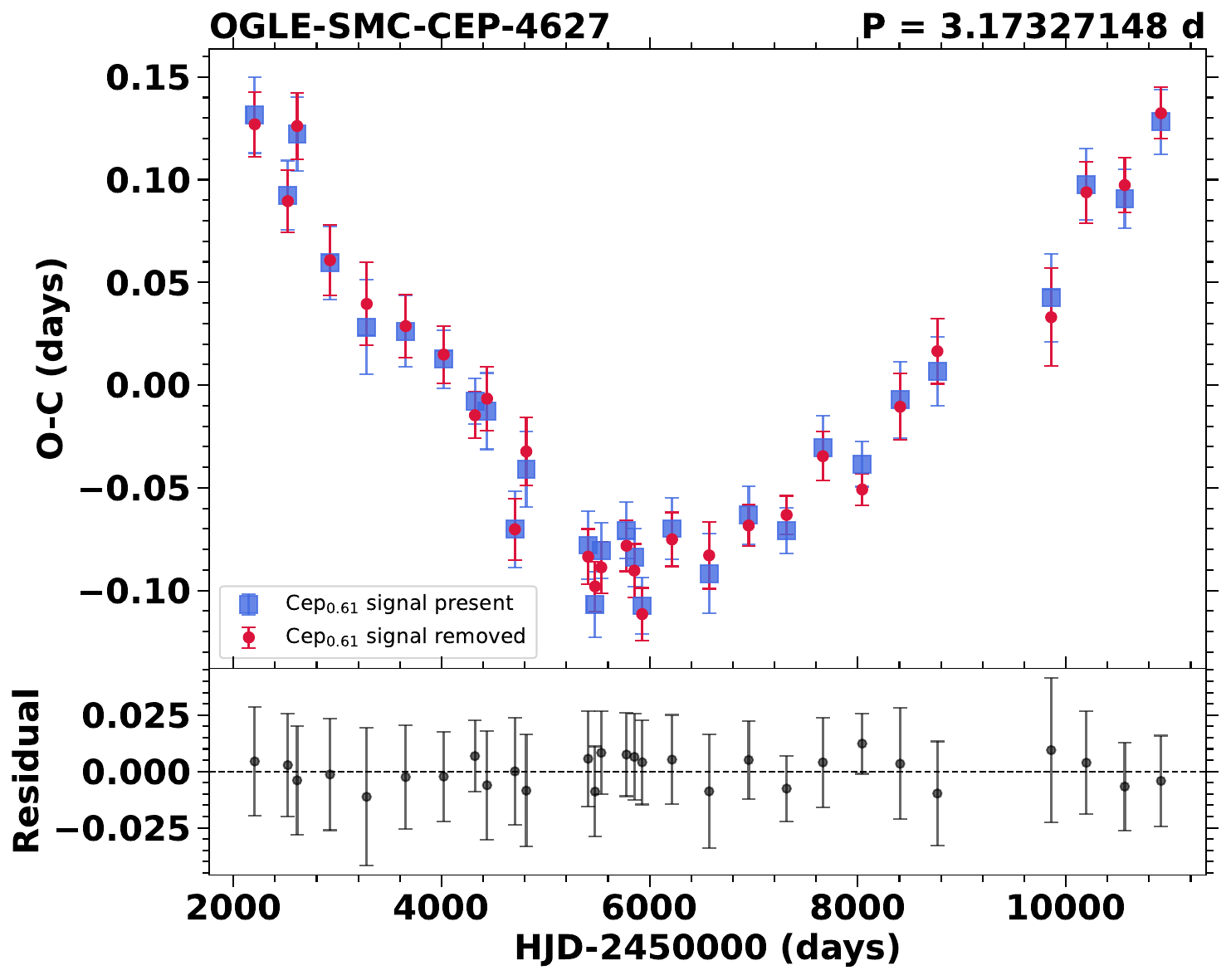}}
{\includegraphics[height=3.5cm,width=0.24\linewidth]{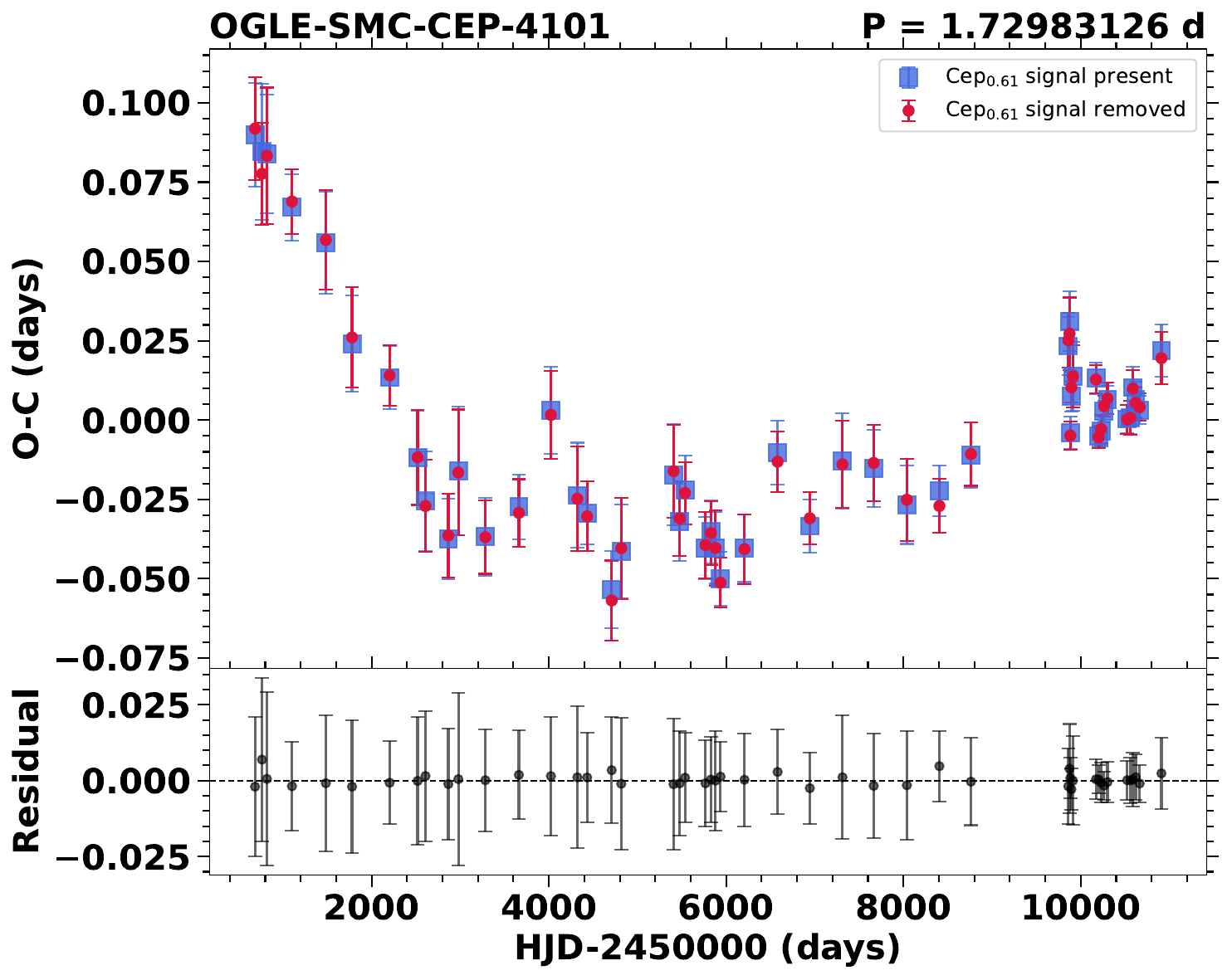}}
{\includegraphics[height=3.5cm,width=0.24\linewidth]{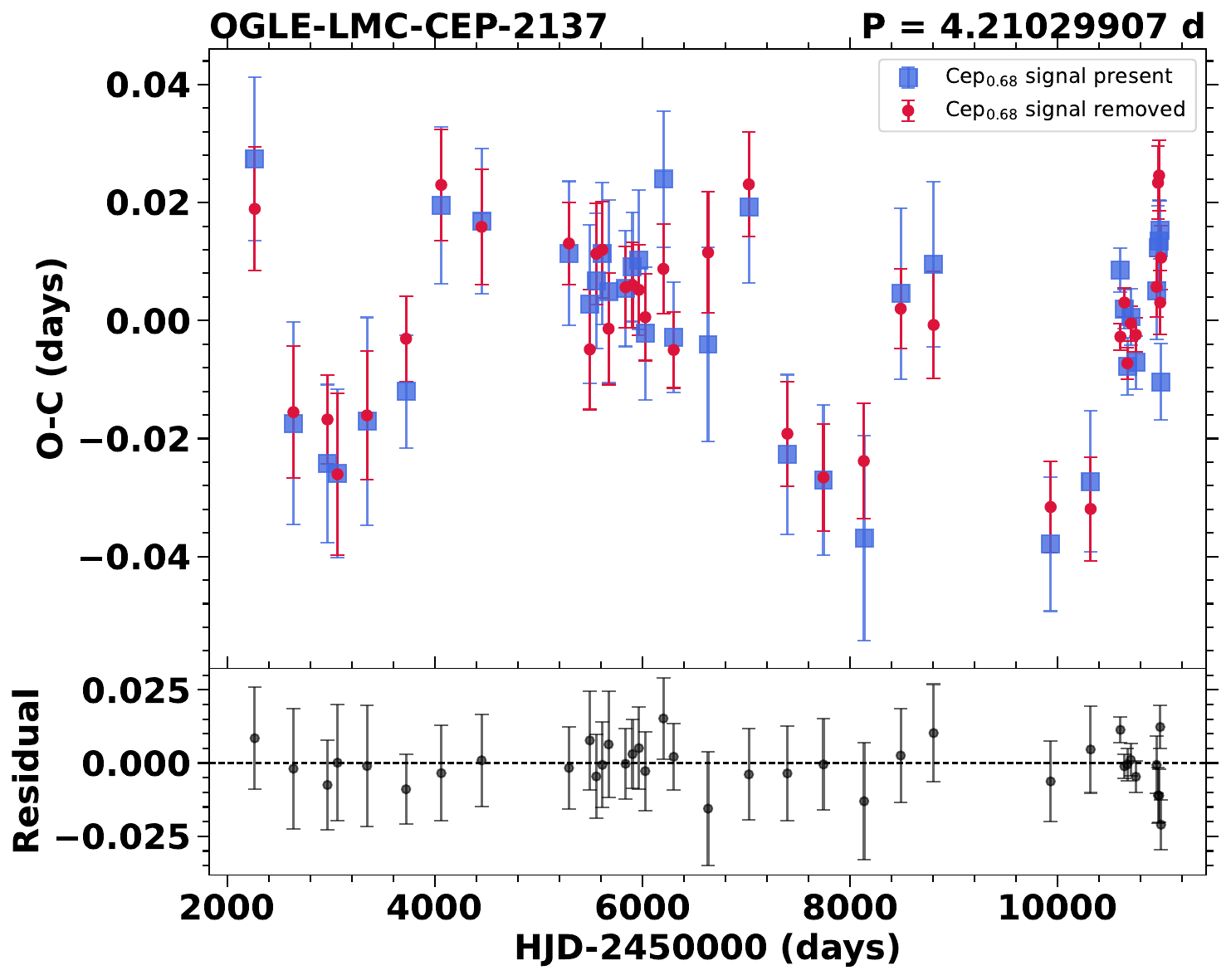}}
{\includegraphics[height=3.5cm,width=0.24\linewidth]{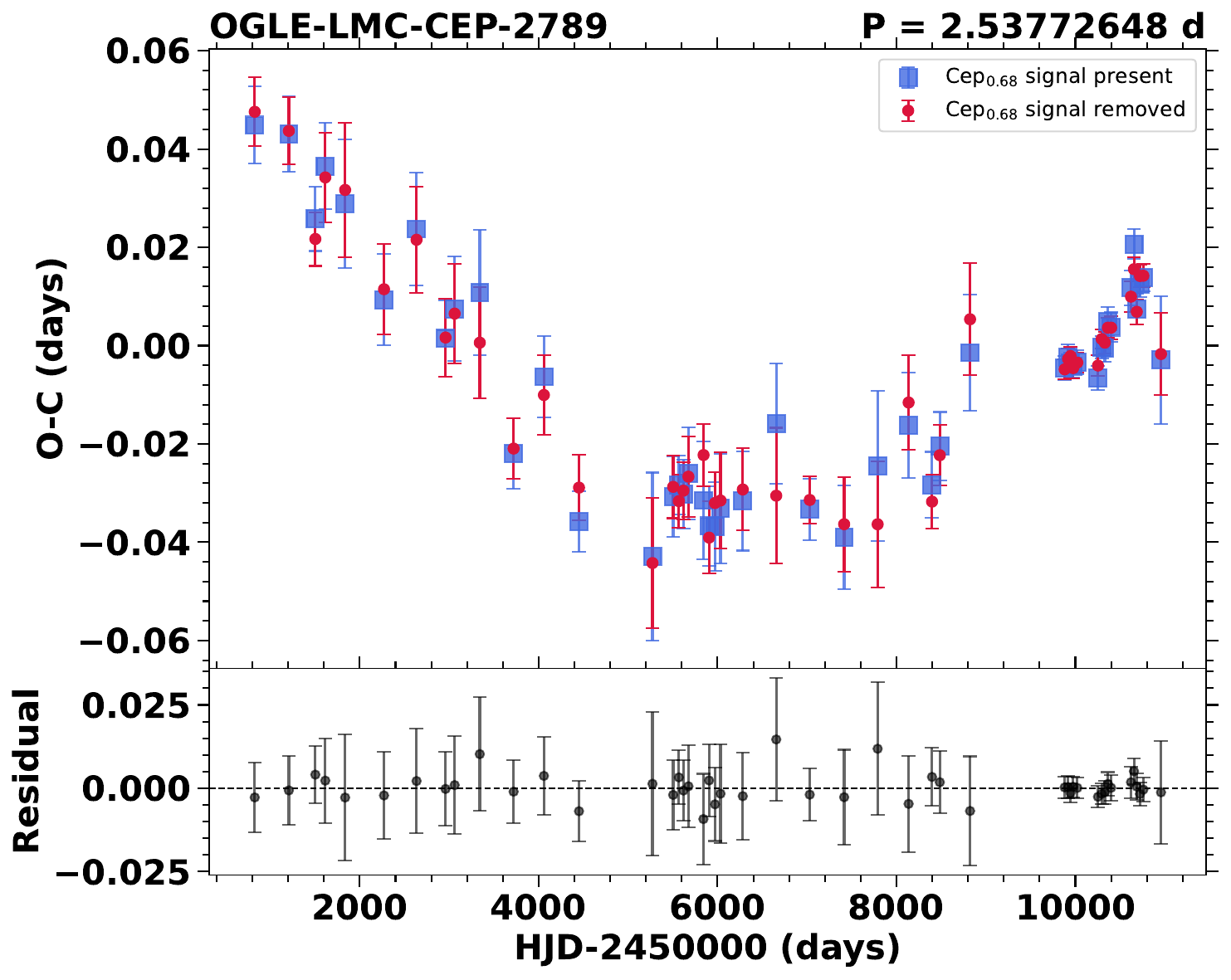}}
{\includegraphics[height=3.5cm,width=0.24\linewidth]{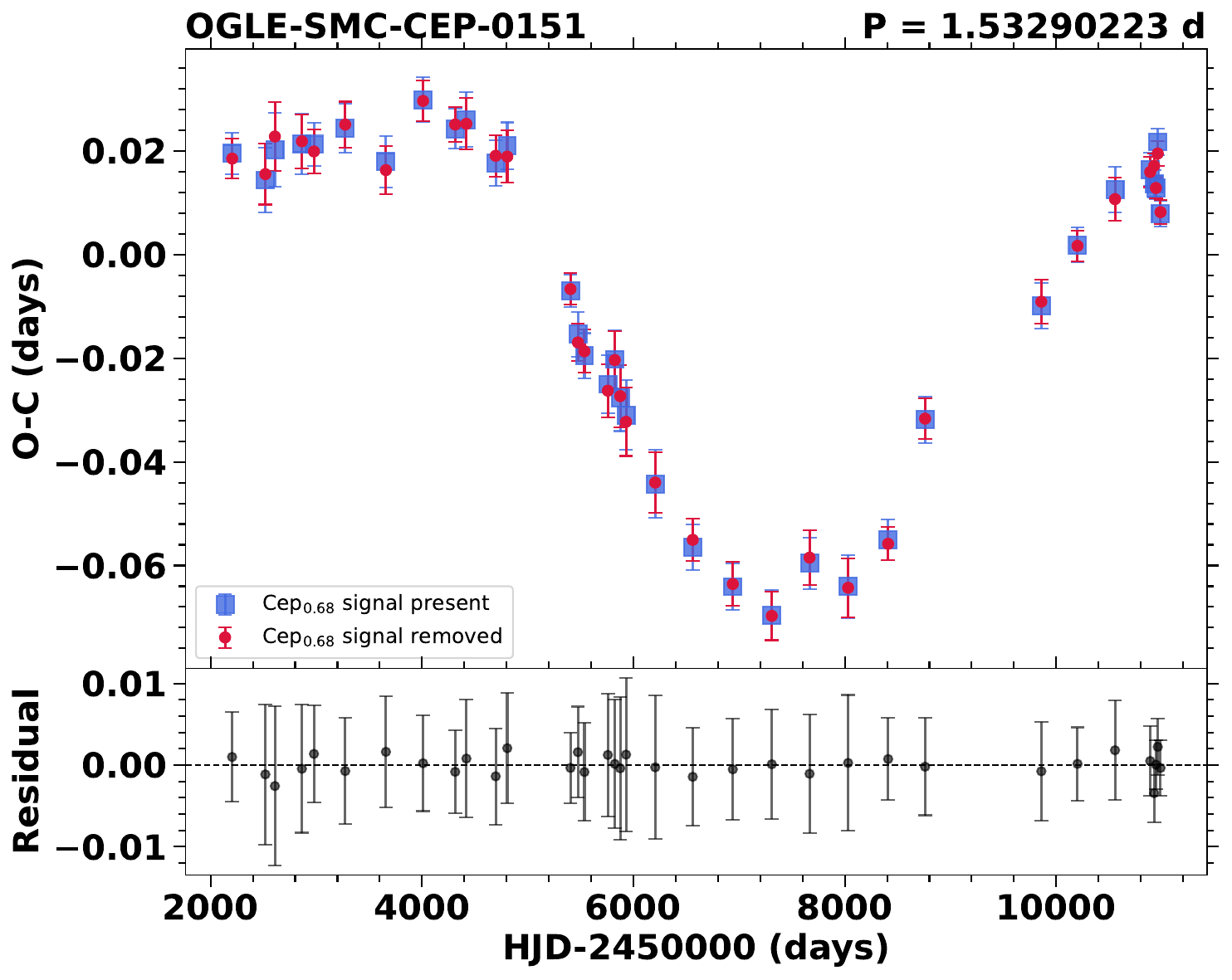}}
{\includegraphics[height=3.5cm,width=0.24\linewidth]{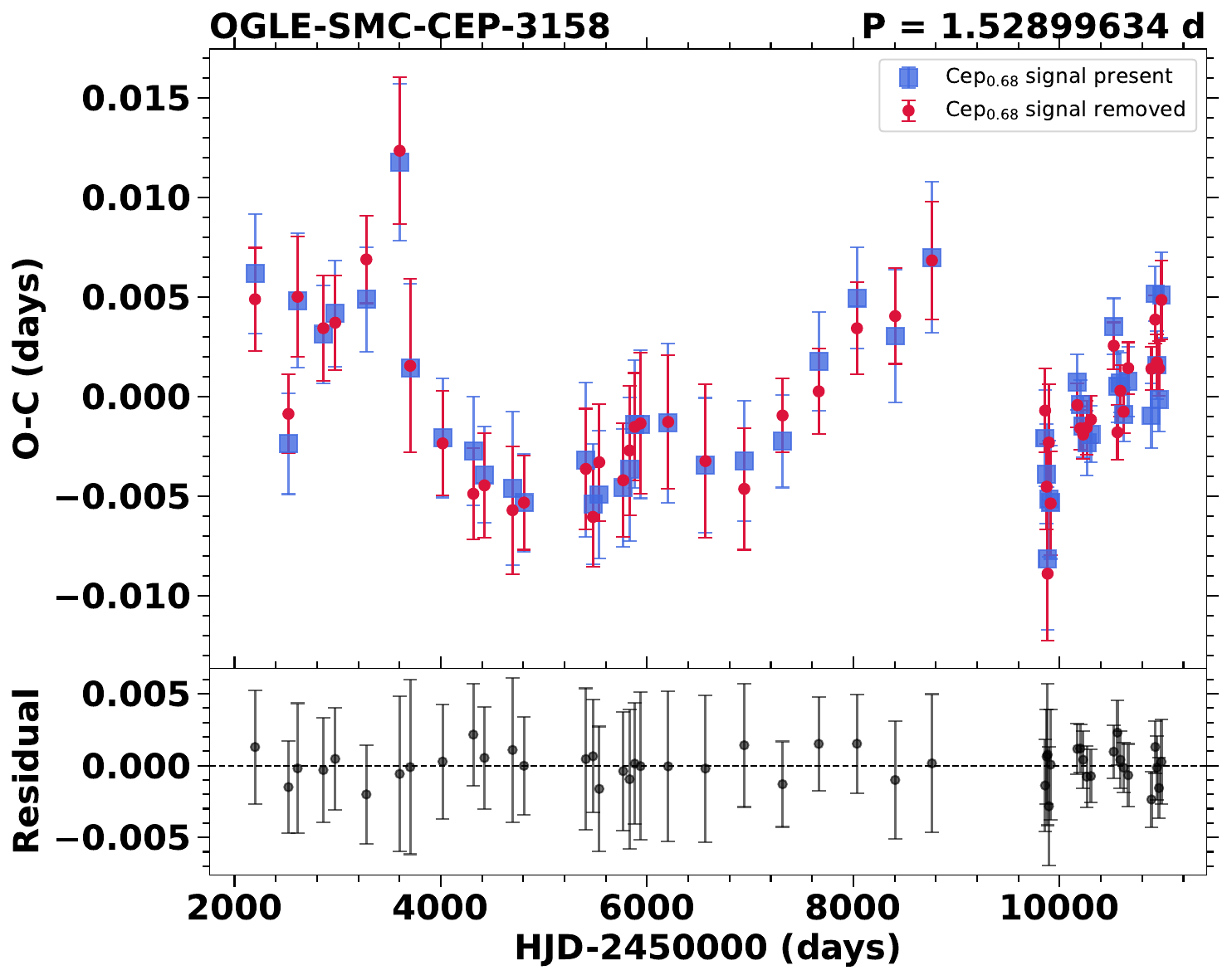}}
\caption{Top panel shows examples of \oc diagrams with \Cepa{} variability present (blue square) over-plotted with the case with the associated variability subtracted (red circle). Bottom panel shows the same for \Cepb{} sample.}
\label{fig:ocplot_justification_examples}
\end{center}
\end{figure*}

To demonstrate this, we selected a total of 8 Cepheids (4 from each MC sample), comprising both variability types ($\Cepa$ and $\Cepb$), with the highest amplitude ratio ($\Ax/\Ao$) among their respective samples. This test case is designed to select only those Cepheids for which the low-amplitude signal is strongest in the sample. In the first case, we computed the \oc diagram in the regular way, as in \PaperI\ and \PaperII. In the second case, we analysed the frequency spectrum of each Cepheid and specifically subtracted the low-amplitude signal before computing the \oc diagram.

In Fig.~\ref{fig:ocplot_justification_examples} we present the differences in the \oc diagrams resulting from the two cases mentioned above. The top panels correspond to the $\Cepa$ sample, and the bottom panels to the $\Cepb$ sample. The blue and red points illustrate the difference between the individual \oc points in the two cases. The marginal differences in both the overall \oc shape and the individual values are clearly depicted. To quantify the differences, we calculate the individual \oc point differences across the time baseline and investigate the residuals. For the $\Cepb$ test candidates, we measure the error-weighted mean of the \oc residuals. This quantity does not exceed $1.3 \times 10^{-4}\,\mathrm{d}$, indicating that the effect does not introduce a significant systematic shift. For the $\Cepa$ sample, the same quantity is at most $12.3 \times 10^{-4}\,\mathrm{d}$. These values correspond to less than 10\% of the typical \oc measurement uncertainty for any individual star in the sample. This confirms that the presence of the additional periodicities does not introduce a statistically significant systematic shift in the period changes.

Hence, we conclude that the removal of the low-amplitude signature from the frequency spectrum has a negligible effect on the resulting \oc diagram. For the remainder of the study, we directly compute \oc diagrams for the Cepheids without individually isolating the low-amplitude signature.

\subsection{Classification}
\label{subsec: Classification}
\begin{figure*}
\begin{center}
{\includegraphics[height=3.5cm,width=0.24\linewidth]{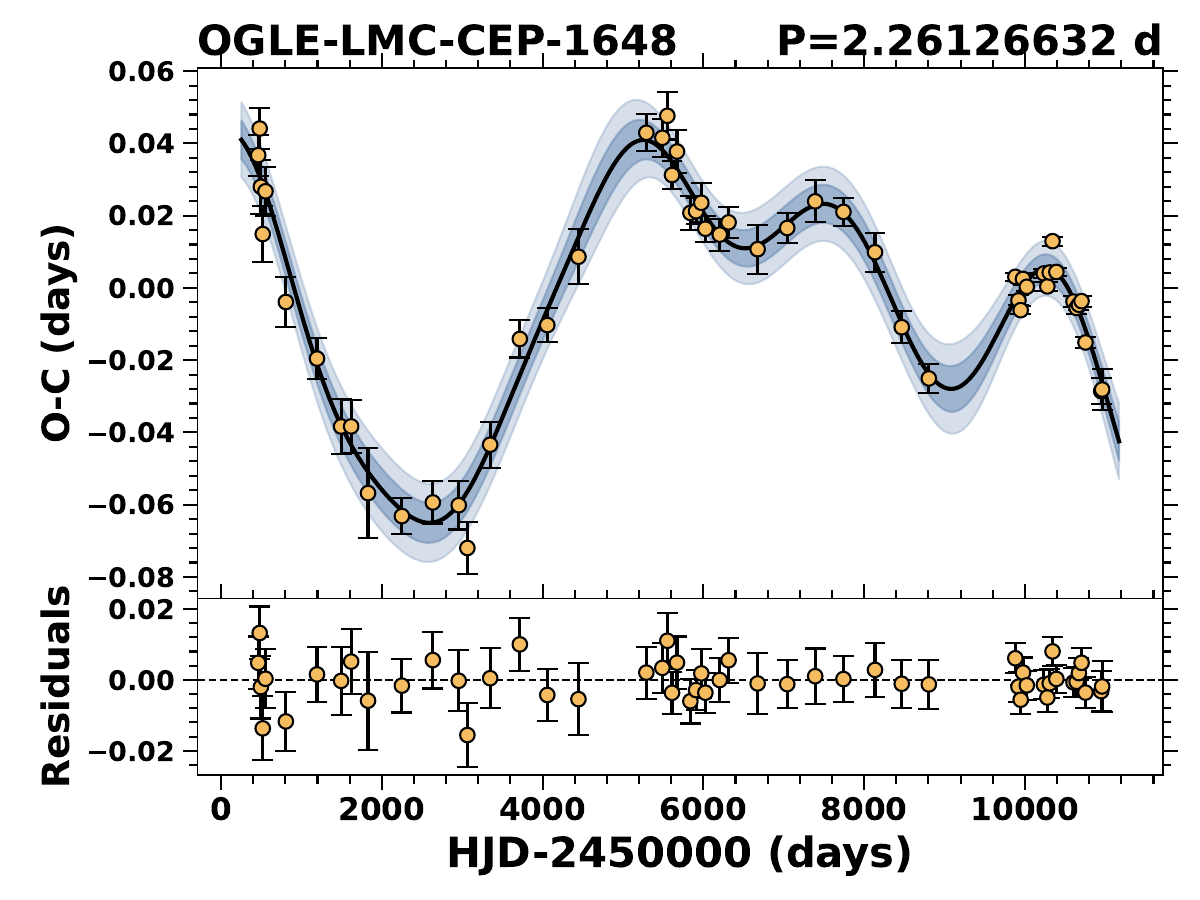}}
{\includegraphics[height=3.5cm,width=0.24\linewidth]{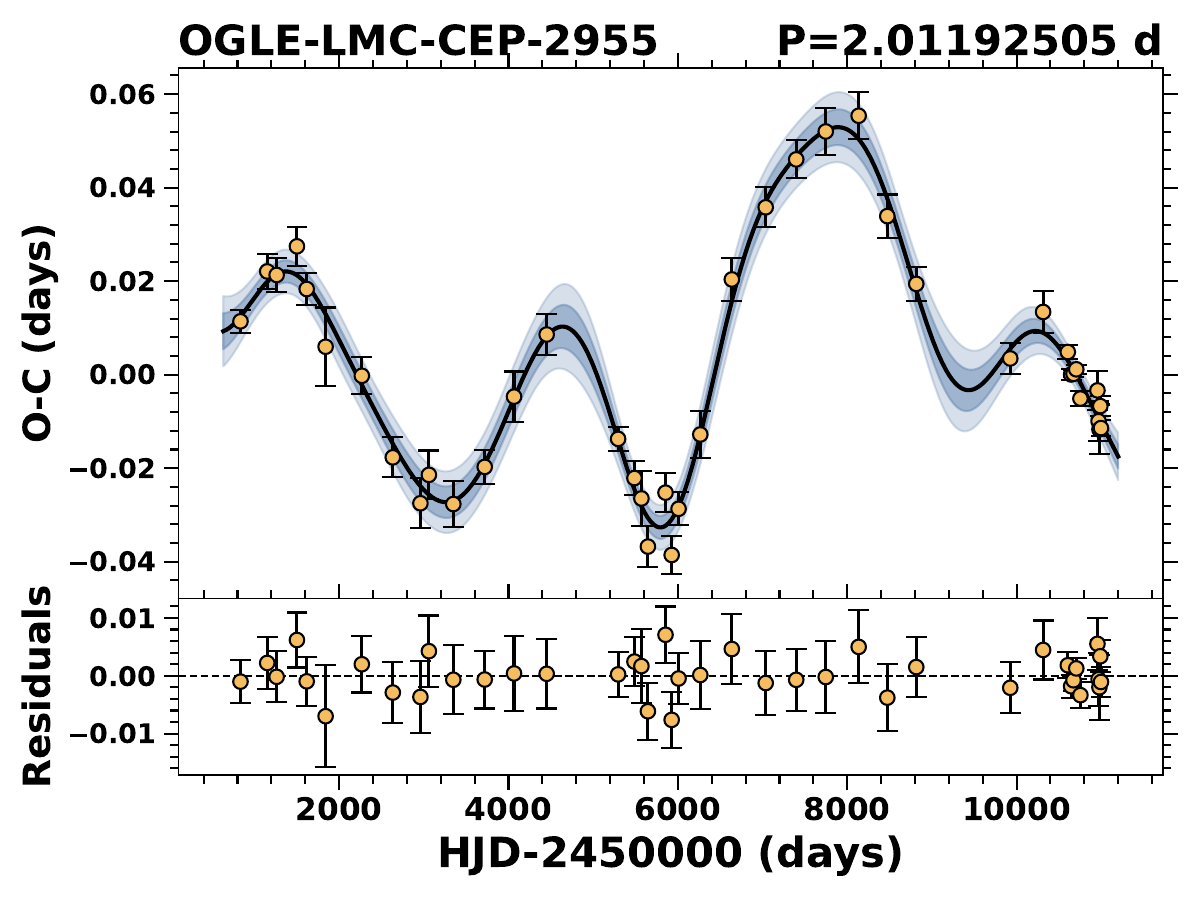}}
{\includegraphics[height=3.5cm,width=0.24\linewidth]{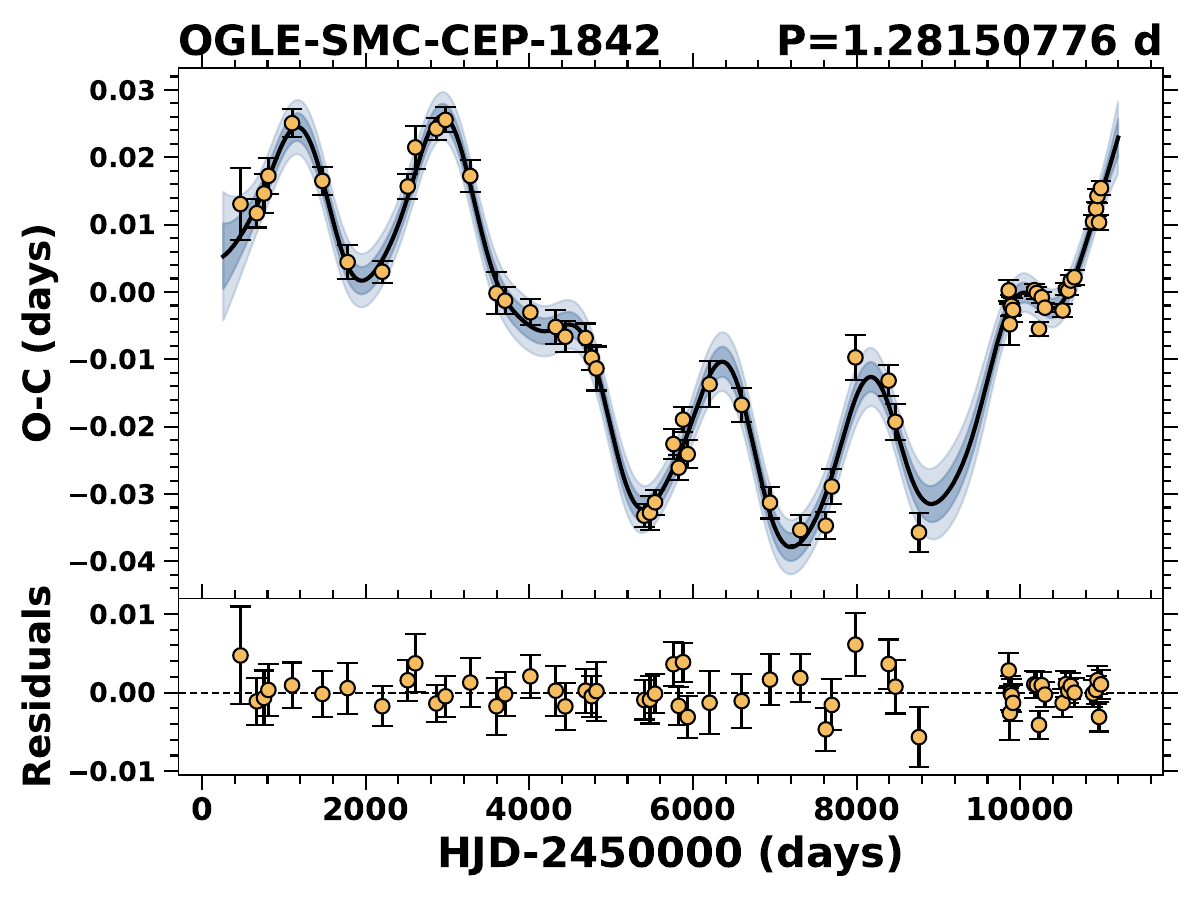}}
{\includegraphics[height=3.5cm,width=0.24\linewidth]{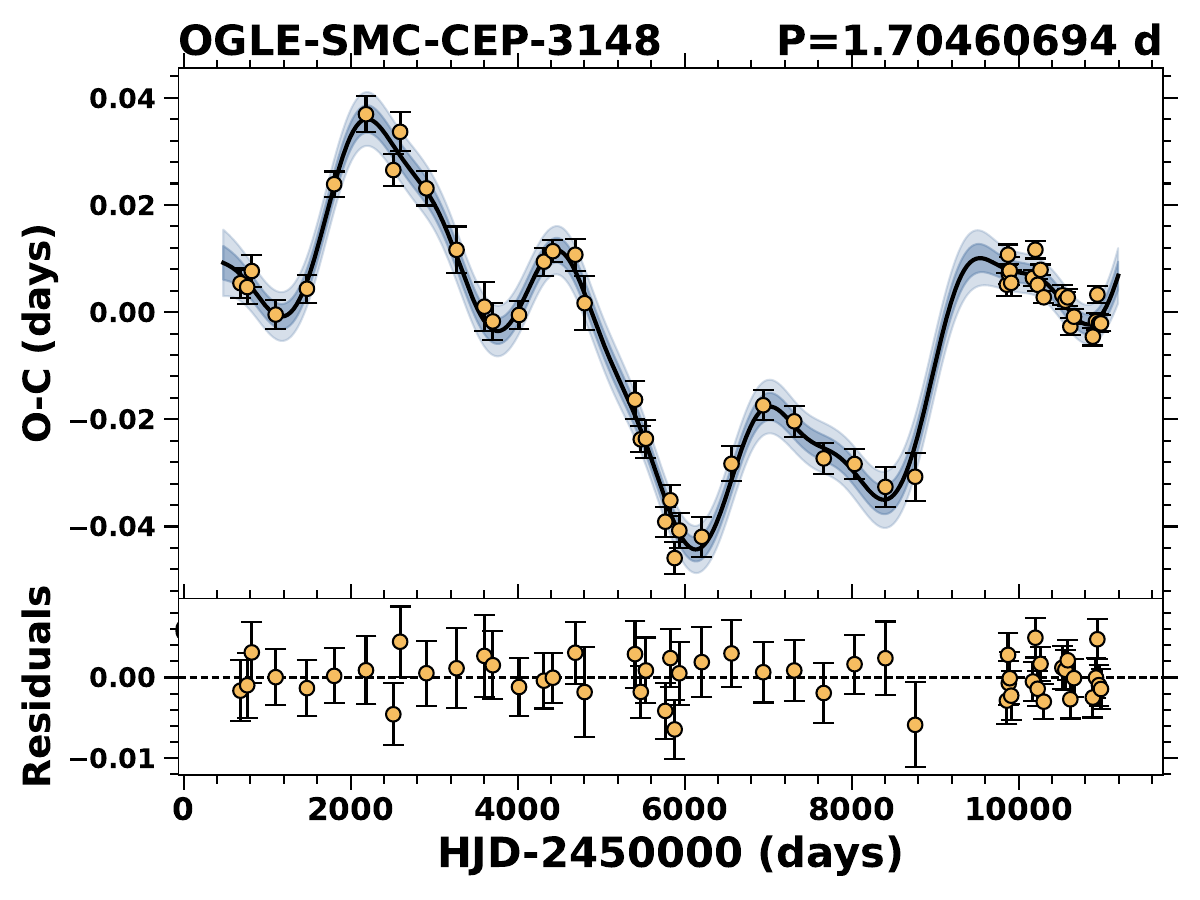}}

{\includegraphics[height=3.5cm,width=0.24\linewidth]{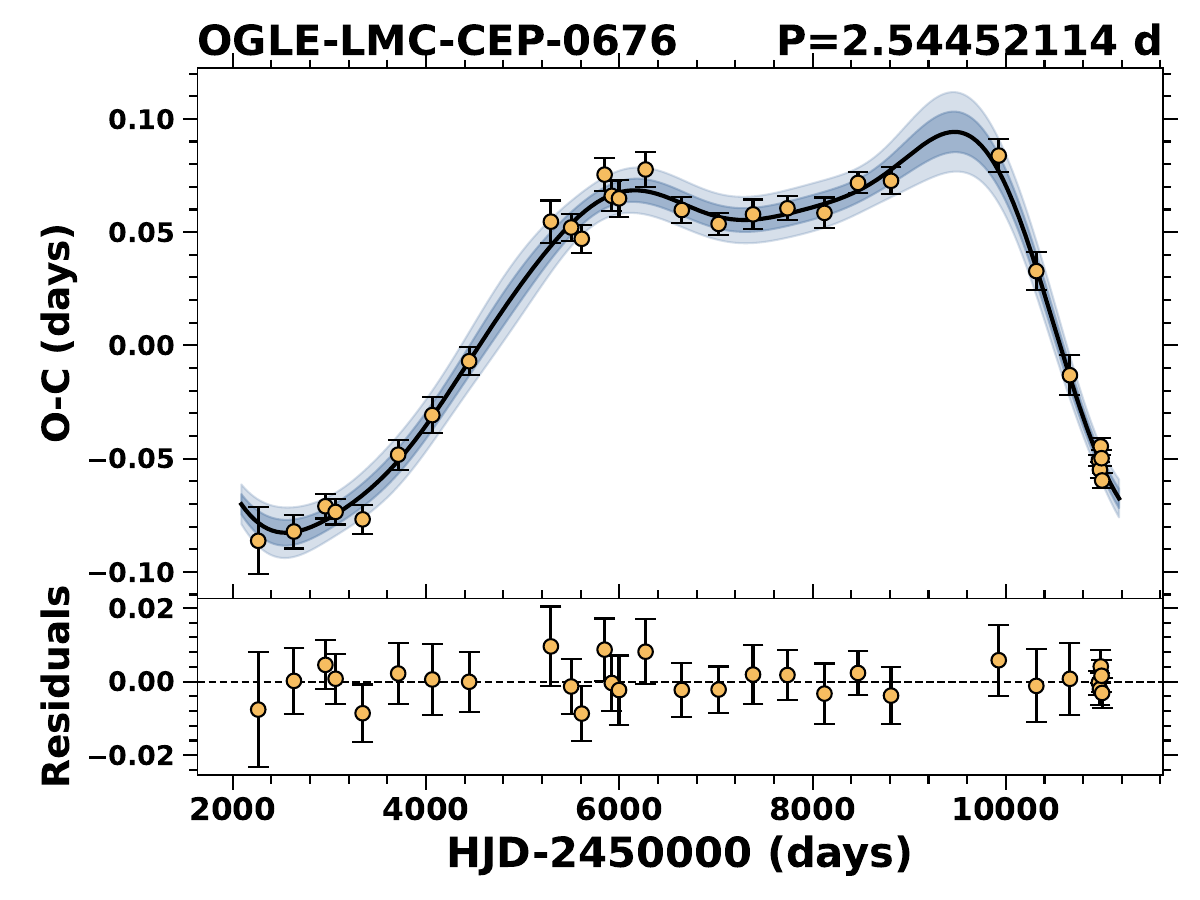}}
{\includegraphics[height=3.5cm,width=0.24\linewidth]{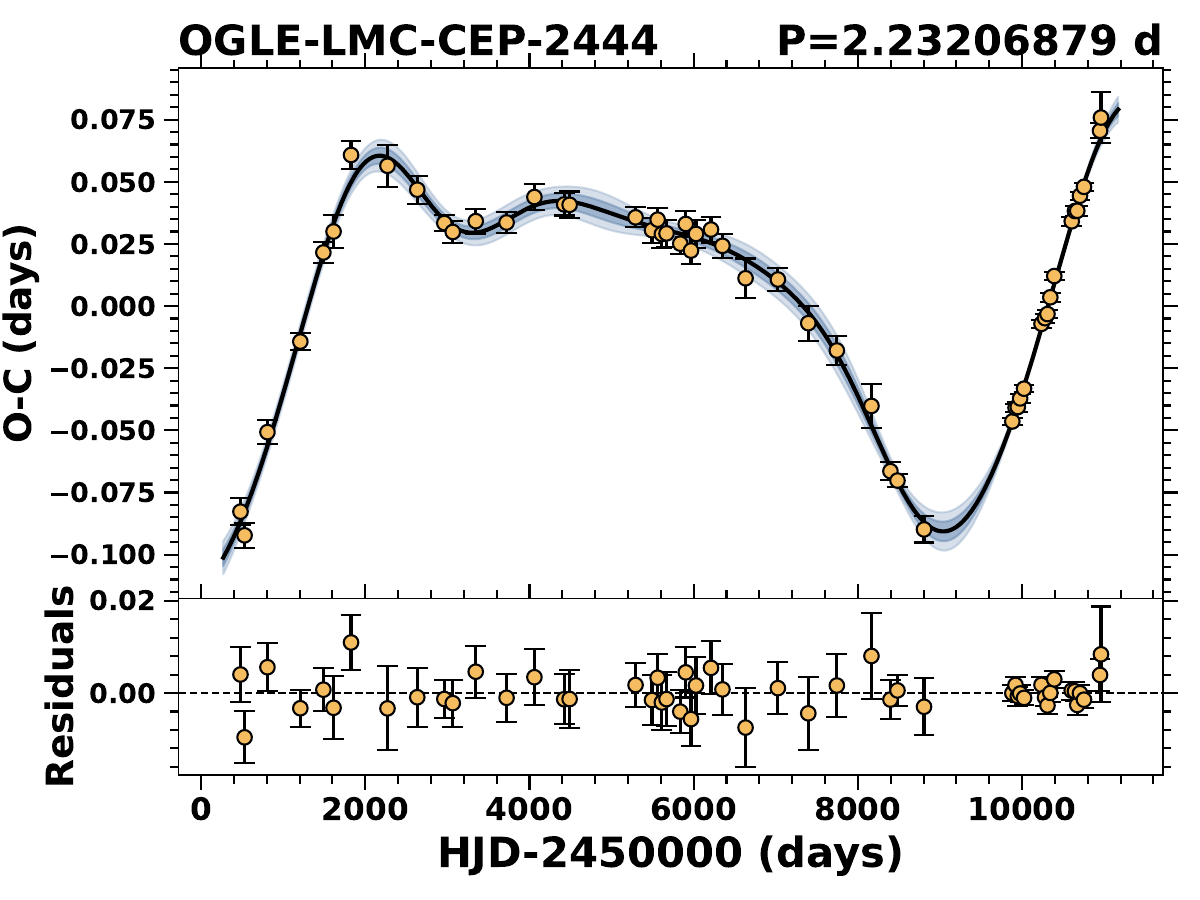}}
{\includegraphics[height=3.5cm,width=0.24\linewidth]{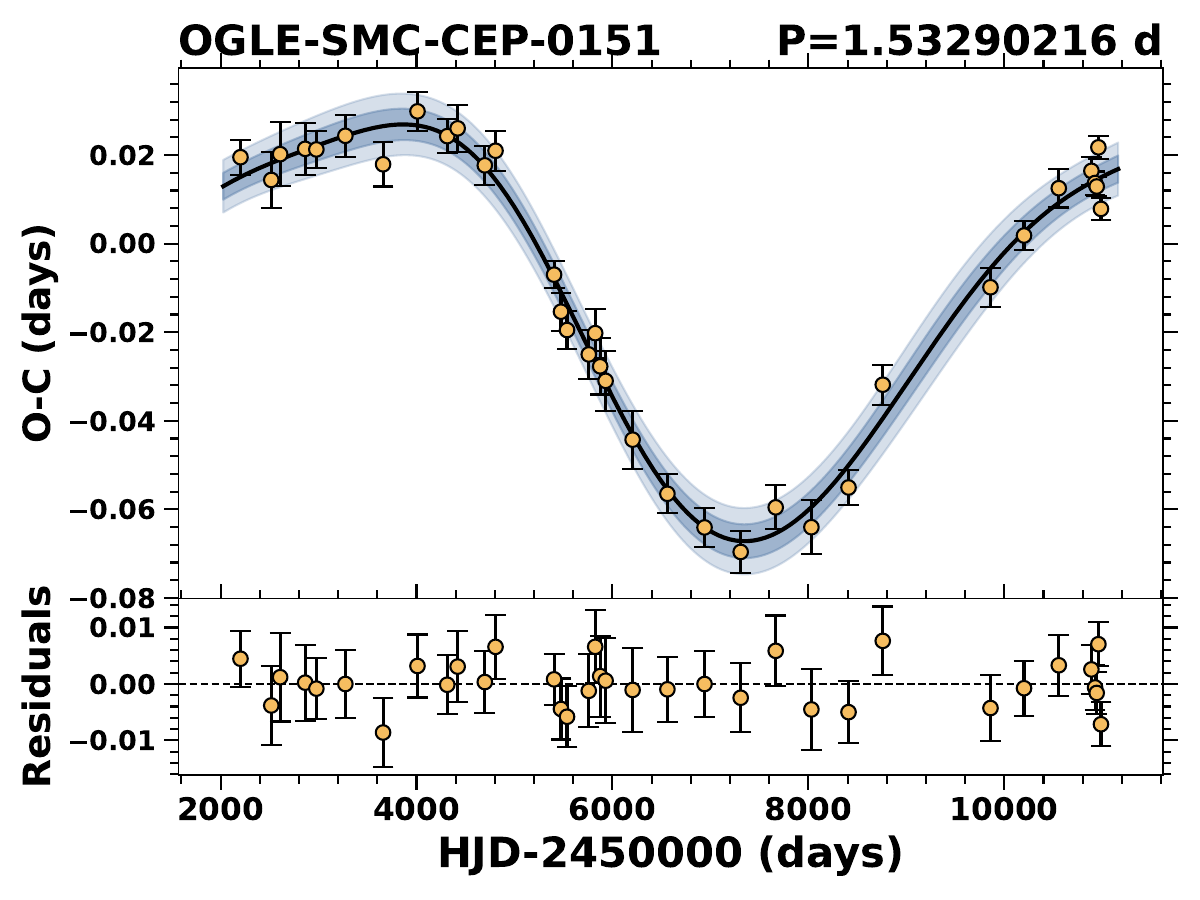}}
{\includegraphics[height=3.5cm,width=0.24\linewidth]{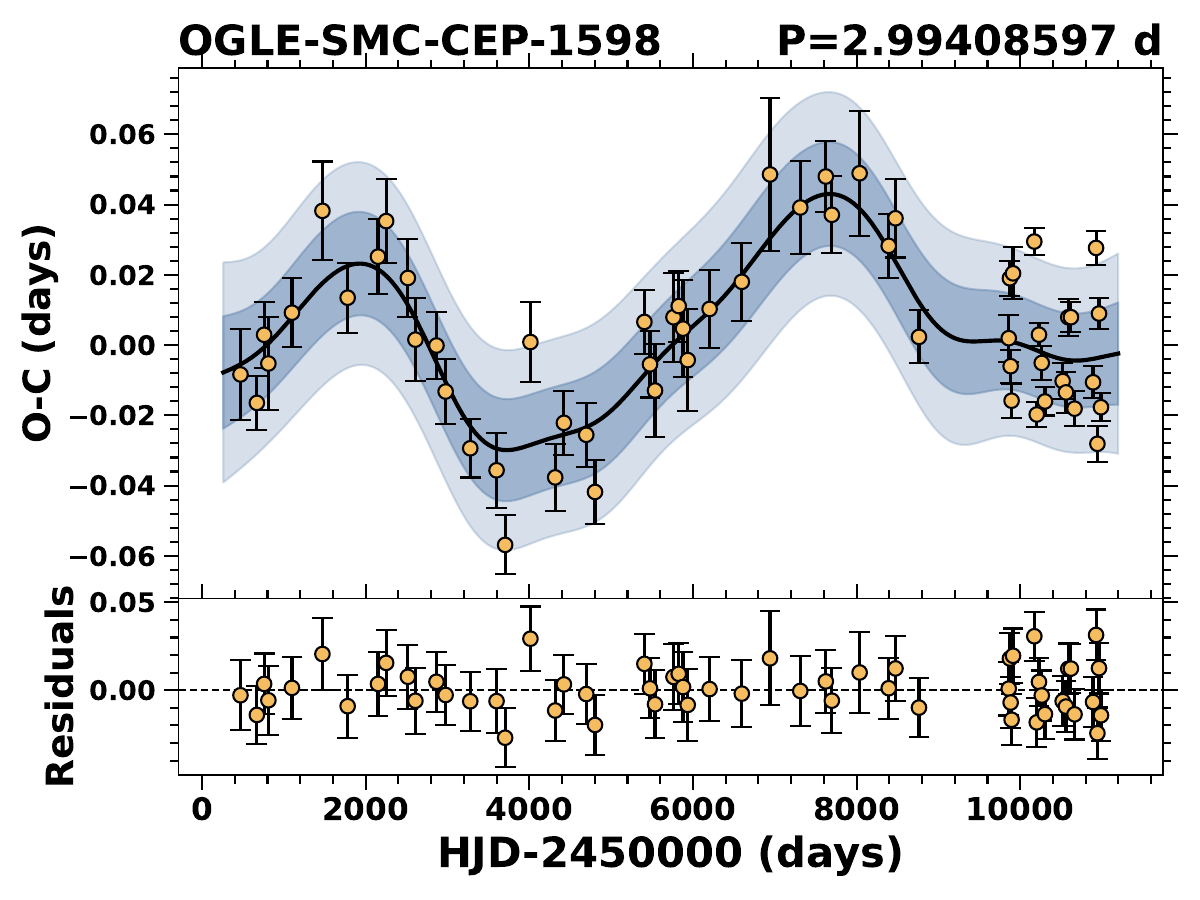}}

\caption{Representative \oc\ diagrams of \Cepa\ (top panel) and \Cepb\ (bottom panel) candidates belonging to class 3. The OGLE identifier and pulsation period are shown at the top of each panel.}

\label{fig:ocplot_sample_class3_examples}
\end{center}
\end{figure*}

We classify the working sample of 449 Cepheids, based on their \oc diagrams, into three main classes, following the methodology described in Sect.~3 of \PaperII. Example \oc diagrams representative of each class are shown in Fig.~\ref{fig:appendix_c1_ocplot_examples} (class 1), Fig.~\ref{fig:appendix_c2_ocplot_examples} (class 2) and Fig.~\ref{fig:ocplot_sample_class3_examples} (class 3) for both \Cepa\ and \Cepb\ samples.

\subsubsection{Class 1 candidates}
\label{subsubsec: Class 1 candidates}

The first are the class 1 candidates, which show a flat \oc variation indicative of negligible period change within the observational baseline, consistent with a constant pulsation period across the OGLE observations. We find 83 class 1 candidates in total, comprising 79 from the \Cepa\ and 4 from the \Cepb\ sample, constituting $\sim$18.5\% of the working sample. Broken down by variability type, this fraction is $\sim$18.4\% among the \Cepa\ group and $\sim$21.1\% among the \Cepb\ group. Additionally, across both galaxies, the LMC shows a notably lower class 1 fraction of $\sim$12.7\% compared to $\sim$25.0\% in the SMC. The complete list of class 1 candidates from this study is presented in Table~\ref{tab:Class_1_list_cepa}.

\subsubsection{Class 2 candidates}
\label{subsubsec: Class 2 candidates}

The second are the class 2 candidates, which exhibit a parabolic \oc shape indicative of a linear period change. However, we caution against a direct evolutionary period-change interpretation. On the $\sim$20~yr OGLE baseline, a parabolic \oc is consistent with secular evolution only for first-crossing candidates, which are characterised by high period-change rates (high enough to be captured within the OGLE time span). Alternatively, a parabolic shape may represent only half a cycle of LTTE signature, pointing to a binary interpretation with an orbital period exceeding $\sim$20~yr. Additionally, a third plausible scenario is of irregular period change with a fluctuation timescale longer than the observational baseline, which can mimic a smooth parabolic shape on a $\sim$20~yr window. In that case, the \oc diagram should gradually reveal its irregular nature in possible future observing seasons. Distinguishing between above-mentioned scenarios requires either a longer observational baseline or radial-velocity follow-up. We recover 27 class 2 candidates in total, comprising 21 from the \Cepa\ and 6 from the \Cepb\ sample, yielding an overall fraction of $\sim$6.0\%. Within the \Cepa\ group this fraction is $\sim$4.9\%, while for the \Cepb\ group it rises to $\sim$31.6\%. The latter is, however, driven by the low-number statistics of the \Cepb\ sample (19 stars) compared to the \Cepa\ sample (430 stars). Again, comparing the two galaxies, the SMC shows a marginally higher class 2 fraction of $\sim$7.1\% compared to $\sim$5.1\% in the LMC, though, given the small numbers involved, this difference is not significant. Among the class 2 additional-variability Cepheids, we find 17 stars ($\sim$63\%) with positive period-change rates consistent with redward evolution across the instability strip (tentative first-crossing candidates). These comprise 14 \Cepa\ stars (5 in the LMC and 9 in the SMC) and 3 \Cepb\ stars (all in the LMC). Ten stars ($\sim$37\%) show negative period change rates, indicating blueward evolution. These comprise 7 \Cepa\ stars (2 in the LMC and 5 in the SMC) and 3 \Cepb\ stars (2 in the LMC and 1 in the SMC). 

Across the combined sample of 17 stars, our first-crossing candidates, the period-change rates ($\dot{P}$) span the range 1.4--26.4 d/Myr, with a median value of 5.75 d/Myr. These values are considerably smaller than those reported by \citet{Rodriguez-Segovia2022MNRAS.509.2885R} for their two LMC first-crossing candidates (OGLE-LMC-CEP-2840: $\dot{P} =$ 30.9 d/Myr and OGLE-LMC-CEP-2132: $\dot{P} =$ 179.8 d/Myr). This difference is expected, since these two Cepheids have pulsation periods longer than all of our tentative candidates. Since the theoretical period-change-rate predictions of \citet{Turner2006PASP..118..410T} show that $\dot{P}$ increases with pulsation period, shorter-period Cepheids are naturally expected to exhibit smaller period-change rates for a given crossing. The list of class 2 candidates, along with their estimated period-change rates, is compiled in Table~\ref{tab:Class_2_list_cepa}.

We note that part of the class 2 variations could be of non-evolutionary origin as well (see the scenarios at the beginning of this subsection), hence the derived rates and fractions given here need caution.

\subsubsection{Class 3 candidates}
\label{subsubsec: Class 3 candidates}

The final are the class 3 candidates (see Table~\ref{tab:Class_3_list_mainpaper}), which are the main focus of this study within our overall low-amplitude-periodicity sample. This class refers to Cepheids showing irregular period changes. There is no known explanation for the origin of this period-change behaviour, as depicted by their \oc diagrams. We find a total of 339 Cepheids in our sample exhibiting irregular period changes (see Table~\ref{tab:data sample}). This sample comprises 330 \Cepa\ and 9 \Cepb\ candidates. The overall fraction of class 3 candidates is $\sim$75.5\%, whereas within each variability subgroup the fractions are $\sim$76.7\% and $\sim$47.4\% for the \Cepa\ and \Cepb\ groups, respectively. Once again, the latter is affected by low-number statistics. The class 3 fraction is markedly higher in the LMC ($\sim$82.3\%) than in the SMC ($\sim$67.9\%).

\begin{figure}
\begin{center}
{\includegraphics[height=10cm,width=0.9\linewidth]{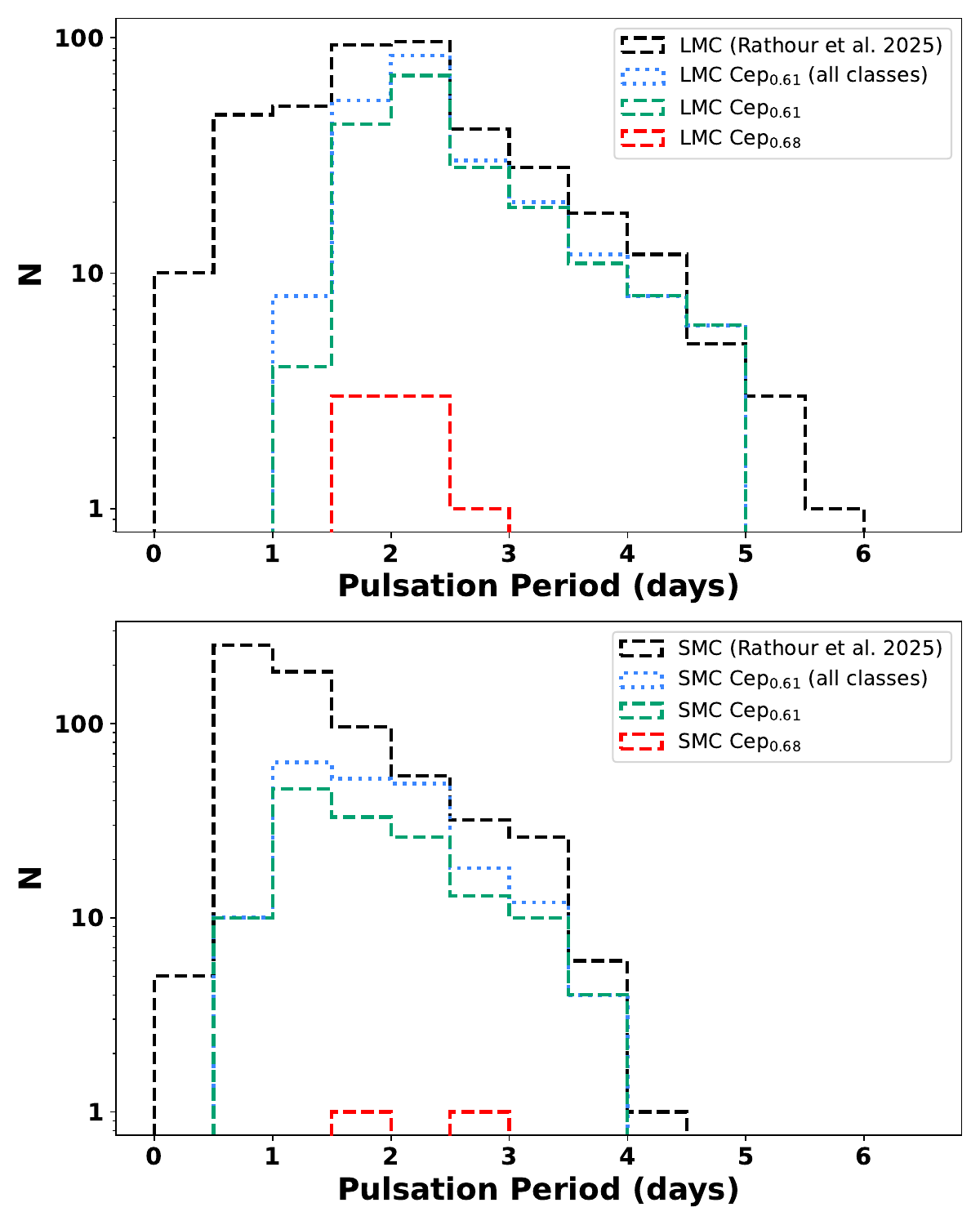}}
\caption{Pulsation period distribution of class 3 candidates in the LMC (top) and SMC (bottom). Black, green, and red dashed histograms correspond to the first-overtone class 3 sample from \PaperII\ and to the \Cepa\, and \Cepb\ class 3 samples of this work, respectively. The blue dotted histogram shows all \Cepa\ stars of the working sample, irrespective of the period change classification. The y-axis is in logarithmic scale.}
\label{fig:Class3_histogram}
\end{center}
\end{figure}

In Fig.~\ref{fig:Class3_histogram} we compare the pulsation-period distribution of our class 3 candidates with that of the first-overtone class 3 sample from \PaperII. In the LMC (top panel), the \PaperII\ class 3 distribution extends from the 0--0.5\,d bin to beyond 5.5\,d and peaks in the broad 1.5--2.5\,d bin ($N=189$). Our \Cepa\ class 3 candidates first appear in the 1--1.5\,d bin ($N=4$) and peak one bin later in the 2--2.5\,d bin ($N=69$), extending out to the 4.5--5\,d bin. The \Cepb\ class 3 candidates in the LMC number only 7 stars across the 1.5--3\,d bins and are too few to describe any distribution. We note that the difference between the \Cepa\ and \PaperII\ distributions is the absence of \Cepa\ stars in the 0--1\,d bins, where the \PaperII\ sample contributes 57 stars, and their near-absence in the 1--1.5\,d bin ($N=4$). Beyond 2.5\,d, both distributions follow a similar declining trend. In the SMC (Fig.~\ref{fig:Class3_histogram}, bottom panel), the \PaperII\ class 3 distribution is strongly concentrated at shorter periods, spanning the 0--4.5\,d range and peaking sharply in the 0.5--1\,d bin ($N=254$) before declining gradually. The \Cepa\ candidates are present but sparse in that dominant bin ($N=10$) and instead peak in the following 1--1.5\,d bin ($N=46$), extending later to the 3.5--4\,d bin. The \Cepb\ class 3 sample in the SMC contains only 2 stars across the 1.5--3\,d bins, and no meaningful distribution can be described. 
In both galaxies, the class 3 candidates follow the period distribution of the overall \Cepa\ working sample (classes 1-3), as shown in Fig.~\ref{fig:Class3_histogram}. The differences with respect to the \PaperII\ distributions simply reflect the period range in which the additional periodicities occur. 

\subsection{Characterisation}
\label{subsec: Characterization}
Following the classification described in Sect.~\ref{subsec: Classification}, we obtain a total of 339 class 3 candidates (LMC: 195 and SMC: 144). These Cepheids simultaneously exhibit additional low-amplitude periodicities and irregular period changes. Subsequently, we characterise their period changes using several statistical tests on the resulting \oc diagrams. Different analyses probe separate features of the period changes. For a detailed methodology, we refer the reader to \PaperII\ (Sect.~4). Here, we provide only a brief summary of each method and then focus on comparing the results for our \Cepa\ and \Cepb\ samples against the larger class 3 sample from \PaperII.

\begin{figure}
\begin{center}
{\includegraphics[height=6cm,width=0.95\linewidth]{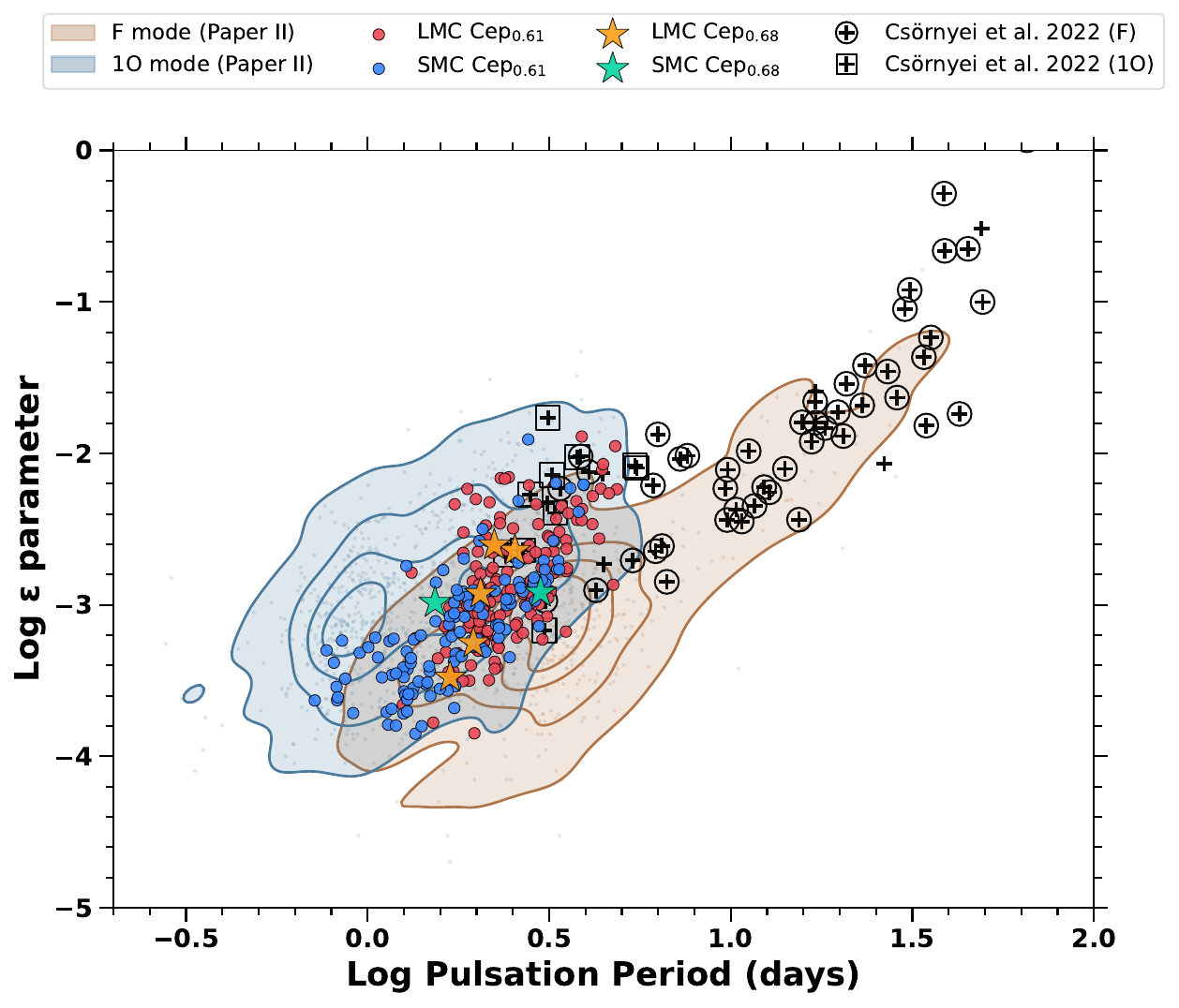}}
{\includegraphics[height=6cm,width=0.95\linewidth]{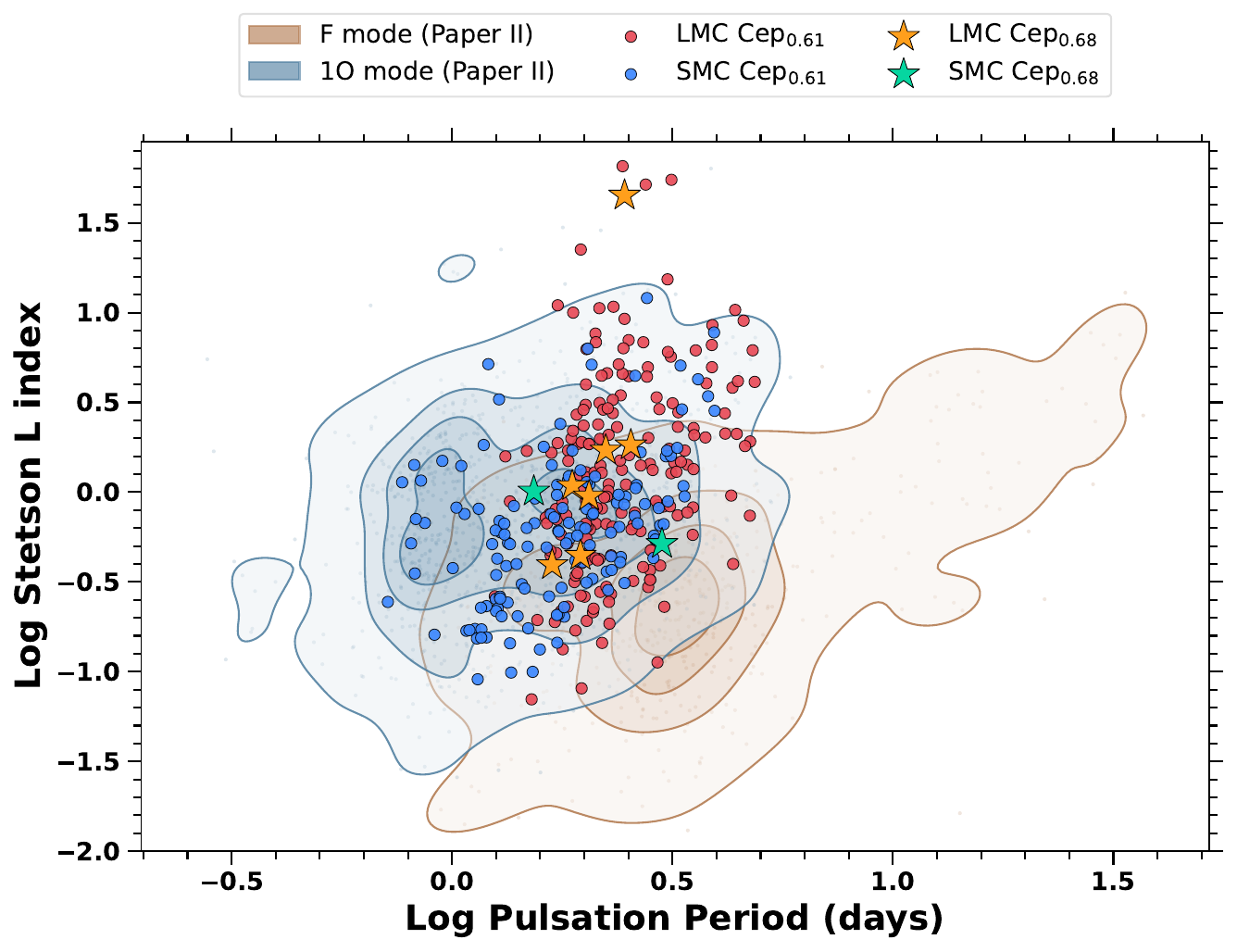}}
\caption{Eddington-Plakidis test based fluctuation parameter $\varepsilon$ (\textit{upper}) and Stetson~$L$ variability index (\textit{lower}) as a function of pulsation period for the class 3 sample of this study. Background density contours show the distributions of fundamental-mode (brown) and first-overtone (blue) Cepheids from \PaperII. Circles mark \Cepa\ and stars mark \Cepb\ candidates from this study, for the LMC (red, orange) and SMC (blue, green). Black plus symbols in the upper panel show the Galactic Cepheid sample of \citet{Csornyei2022MNRAS.511.2125C}: encircled for fundamental-mode, boxed for first-overtone, and bare for uncertain mode classification.}
\label{fig:EP_plot}
\end{center}
\end{figure}

The first method is the Eddington-Plakidis test \citep{Eddington1929MNRAS..90...65E}, which exploits all pairwise differences among the \oc\ measurements rather than their individual values. It quantifies the pulsation-period irregularity by examining how the squared \oc\ differences between all observation pairs, $\langle u^2 \rangle$, grow with their cycle separation $x = |t_j - t_i|/P$. For a star with a stochastically wandering period, this growth is linear,
\begin{equation}
\label{eq:EP_equation}
\langle u^2 \rangle (x) = 2\alpha^2 + x\,\varepsilon^2,
\end{equation}
where $\alpha$ is the measurement noise and $\varepsilon$ is the mean period fluctuation over a pulsation cycle. A significant positive slope in Eq.~\ref{eq:EP_equation} constitutes a detection of period irregularity, which is encapsulated in the $\varepsilon$ parameter.

Figure~\ref{fig:EP_plot} (upper panel) places our results in the context of the full F- and 1O-mode Cepheid populations from \PaperII\ (shown as brown and blue contours, respectively). Since our sample consists only of first-overtone candidates, we split the larger class 3 sample from \PaperII\ by pulsation mode, merging the LMC and SMC stars within each mode, for comparison. The \Cepa\ and \Cepb\ stars from both galaxies lie fully within the 1O-mode contours across the period range they occupy ($\log P_{1\rm O}\approx0.0$--$0.7$\,d). The estimated $\varepsilon$ values span the ranges $\log_{10}\varepsilon \in [-3.85, -1.89]$ for the detected \Cepa\ and $[-3.48, -2.61]$ for the detected \Cepb\ candidates. The median $\log_{10}\varepsilon = -3.01$ for \Cepa\ and $-2.93$ for \Cepb\ is in excellent agreement with the median of the general 1O class 3 Cepheid population from \PaperII\ ($\log_{10}\varepsilon_{\rm med} = -2.93$). No systematic offset in $\varepsilon$ is apparent between the additional-variability stars and the bulk 1O population at corresponding periods. For comparison, we also include the Galactic, predominantly fundamental-mode, Cepheid sample of \citet{Csornyei2022MNRAS.511.2125C}; these occupy longer periods and substantially higher $\varepsilon$ values. The previously established period dependence of the irregularity amplitude is again supported by our sample. Consequently, we note that the presence of \Cepa- or \Cepb-type additional low-amplitude periodicity is not associated with any systematic change (enhancement or suppression) in the $\varepsilon$ parameter.

The second method estimates the Stetson $L$ index \citep{Stetson1996PASP..108..851S}, which measures the temporal correlation of the \oc\ points. When the pulsation period wanders irregularly, the \oc\ diagram drifts steadily and consecutive points cluster on the same side of the mean, which is captured by large $L$-index values. In the ideal scenario, for a strictly periodic star, the \oc\ points should scatter randomly around zero, and hence the $L$ index would remain low. Conversely, a higher $L$ index indicates stronger period irregularity.

Similar to the previous result, here we compare the Stetson-test estimates ($\log L$) with the overall sample from \PaperII. Both \Cepa\ and \Cepb\ stars lie well within the 1O-mode contours across their occupied period range (see Fig.~\ref{fig:EP_plot}, lower panel), and the combined median $\log L = -0.09$ agrees well with the \PaperII\ 1O class 3 median of $-0.15$. Consequently, the current sample distribution also lies systematically above the F-mode contours, reinforcing the empirical finding that first-overtone stars show a higher degree of irregularity than fundamental-mode Cepheids.

\begin{figure*}
\begin{center}
{\includegraphics[height=10cm,width=0.48\linewidth]{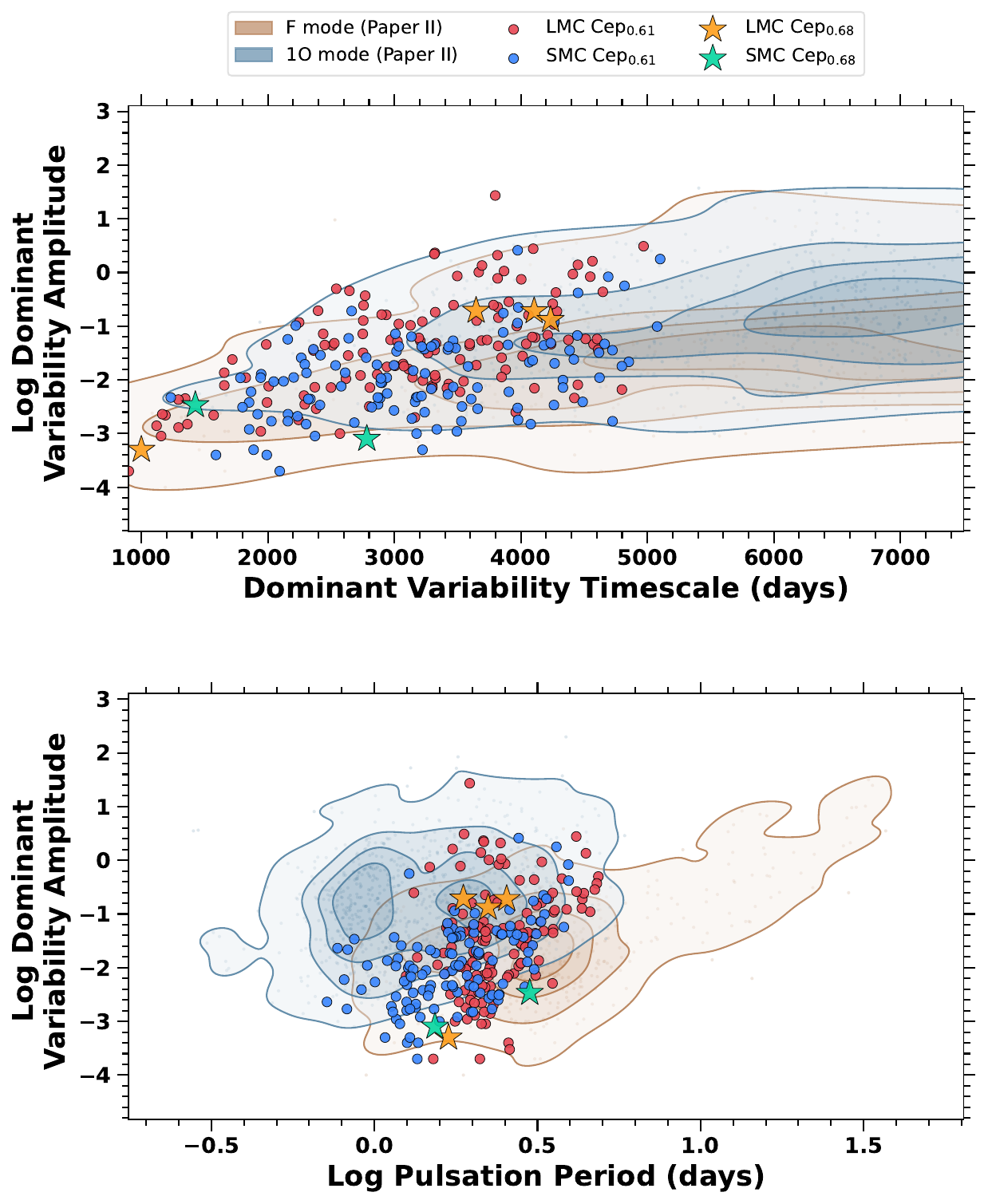}}
{\includegraphics[height=10cm,width=0.48\linewidth]{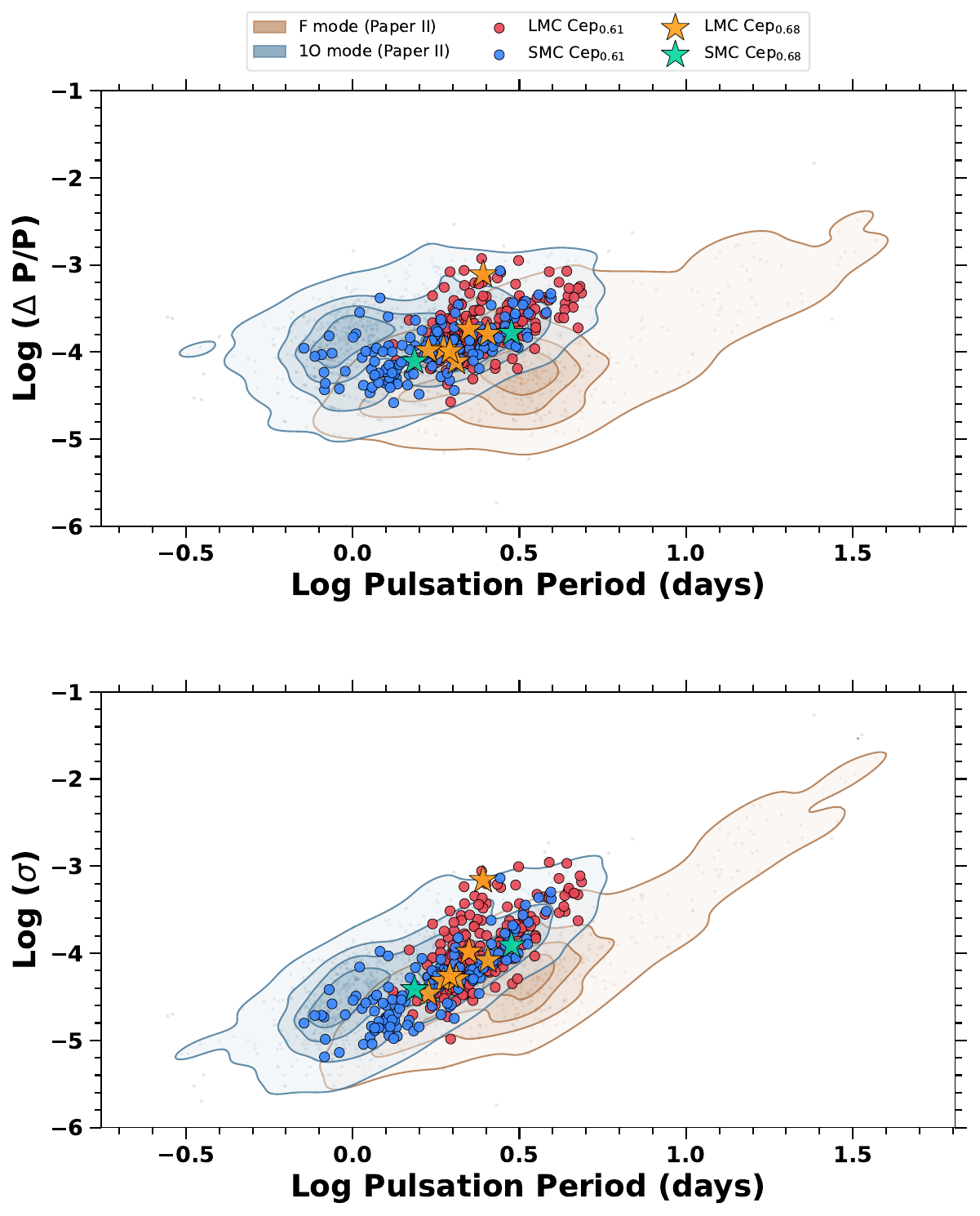}}
\caption{Distribution of log dominant variability amplitude as a function of the dominant variability timescale (upper-left) and pulsation period (lower-left), derived from the wavelet analysis. The right panels show the instantaneous period method derived $\log(\Delta P/P)$ (upper right) and period scatter $\log(\sigma)$ (lower right) as a function of pulsation period. Background density contours and symbol conventions follow from Fig.~\ref{fig:EP_plot}.}
\label{fig:TF_plot}
\end{center}
\end{figure*}

In the third method, to characterise the timescales and amplitudes of the \oc\ variations, we apply a wavelet-based time-frequency analysis, as in \PaperII\ (see Sect.~4.2 therein). In brief, the \oc\ diagram of each star is first fitted with a Gaussian Process Regression (GPR) model, and the resulting profile is analysed with a Continuous Wavelet Transform \citep[CWT;][]{Farge1992AnRFM..24..395F,Torrence1998BAMS...79...61T} using a Morlet wavelet \citep{Morlet1982Geop...47..203M,Torrence1998BAMS...79...61T}. The dominant variability timescale and its associated amplitude are identified from the wavelet power spectrum. In essence, the GPR turns the scattered \oc points into a continuous curve, and the wavelet transform measures how strongly, and on which timescales, that curve fluctuates at each moment of the observing baseline. We caution that the dominant periodicity estimates should be interpreted as characteristic timescales of irregular variability rather than strict periodicities.

In Fig.~\ref{fig:TF_plot} we show the estimated dominant variability amplitude as a function of timescale (upper left panel) and pulsation period (lower left panel). Both \Cepa\ and \Cepb\ stars follow the same amplitude--timescale relation as the larger \PaperII\ class 3 sample, populating its shorter-timescale, lower-amplitude end. The dominant variability timescales of our sample span $\sim$1000--5000\,d, with a median of $\sim$3200\,d, well below the \PaperII\ 1O-mode median of $\sim$6600\,d. The dominant variability amplitudes span $\log_{10}$-amplitude $\in [-3.7, 1.4]$, with a median of $-1.73$\,dex. This lies $\sim$0.9\,dex below the \PaperII\ 1O-mode median ($-0.80$) and close to the F-mode median ($-1.62$), reflecting the systematically shorter dominant timescales of our sample rather than a departure from the class 3 sequence. Overall, the wavelet analysis places our stars on the general class 3 amplitude--timescale relation, at the short-timescale end of the 1O-mode population.

The fourth and final method estimates the instantaneous period $P(t)$ directly from the \oc\ diagram using two parallel approaches, following \PaperII\ (see their Sects.~4.3 and 5.6). In the first, the \oc\ is fitted with a polynomial of order $k$, selected by minimising the Akaike information criterion \citep[AIC;][]{Akaike1974ITAC...19..716A} and the Bayesian information criterion \citep[BIC;][]{Schwarz1978_10.1214/aos/1176344136} values (with a minimum degree of 3, since degrees 1 and 2 correspond to class 1 and class 2 behaviour by construction), and the instantaneous period is derived analytically from its derivative:
\begin{equation}
\label{eq:IP_equation}
P(t) = P_a \left(\frac{\mathrm{d}(\mathrm{O-C})}{\mathrm{d}t} + 1\right),
\end{equation}
where $P_a$ denotes the period used to construct the \oc\ diagram. In the second approach, a GPR model is fitted to the \oc\ diagram, and $P(t)$ is obtained from its numerical derivative. Two summary statistics are recorded: $\Delta P / P$, the normalised peak-to-peak range of the instantaneous-period deviations $P(t) - P_a$, and $\sigma$, the standard deviation of those deviations. Both estimates quantify the amplitude of pulsation-period fluctuations over the observed baseline. Following \PaperII, the GPR-based instantaneous period is the preferred estimator owing to its flexibility in handling the irregular class 3 \oc\ diagrams, while the best-fit polynomial serves as a consistency cross-check.

We present the above quantities in Fig.~\ref{fig:TF_plot}, showing $\log(\Delta P/P)$ (upper right panel) and $\log\sigma$ (lower right panel) as functions of pulsation period. In both panels, our combined class 3 sample lies well within the \PaperII\ class 3 1O-mode contours across the occupied period range, relative to the larger class 3 sample. Our sample has median values of $\log(\Delta P/P)=-3.86$ and $\log\sigma=-4.16$. Both quantities show a statistically significant positive correlation with pulsation period (Spearman $r = 0.62$ and $0.80$, respectively; $p \ll 0.001$), confirming that larger period fluctuations occur at longer pulsation periods, consistent with the trend reported in \PaperII.

\section{Discussion}
\label{sec: Discussion} 

\subsection{Incidence rate}
\label{subsec: Incidence rate}

We now discuss the incidence rates of period-change classes among the low-amplitude variability Cepheid population and place them in the context of the broader Magellanic Cloud Cepheid sample studied by \PaperII. The incidence rates of irregular period changes (class 3) for the first-overtone samples of this work, of \PaperII\,and of the two combined are summarised in Table~\ref{tab:Class_3_incidence_rate}.

\begin{table*}
\caption{Class 3 incidence rates for the first-overtone samples of this work, of \PaperII, and of the two combined.}
\label{tab:Class_3_incidence_rate}
\centering\small
\begin{tabular}{l|ccc|ccc}
\hline\hline
\textbf{Galaxy} & \multicolumn{3}{c|}{\textbf{Class 3 candidates}} & \multicolumn{3}{c}{\textbf{Incidence Rate (\%)}} \\
 & \textbf{Cep$_{0.61}$} & \textbf{Cep$_{0.68}$} & \textbf{Paper II} & \textbf{This work} & \textbf{Paper II} & \textbf{Combined} \\
\hline\hline
\textbf{LMC 1O} & 188 & 7 & 405  & $82.3\pm2.5$ & $64.3\pm1.9$ & $69.2\pm1.6$ \\
\textbf{SMC 1O} & 142 & 2 & 659  & $67.9\pm3.2$ & $70.6\pm1.4$ & $70.1\pm1.4$ \\
\hline
\textbf{Total}  & 330 & 9 & 1064 & $75.5\pm2.0$ & $68.1\pm1.2$ & $69.7\pm1.0$ \\
\hline
\end{tabular}
\tablefoot{All rows refer to first-overtone Cepheids only. Combining both pulsation modes and both galaxies, the overall \PaperII\ incidence rate of $33.5\pm0.7$\% rises to $37.2\pm0.7$\% after adding the new \Cepa\ and \Cepb\ candidates from this work.}
\end{table*}

The class 1 candidates, exhibiting negligible period change with flat \oc\ diagrams, constitute 83 stars in our working sample, corresponding to an overall incidence rate of $\sim$18.5$\pm$1.8\%. The incidence rates within the \Cepa\ and \Cepb\ samples are $\sim$18.4$\pm$1.9\% and $\sim$21.1$\pm$9.4\%, respectively. The difference between the two subtypes is small and not statistically significant (within 0.3$\sigma$), the larger \Cepb\ uncertainty reflecting its small sample. Compared to \PaperII, which reported a class 1 fraction of $\sim$56.3$\pm$0.7\% across all pulsation modes and both galaxies, the considerably lower fraction in the present study is expected, as our sample consists exclusively of first-overtone Cepheids with additional low-amplitude variability. Restricting the \PaperII\ comparison to its first-overtone sample yields a class 1 fraction of $\sim$17.1$\pm$1.0\%, in excellent agreement with the overall incidence rate in this work, confirming a consistent class 1 incidence rate among first-overtone Cepheids regardless of the type of additional variability. Split by galaxy, the class 1 incidence rate in this work is $\sim$12.7$\pm$2.2\% for the LMC and $\sim$25.0$\pm$3.0\% for the SMC, a difference significant at the 3.4$\sigma$ level. This trend is opposite to that in the \PaperII\ first-overtone results, where the LMC incidence rate ($\sim$19.8$\pm$1.6\%) exceeds that of the SMC ($\sim$15.2$\pm$1.1\%). Combining the class 1 candidates from both studies across the first-overtone population yields an updated incidence rate of $\sim$17.4$\pm$0.8\% (350 candidates in total), consistent with the \PaperII\ first-overtone value of $\sim$17.1$\pm$1.0\% and confirming a stable class 1 incidence rate among first-overtone Cepheids in the Magellanic Clouds.

The class 2 candidates amount to 27 stars in our sample, yielding an overall incidence rate of $\sim$6.0$\pm$1.1\%, with sub-type incidence rates of $\sim$4.9$\pm$1.0\% for \Cepa\ and $\sim$31.6$\pm$10.7\% for \Cepb, the latter affected by low-number statistics, as before. \PaperII\ reported a class 2 overall incidence rate of $\sim$10.2$\pm$0.5\% across all pulsation modes, higher than that found here. However, the more meaningful comparison is with the \PaperII\ first-overtone sample, which yields $\sim$14.9$\pm$0.9\%, substantially higher than the $\sim$6.0$\pm$1.1\% found in the present work. Unlike the class 1 comparison, where the two first-overtone rates were in excellent agreement, the class 2 rate in our sample is notably suppressed. Split by galaxy, the LMC and SMC class 2 rates are $\sim$5.1$\pm$1.4\% and $\sim$7.1$\pm$1.8\%, respectively, a small difference, in contrast to the pronounced galaxy-wise asymmetry seen in class 1. Combining the class 2 candidates from both studies across the first-overtone population yields an updated incidence rate of $\sim$12.9$\pm$0.7\% (260 candidates in total), lower than the \PaperII\ first-overtone rate of $\sim$14.9$\pm$0.9\%.

An important empirical constraint on the nature of the irregular period-change phenomenon is its incidence rate. In \PaperII\, we reported a class 3 incidence rate of $\sim$33.5$\pm$0.7\% across both pulsation modes and both galaxies, based on a total of 1585 class 3 Cepheids. Restricted to its first-overtone sample, the appropriate reference for this work, the rate is considerably higher, $\sim$68.1$\pm$1.2\%, as fundamental-mode Cepheids are far less prone to irregular period changes. As highlighted earlier, however, those rates were likely underestimated due to the omission of Cepheids exhibiting additional low-amplitude variability. The present study recovers 339 new class 3 candidates (195 in the LMC and 144 in the SMC), increasing the total Magellanic class 3 sample to 1924 stars. Within our sample alone, the class 3 incidence rate is $\sim$75.5$\pm$2.0\%; separately within the \Cepa\ and \Cepb\ subsamples, the rates are $\sim$76.7$\pm$2.0\% (330 of 430) and $\sim$47.4$\pm$11.5\% (9 of 19), respectively, the latter subsample subject to large uncertainty from low-number statistics. Compared with the \PaperII\ first-overtone rate, our incidence is higher at the $\sim$3.2$\sigma$ level, suggesting that first-overtone Cepheids with low-amplitude variability are moderately more prone to irregular period changes than the general first-overtone population. Split by galaxy, the class 3 incidence rates in this work are $\sim$82.3$\pm$2.5\% for the LMC and $\sim$67.9$\pm$3.2\% for the SMC, a difference significant at the 3.5$\sigma$ level. This is opposite in sense to the \PaperII\ first-overtone trend, where the SMC ($\sim$70.6$\pm$1.4\%) exceeds the LMC ($\sim$64.3$\pm$1.9\%), mirroring the galaxy-wise reversal seen in the class 1 rates. Considering the whole sample of 1O Cepheids, both with and without additional periodicities, the combined incidence rate of class 3 candidates becomes $\sim$69.7$\pm$1.0\% and seems not to depend on environment, with the corresponding rates for the LMC and SMC being $\sim$69.2$\pm$1.6\% and $\sim$70.1$\pm$1.4\%, respectively.

If, on the other hand, we consider the incidence rates of Cepheids with additional periodicities (both \Cepa\ and \Cepb) among class 3 candidates, again combining results from the present analysis and \PaperII, we find a significantly higher rate for the LMC, $\sim$32.5$\pm$1.9\%, compared to $\sim$17.9$\pm$1.3\% for the SMC. Notably, \cite{Smolec2023MNRAS.519.4010S} demonstrated that additional signals remain more frequent in the LMC, a real empirical effect that they attributed to more efficient mode excitation at higher metallicity. The conditions favouring the excitation of these modes may thus also raise the incidence of irregular period changes among their hosts.

\subsection{Cycle-to-cycle variation}
\label{subsec: Cycle-to-cycle variation} 
A pulsating star does not always behave like a perfect metronome. Each pulsation cycle may be slightly shorter or longer than the average, and these small, seemingly random deviations accumulate over time to produce apparent period changes in the \oc\ diagram even when the underlying mean period is perfectly stable. This intrinsic stochasticity is referred to as cycle-to-cycle (C2C) variation and has long been suspected \citep{Eddington1929MNRAS..90...65E}, but remained largely unquantified until modern space telescopes delivered the combination of photometric precision, temporal continuity, and high cadence needed to resolve it directly \citep[see, e.g.,][]{Derekas2012MNRAS.425.1312D,Plachy2017MNRAS.465..173P,benko2019MNRAS.485.5897B,Plachy2021ApJS..253...11P}.

\begin{figure}
\begin{center}
{\includegraphics[height=6cm,width=0.95\linewidth]{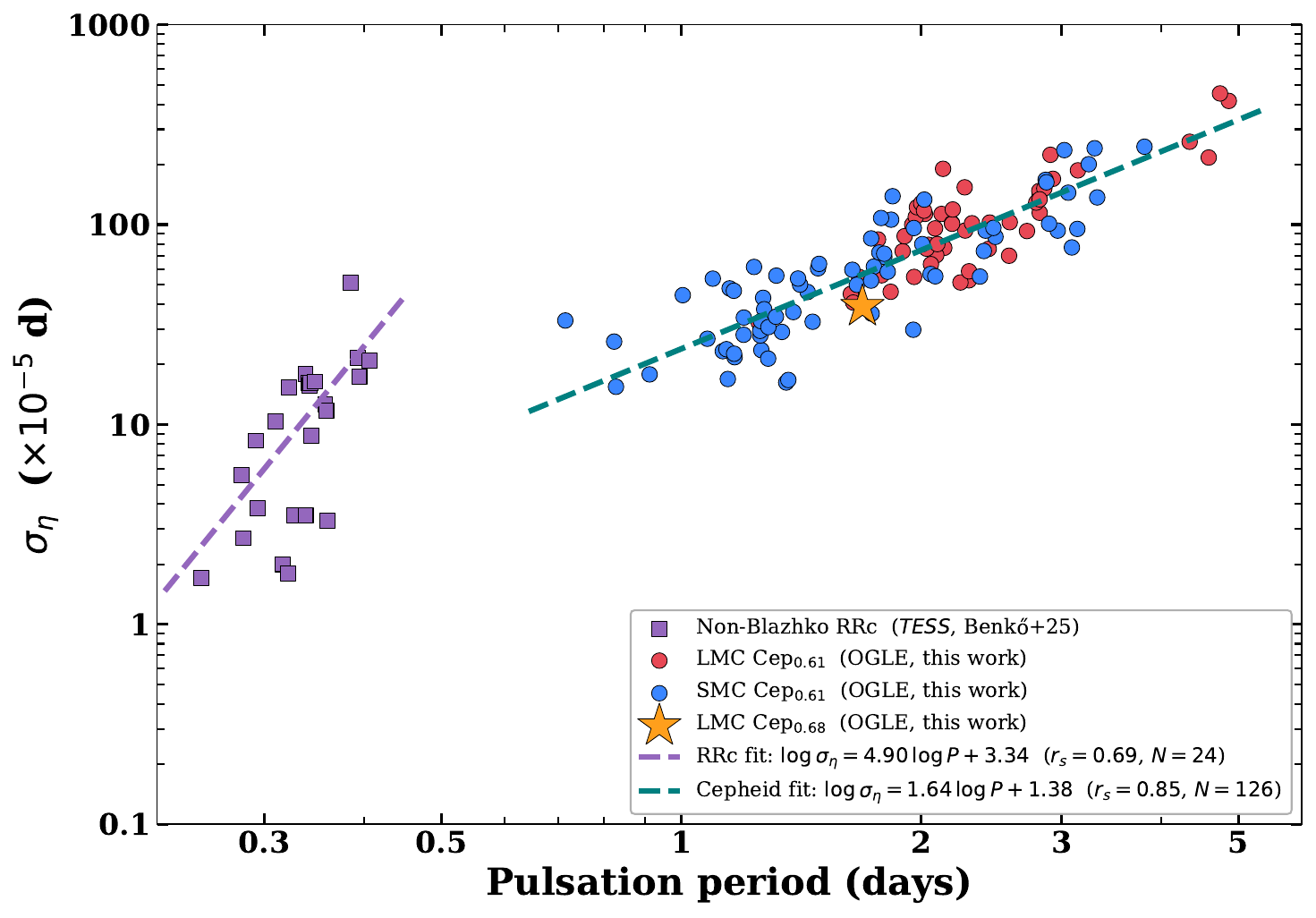}}
\caption{Comparison of cycle-to-cycle (C2C) period variation strength $\sigma_\eta$ for non-Blazhko RRc stars \citep{Benko2025A&A...697A.154B} with OGLE first-overtone Cepheids (class 3 with preferred M2 or M4 model) from this study.}
\label{fig:rrc_vs_cep}
\end{center}
\end{figure}

Recent studies \citep{benko2019MNRAS.485.5897B,Benko2025A&A...697A.154B,Benko2026A&A...705L...8B} have placed C2C variation on a quantitative footing by applying the statistical framework of \citet{Koen2006MNRAS.365..489K} to Kepler and TESS \oc\ diagrams of RR Lyrae stars (including Blazhko and non-Blazhko stars). In this approach, the observed \oc\ scatter is decomposed into three components: measurement noise ($\sigma_{e}$), a smooth secular drift ($\sigma_{\xi}$), and pure C2C jitter ($\sigma_{\eta}$). These combine into four models: M1 includes the measurement noise only ($\sigma_{e}$); M2 adds the C2C jitter ($\sigma_{e}+\sigma_{\eta}$); M3 instead adds the secular drift ($\sigma_{e}+\sigma_{\xi}$); and M4 includes all three components ($\sigma_{e}+\sigma_{\eta}+\sigma_{\xi}$) (see Appendix~\ref{appendix:Koen model Distribution} for a detailed description). The models are ranked by their AIC/BIC probabilities to identify which best describes each star. We applied the same \citet{Koen2006MNRAS.365..489K} framework to the Hertzsprung-binned \oc\ diagrams of our \Cepa\ and \Cepb\ samples in the LMC and SMC, the first application of this framework to Cepheids with additional low-amplitude variability. Because each of our \oc\ data points represents the mean epoch of $\sim$50--100 pulsation cycles, within-bin C2C fluctuations are absorbed into $\sigma_{e}$ rather than $\sigma_{\eta}$, so the $\sigma_{\eta}$ values we recover are conservative lower bounds on the true C2C jitter amplitude.

Across the full sample of 339 class 3 candidates analysed (\Cepa: 330 and \Cepb: 9), a total of 126 (37\%) Cepheids are best described by models incorporating C2C variation (M2 or M4 models; see Appendix~\ref{appendix:Koen model Distribution}). For the \Cepa\ sub-sample, 125 out of 330 stars (38\%) show significant C2C jitter, with 52 of 188 in the LMC (28\%) and 73 of 142 in the SMC (51\%). The \Cepa\ sub-sample shows a mean value of $\langle\sigma_{\eta}\rangle = 89.1 \times 10^{-5}$~d. For the \Cepb\ sub-sample, only a single detection is found among the nine stars in the combined LMC and SMC sample (OGLE-LMC-CEP-2432, $\sigma_{\eta} = 38.9 \times 10^{-5}$~d); the remaining eight stars prefer model M3, and the sample is too small to draw any statistical conclusions about C2C behaviour in this sub-type. The overall mean C2C amplitude across the full Cepheid sample is $\langle\sigma_{\eta}\rangle = 88.7 \times 10^{-5}$~d, roughly seven times larger than the value measured for non-Blazhko RRc stars by \citet{Benko2025A&A...697A.154B}, indicating a substantially higher degree of intrinsic period instability in longer-period classical pulsators. As shown in Fig.~\ref{fig:rrc_vs_cep}, this contrast is not a uniform offset but reflects a fundamentally different dependence of $\sigma_{\eta}$ on pulsation period between the two stellar classes. For the 24 non-Blazhko RRc stars reported by \citet{Benko2025A&A...697A.154B}, a linear fit in $\log$--$\log$ space yields
\begin{equation}
\label{eq:RRc_fit}
    \log \sigma_\eta = 4.90 \log P + 3.34 \;,
\end{equation}

\noindent whereas the same fit for our combined \Cepa\ and \Cepb\ sample of 126 M2/M4 Cepheids gives
\begin{equation}
\label{eq:Cep_fit}
    \log \sigma_\eta = 1.64 \log P + 1.38 \;,
\end{equation}
where $\sigma_{\eta}$ is in units of $10^{-5}$~d and $P$ is the pulsation period in days. Both relations are statistically significant, with the Cepheid fit yielding a tighter correlation ($r_s = 0.85$) over a broader pulsation-period baseline than the RRc sample ($r_s = 0.69$). Their slopes also differ substantially: the RRc relation is nearly three times steeper (slope $4.90$) than our Cepheid fit (slope $1.64$), showing that $\sigma_{\eta}$ rises far more rapidly with period in RR Lyrae stars than in classical Cepheids. The substantially higher amplitudes found for the Cepheids are qualitatively consistent with the expectation that more luminous pulsators with deeper convective envelopes produce stronger stochastic jitter, while the shallower period dependence suggests that stellar structure, rather than pulsation timescale alone, governs the magnitude of the C2C variation. We also note a contrast in the preferred models as in non-Blazhko RR Lyrae sample of \citet{Benko2025A&A...697A.154B}, the C2C model M2 is optimal for majority of stars, whereas most of our class 3 Cepheids prefer the M3 model. At least part of this difference likely reflects the different time baselines and the samplings of the two photometry datasets, to which the jitter and the drift terms may not be equally sensitive.

Finally, across our class 3 sample no star prefers model M1 (the noise-only model), which further confirms that the \oc\ variations in our sample cannot be attributed to measurement noise alone and are therefore physical in origin.

\subsection{Amplitude dependence of the irregular period changes}
\label{subsec: Amplitude dependence} 
In \PaperII, working on a large sample of Magellanic Cepheids, we previously established two important empirical properties of the irregular period changes. The first is that the irregularity grows with pulsation period, a trend also seen in Galactic Cepheids \citep{Csornyei2022MNRAS.511.2125C}. This is further confirmed by our smaller class 3 sample in this work, where all characterisation measures show significant positive correlations with pulsation period (see Sect.~\ref{subsec: Characterization}). The second empirical property is that at a given pulsation period the phenomenon is systematically stronger in first-overtone Cepheids than in fundamental-mode Cepheids. The physical origin of these two dependences remains unclear. One suspect for the stochastic perturbations is the convective envelope, which can disturb the pulsation directly or distort the observed light curve through hot spots \citep[see, e.g.,][]{Neilson2014A&A...563L...4N}. A useful framework was developed by \citet{Avelino2020MNRAS.492.4477A} and \citet{Cunha2020MNRAS.499.4687C}, who modelled pulsating stars as coherently driven oscillators subject to random perturbations. In their model, each perturbation shifts the pulsation phase by an amount inversely proportional to the oscillation amplitude, and the accumulated phase wander in the \oc\ diagram grows as a random walk with an rms proportional to the ratio of the perturbation strength to the oscillation amplitude. Our sample is well suited to test this prediction, since \citet{Smolec2023MNRAS.519.4010S} provide, for every star, not only the amplitude of the dominant first-overtone mode, $A_{\rm 1O}$, but also the amplitude of the additional low-amplitude periodicity, $A_{\rm x}$.

\begin{figure*}
\begin{center}
{\includegraphics[height=9cm,width=0.95\linewidth]{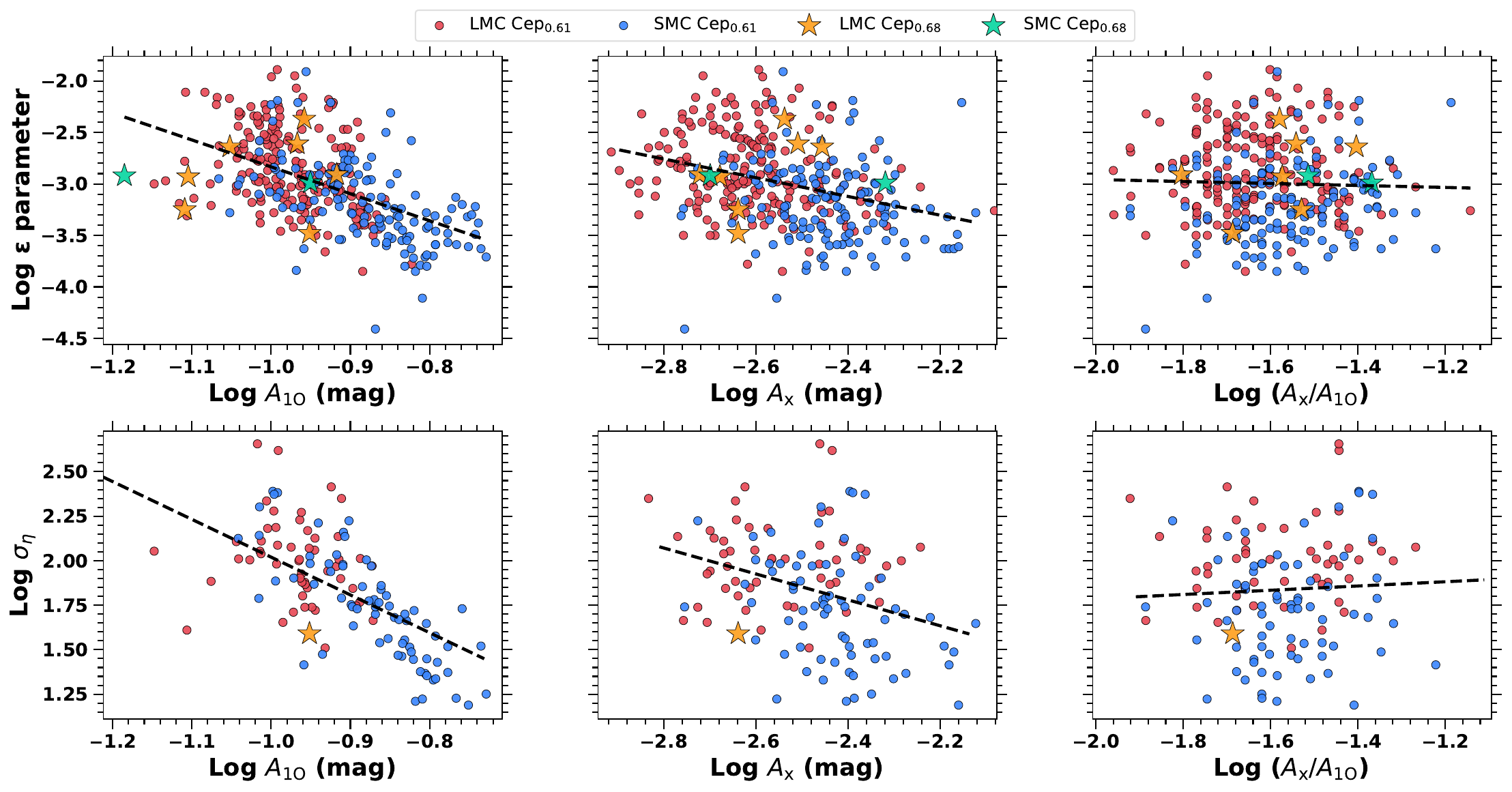}}
\caption{Strength of the irregular period changes as a function of the pulsation amplitude $A_{\rm 1O}$ (left) and the additional-signal amplitude $A_{\rm x}$ (middle), and their ratio $A_{\rm x}/A_{\rm 1O}$ (right),
for the $\varepsilon$ parameter of the class 3 sample (top) and the C2C jitter $\sigma_\eta$ (bottom). Dashed lines show linear fits in the log--log plane.}
\label{fig:amp_irregularity}
\end{center}
\end{figure*}

In Fig.~\ref{fig:amp_irregularity} we show the comparison of the $\varepsilon$ parameter and the C2C jitter $\sigma_\eta$ (available only for the 126 M2/M4 candidates) with the amplitude of the first-overtone mode ($A_{\rm 1O}$) and the amplitude of the additional periodicity ($A_{\rm x}$), for the overall class 3 sample presented in this work. All four panels show significant anticorrelations, with Spearman rank coefficients of $r_s = -0.57$ and $-0.36$ for $\varepsilon$ against $A_{\rm 1O}$ and $A_{\rm x}$, respectively, and $r_s = -0.71$ and $-0.34$ for $\sigma_\eta$. However, both the amplitudes and the irregularity measures depend on the pulsation period, so these correlations may simply be inherited from their common period trends. To separate the two effects, we compute a period-controlled coefficient, $r_{s,P}$, obtained by removing the period trend from both quantities and correlating the residuals. A significant $r_{s,P}$ indicates a genuine amplitude dependence at fixed pulsation period. The anticorrelations with the pulsation amplitude persist at fixed period ($r_{s,P} = -0.34$ for $\varepsilon$ and $-0.46$ for $\sigma_\eta$, both $p \ll 0.001$), with an observed scaling $\sigma_\eta \propto A_{\rm 1O}^{-0.8 \pm 0.2}$, and the remaining characterisation measures behave similarly ($r_{s,P} = -0.23$ to $-0.32$). In contrast, the correlations with the additional-signal amplitude vanish ($r_{s,P} = -0.05$ for $\varepsilon$ and $+0.14$ for $\sigma_\eta$, both not significant), and the relative strength $A_{\rm x}/A_{\rm 1O}$ shows no robust independent correlation with any of the measures (see Fig.~\ref{fig:amp_irregularity} right panels).

The strength of the irregularity therefore tracks the amplitude of the dominant pulsation mode, and not that of the additional periodicity. This is exactly what the results outlined above predict. The measured scaling, $\sigma_\eta \propto A_{\rm 1O}^{-0.8 \pm 0.2}$, matches the inverse-amplitude phase wander expected for a coherently driven pulsator perturbed by random kicks of fixed strength \citep{Avelino2020MNRAS.492.4477A, Cunha2020MNRAS.499.4687C}, and extends this picture to classical Cepheids. In this interpretation, the pulsation amplitude measures the robustness of the limit cycle. High-amplitude pulsators resist the stochastic perturbations, while low-amplitude ones accumulate phase wander more rapidly. The same reasoning explains the second empirical property noted above, because fundamental-mode Cepheids pulsate with systematically larger amplitudes than the overtones. A lower stability of first-overtone pulsation was also noted by \citet{Suveges2018a}. The absence of an independent $A_{\rm x}$ correlation argues against the additional periodicity being a driver of the irregularity, in line with the lack of relations between the primary and secondary-mode amplitudes reported by \citet{Suveges2018b}. We instead interpret its presence as a marker of the underlying pulsation state. \citet{Dziembowski2016CoKon.105...23D} regards the coexistence of the non-radial modes with the radial overtone as a form of double-mode pulsation, arising in circumstances similar to those of radial double-mode pulsators. In such a state, the dominant mode does not fully saturate the driving. A weakly saturated first-overtone limit cycle would then do two things at once: admit the additional modes, and respond more strongly to the stochastic perturbations. This explains why the presence of the additional periodicity accompanies an elevated incidence of irregular period changes (75.5$\pm$2.0\% in our current sample versus 68.1$\pm$1.2\% for the general first-overtone population; (Sect.~\ref{subsec: Incidence rate}), while its strength leaves no imprint. The frequently non-stationary behaviour of both the additional signals and the radial overtone in these stars \citep{Smolec2023MNRAS.519.4010S} fits the same picture.

\section{Conclusions}
\label{sec: Conclusions} 

In this study we analysed a sample of 544 Cepheids manifesting low-amplitude variability (516 \Cepa\ and 28 \Cepb) for their associated period-change behaviour. All of these are first-overtone stars in both the LMC and SMC. This sample is complementary to that of \PaperII, where we analysed a large sample of single-mode Magellanic Cepheids and deliberately excluded stars with additional low-amplitude variability. The present work targets precisely those excluded Cepheids. After sample cuts, overall in the working sample, we find that 449 Cepheids belong to three different period-change classes. Our main findings are summarised as follows:
\begin{itemize}
\item We identify 83 class 1 Cepheids with flat \oc\ diagrams indicative of negligible period change, yielding an incidence rate of $\sim$18.5$\pm$1.8\%. This is in excellent agreement with the \PaperII\ first-overtone rate of $\sim$17.1$\pm$1.0\%.
\item A total of 27 class 2 Cepheids exhibit parabolic \oc\ shapes, with an incidence rate of $\sim$6.0$\pm$1.1\%. Combining these with the previous \PaperII\ detections, the updated combined first-overtone class 2 incidence rate is $\sim$12.9$\pm$0.7\%. Several of these may be candidates for secular period change, their upward-parabolic \oc\ diagrams tentatively marking first-crossing candidates.
\item The majority of our sample, 339 Cepheids, are class 3 candidates, simultaneously exhibiting low-amplitude variability and irregular period changes, at an incidence rate of $\sim$75.5$\pm$2.0\%. These candidates increase the total Magellanic class 3 sample to 1924 stars, and the updated combined first-overtone class 3 incidence rate is $\sim$69.7$\pm$1.0\%.
\item Various characterisation tests on the class 3 candidates all indicate a coherent picture of increasing irregular period-change fluctuations with pulsation period, a trend in excellent agreement with the behaviour of the \PaperII\ first-overtone sample and consistently elevated above the \PaperII\ fundamental-mode Cepheids. This indicates that period instability is an intrinsic property that intensifies towards longer periods, irrespective of the presence of \Cepa- or \Cepb-type low-amplitude variability.
\item We detect significant C2C period variation in 37\% (126 Cepheids) of our class 3 sample, with a mean jitter amplitude $\langle\sigma_{\eta}\rangle = 88.7 \times 10^{-5}$~d, $\sim$7 times larger than that measured in non-Blazhko RRc stars \citep{Benko2025A&A...697A.154B}. We also find that $\sigma_{\eta}$ scales empirically with pulsation period, with a shallower slope than in RRc stars.
\item At fixed pulsation period, the strength of the irregular period changes anticorrelates with the pulsation amplitude across all our characterisation measures, most strongly for the C2C jitter, which scales as $\sigma_\eta \propto A_{\rm 1O}^{-0.8 \pm 0.2}$. The amplitude of the additional periodicity shows no independent correlation. This behaviour matches the inverse-amplitude response expected for a coherently driven pulsator perturbed by stochastic kicks \citep{Avelino2020MNRAS.492.4477A, Cunha2020MNRAS.499.4687C}, and suggests that the additional periodicity marks a weakly saturated pulsation state rather than driving the irregularity.
\end{itemize}

The empirical picture of irregular period changes in classical Cepheids is now substantially richer following our previous work in \PaperII\ and this study. For class 3 candidates we have provided comprehensive incidence rates, their pulsation-period distribution, and a multi-method characterisation of the irregular period-change amplitude, probing both a large sample of purely radial-mode Cepheids (F-mode and 1O-mode) and the niche sub-sample of 1O Cepheids hosting low-amplitude signals associated with non-radial modes (\Cepa) or additional variability (\Cepb). The physical origin of these irregular period changes, however, remains unresolved. With a statistically significant sample, we have established a positive correlation between irregular period-change strength and pulsation period, a strong empirical constraint that Cepheid pulsation models should reproduce. We plan to extend our investigation to longer-period regimes by analysing irregular period changes in Type-II Cepheids using OGLE Magellanic photometry. Such targets have long been known to undergo abrupt and erratic period changes \citep[see, e.g.,][]{Parenago1956PZ.....11..236P, Percy1997PASP..109..264P, Smolec2018MNRAS.481.3724S}, and their inclusion will provide a more complete view of the phenomenon across the instability strip. Finally, the high-cadence, continuous photometry from TESS \citep{Ricker2015JATIS...1a4003R} and the forthcoming PLATO mission \citep{Rauer2025ExA....59...26R} offer promising prospects for resolving C2C period jitter directly in individual pulsation cycles, extending this analysis to a broader census of Cepheids. 

\section{Data availability}
\label{sec: Data availability} 
The full versions of Tables~\ref{tab:Class_1_list_cepa}, \ref{tab:Class_2_list_cepa}, and \ref{tab:Class_3_list_mainpaper} are available in electronic form at the CDS via \url{http://cdsweb.u-strasbg.fr/cgi-bin/qcat?J/A+A/}, and at Zenodo: \url{https://doi.org/10.5281/zenodo.22644776}.

\begin{acknowledgements}
We thank all the OGLE observers for their contribution to the collection of the photometric data over the decades. RSR, RS, and OZ acknowledge support from the National Science Centre, Poland, OPUS project 2024/53/B/ST9/02630. RSR, VH, and NN acknowledge the support of the French Agence Nationale de la Recherche (ANR) under grant ANR-23-CE31-0009-01 (Unlock-pfactor). IS acknowledges support from the National Science Centre, Poland, grant no.\ 2022/45/B/ST9/00243. The OGLE project has received funding from the Polish National Science Centre grant OPUS-28 2024/55/B/ST9/00447
awarded to AU. This research made use of the SIMBAD and VIZIER databases at CDS, Strasbourg (France) and the electronic bibliography maintained by the NASA/ADS system.

\end{acknowledgements}

\bibliographystyle{aa}
\bibliography{aa63195-26}

\begin{appendix}
\nolinenumbers

\section{Class 1 candidates}
\begin{figure*}
\begin{center}
{\includegraphics[height=3.5cm,width=0.24\linewidth]{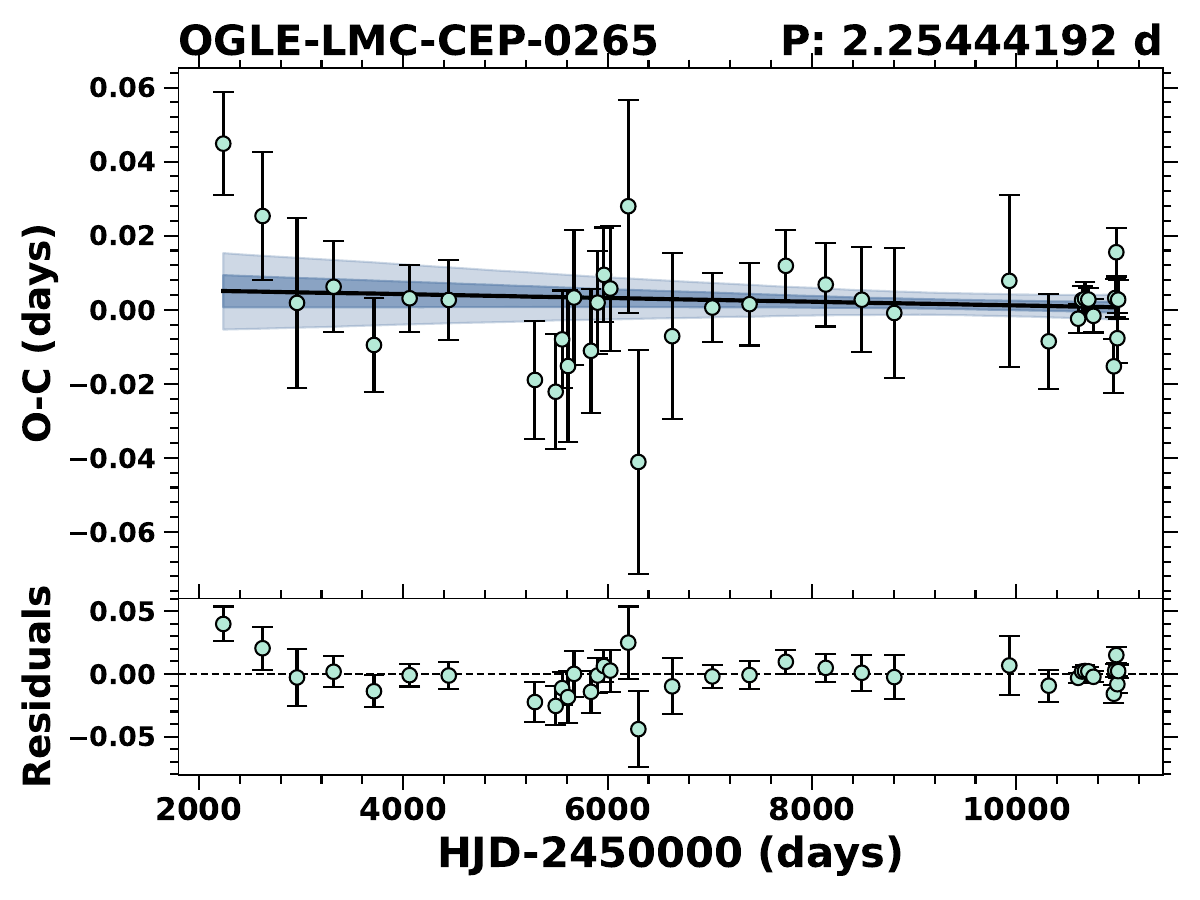}}
{\includegraphics[height=3.5cm,width=0.24\linewidth]{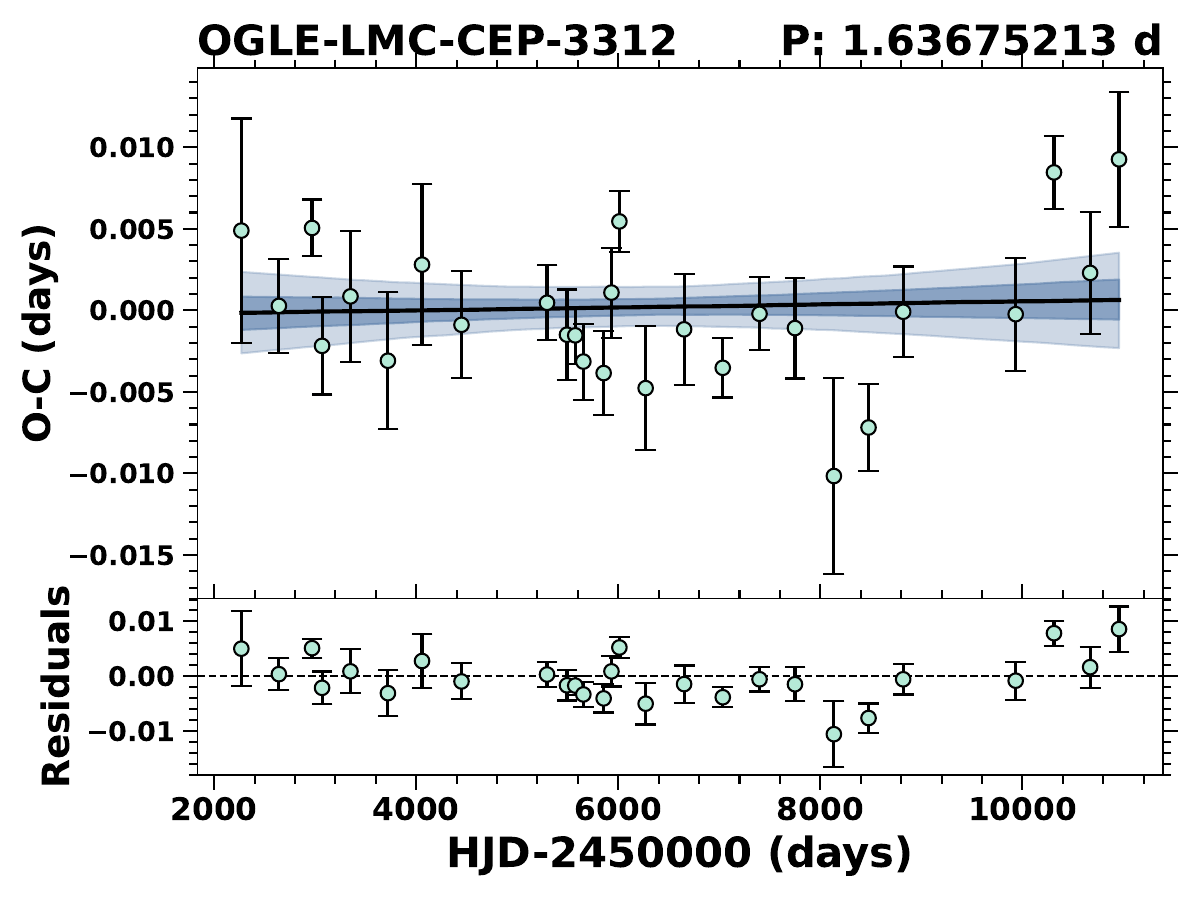}}
{\includegraphics[height=3.5cm,width=0.24\linewidth]{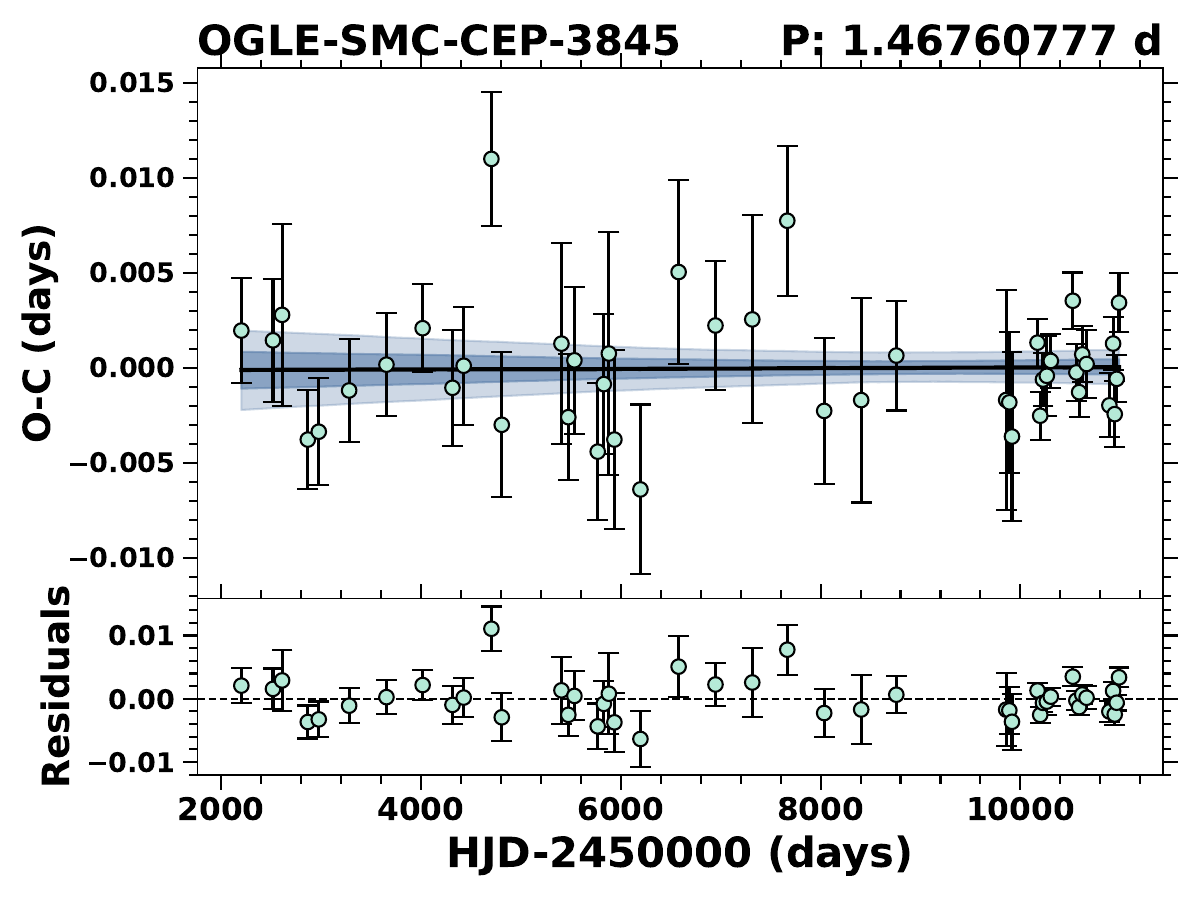}}
{\includegraphics[height=3.5cm,width=0.24\linewidth]{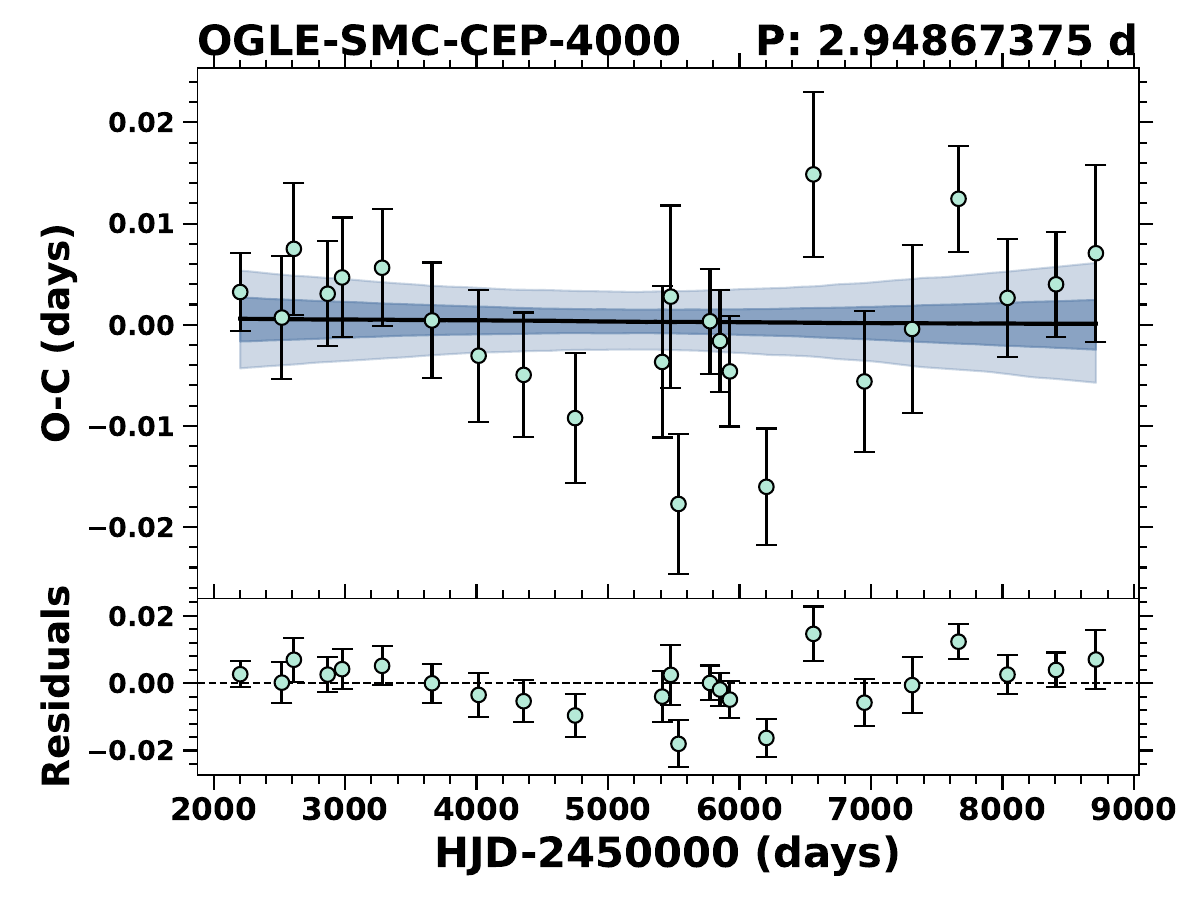}}

{\includegraphics[height=3.5cm,width=0.24\linewidth]{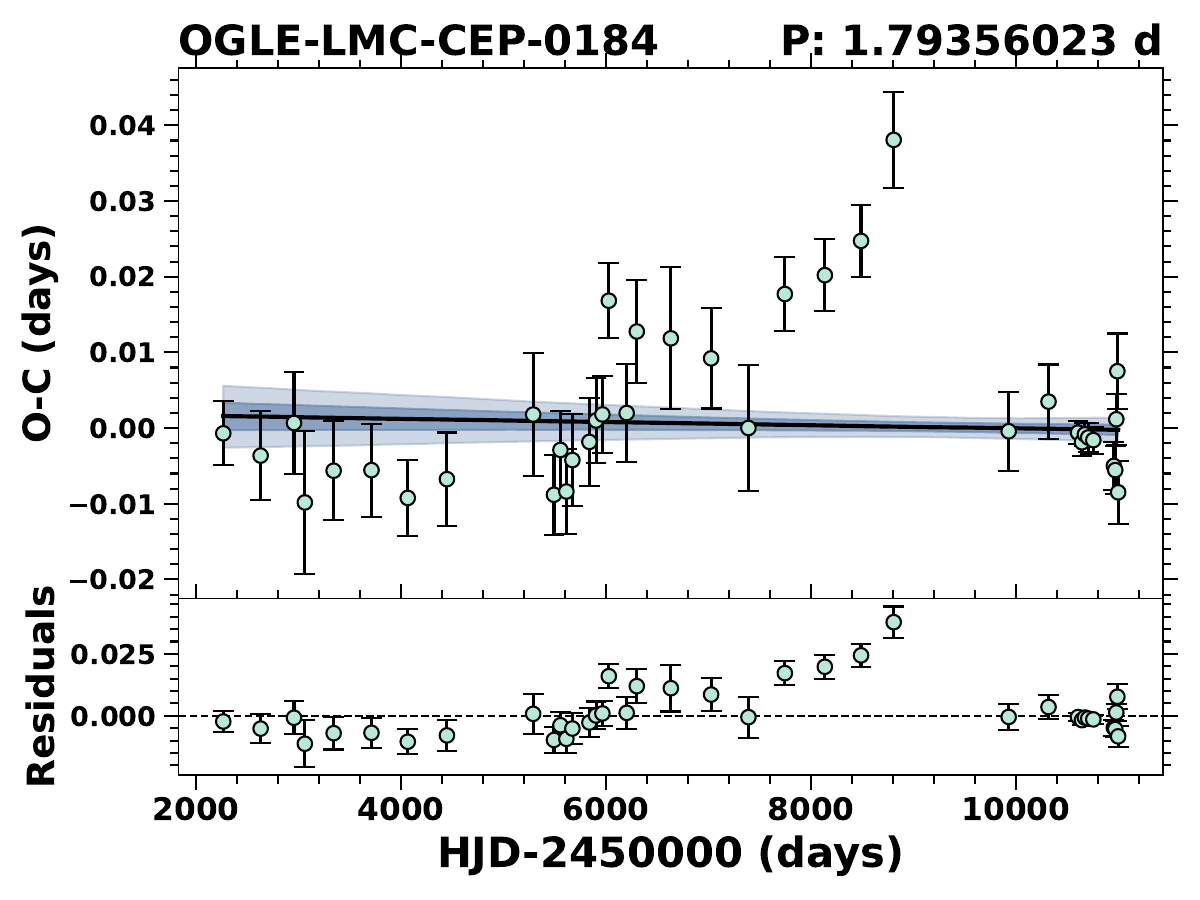}}
{\includegraphics[height=3.5cm,width=0.24\linewidth]{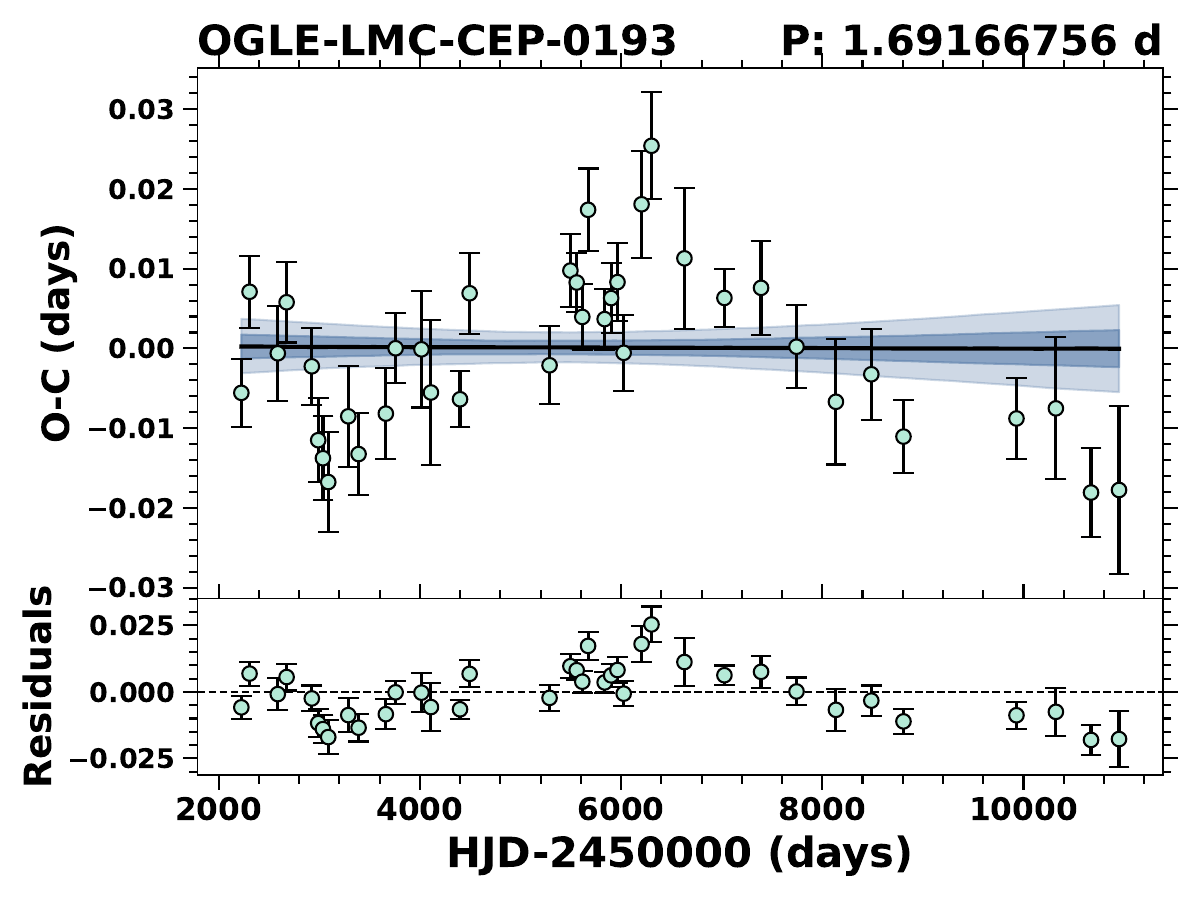}}
{\includegraphics[height=3.5cm,width=0.24\linewidth]{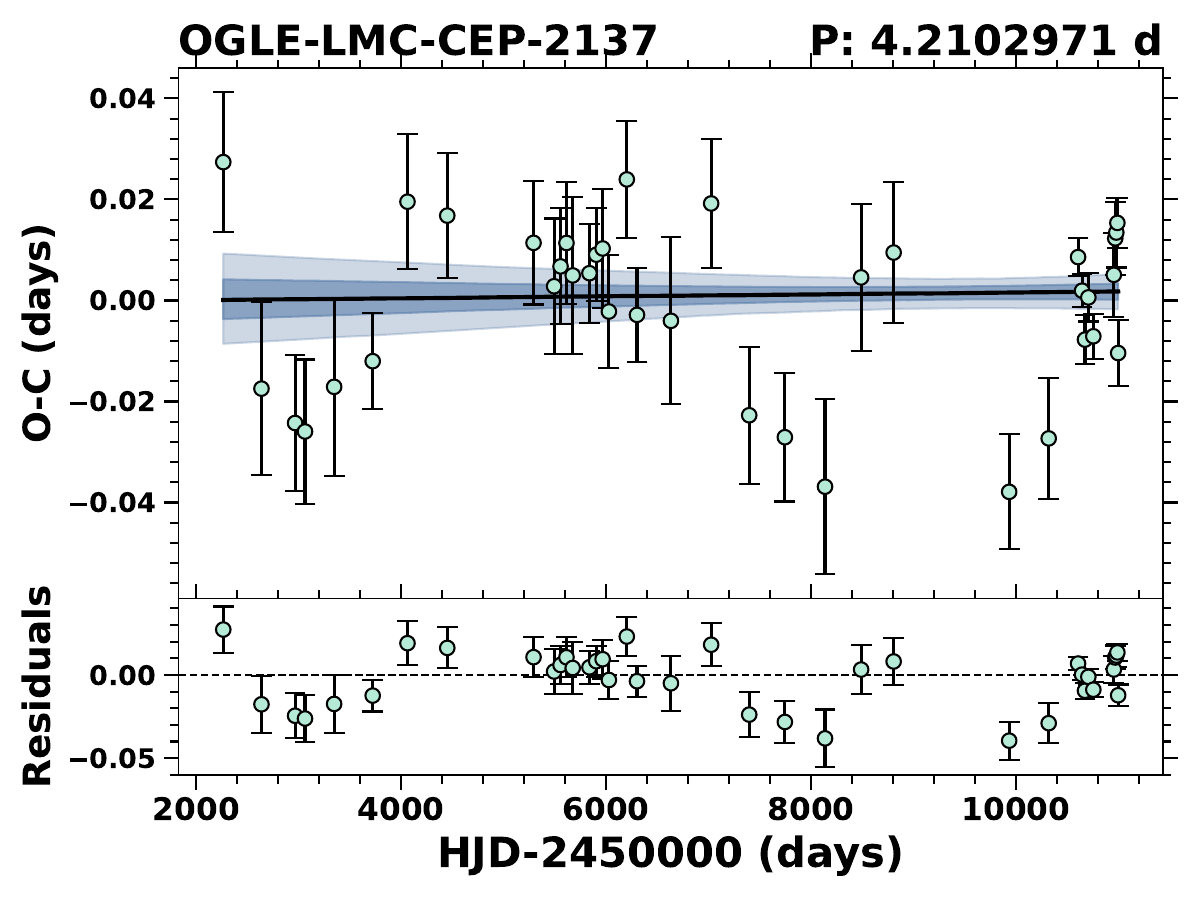}}
{\includegraphics[height=3.5cm,width=0.24\linewidth]{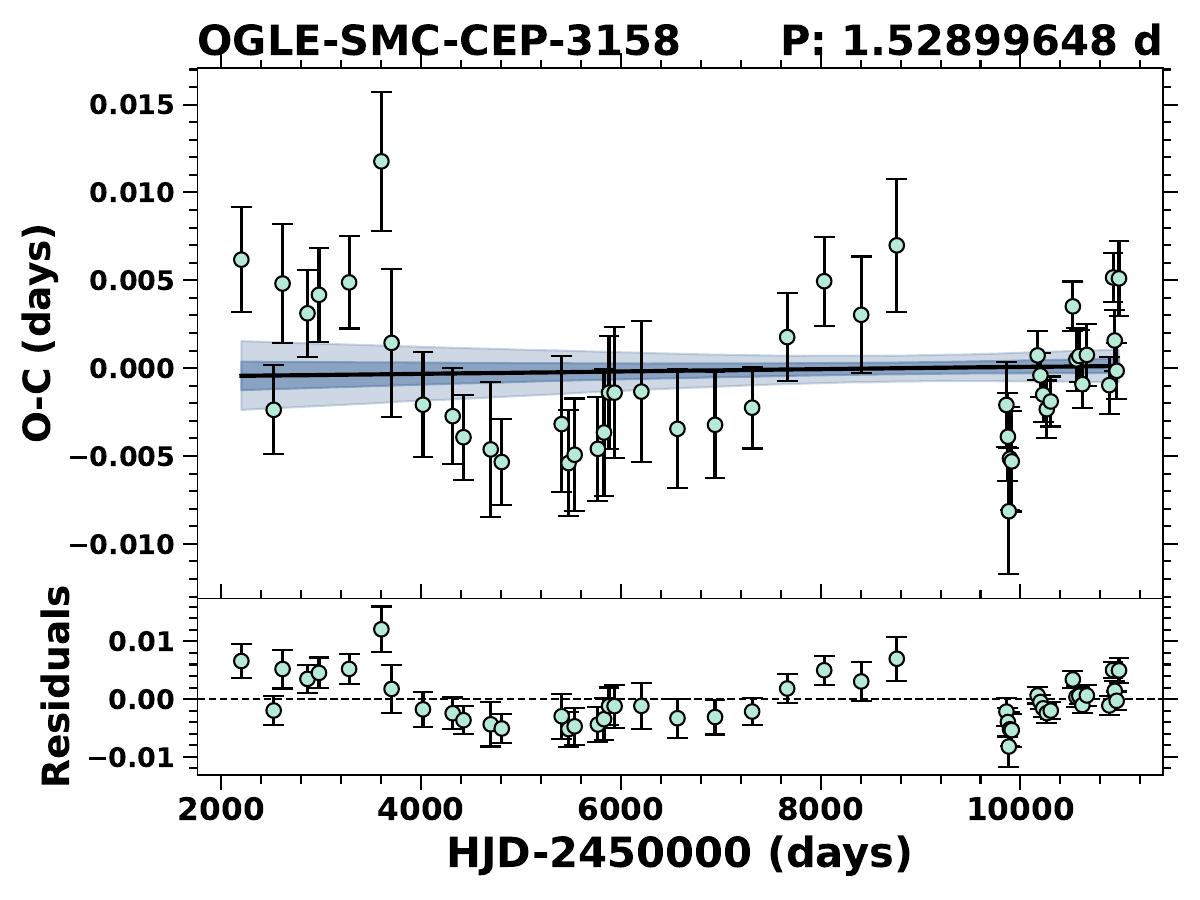}}

\caption{Representative \oc\ diagrams of \Cepa\ (top panel) and \Cepb\ (bottom panel) candidates belonging to class 1. The OGLE identifier and pulsation period are shown at the top of each panel.}

\label{fig:appendix_c1_ocplot_examples}
\end{center}
\end{figure*}

Table~\ref{tab:Class_1_list_cepa} lists the class 1 candidates along with their pulsation periods. Representative \oc\ diagrams are shown in Fig~\ref{fig:appendix_c1_ocplot_examples}.

\begin{table}[ht]
\caption{Class 1 Cepheid candidates.}
\label{tab:Class_1_list_cepa}
\centering\footnotesize
\begin{tabular}{lr}
\hline
\hline
\textbf{OGLE ID} & \textbf{Pulsation period (d)} \\
\hline
\textbf{LMC \Cepa\ sample} & \\
\hline
OGLE-LMC-CEP-0071 & 1.38588763 \\
OGLE-LMC-CEP-0221 & 2.21645283 \\
OGLE-LMC-CEP-0265 & 2.25444192 \\
\ldots & \ldots \\
\ldots & \ldots \\

\hline
\textbf{SMC \Cepa\ sample} & \\
\hline
OGLE-SMC-CEP-0348 & 2.06688185 \\
OGLE-SMC-CEP-0447 & 1.26514836 \\
OGLE-SMC-CEP-0466 & 1.13176248 \\
\ldots & \ldots \\
\ldots & \ldots \\

\hline
\textbf{LMC \Cepb\ sample} & \\
\hline
OGLE-LMC-CEP-0184 & 1.79356023 \\
OGLE-LMC-CEP-0193 & 1.69166756 \\
OGLE-LMC-CEP-2137 & 4.21029710 \\
\hline

\textbf{SMC \Cepb\ sample} & \\
\hline
OGLE-SMC-CEP-3158 & 1.52899648 \\

\hline

\hline
\end{tabular}
\tablefoot{The two columns list the star's ID and its pulsation period in days. Only first few entries are shown and the full table is available at Zenodo ( see Sect.~\ref{sec: Data availability})}
\end{table}

\section{Class 2 candidates}
\begin{figure*}
\begin{center}

{\includegraphics[height=3.5cm,width=0.24\linewidth]{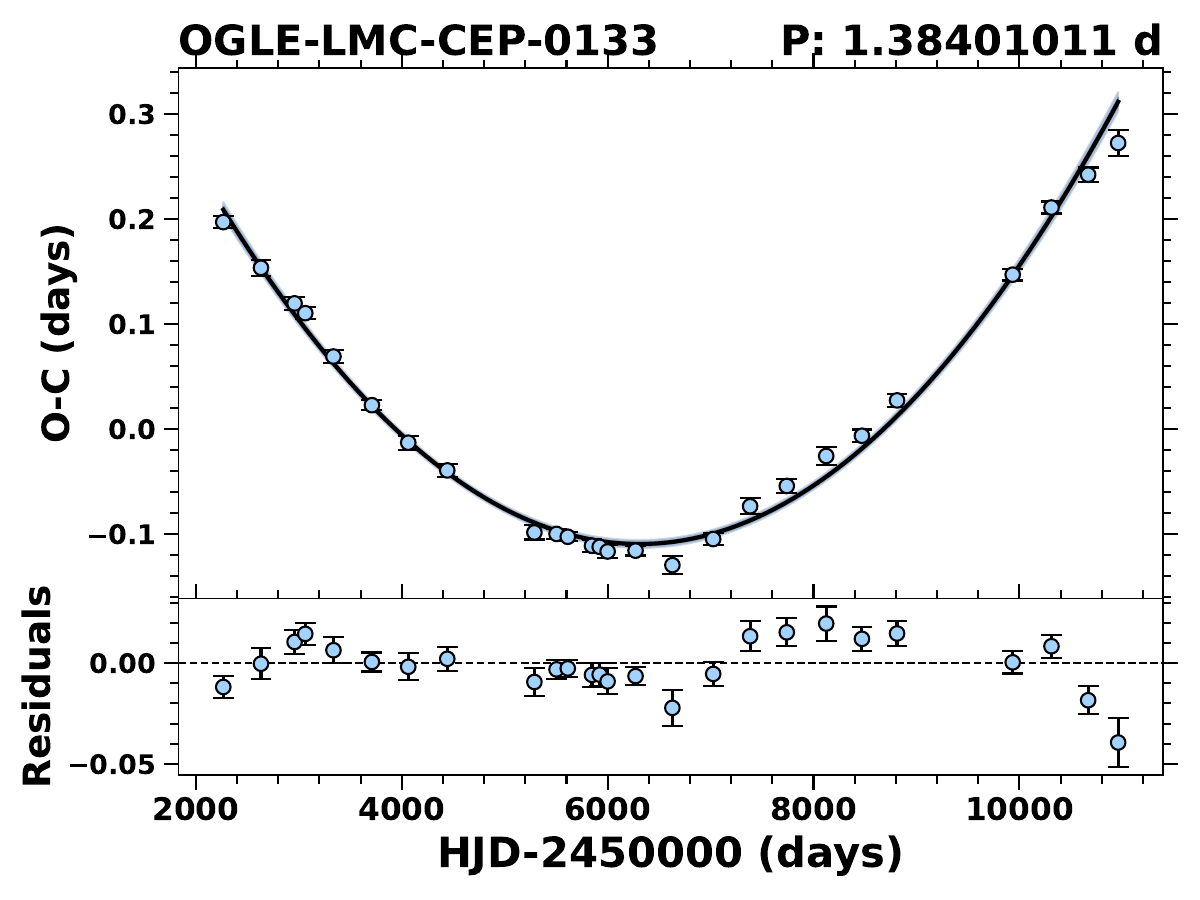}}
{\includegraphics[height=3.5cm,width=0.24\linewidth]{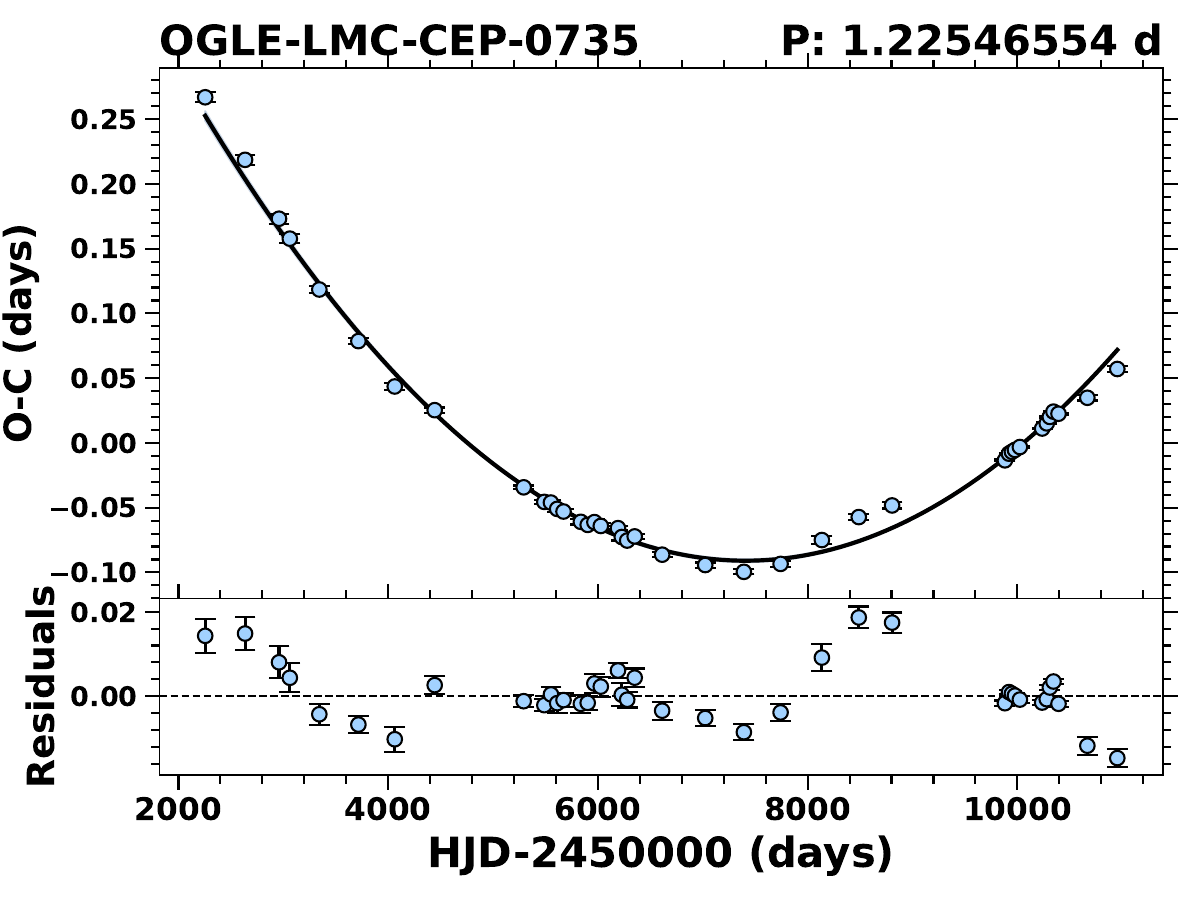}}
{\includegraphics[height=3.5cm,width=0.24\linewidth]{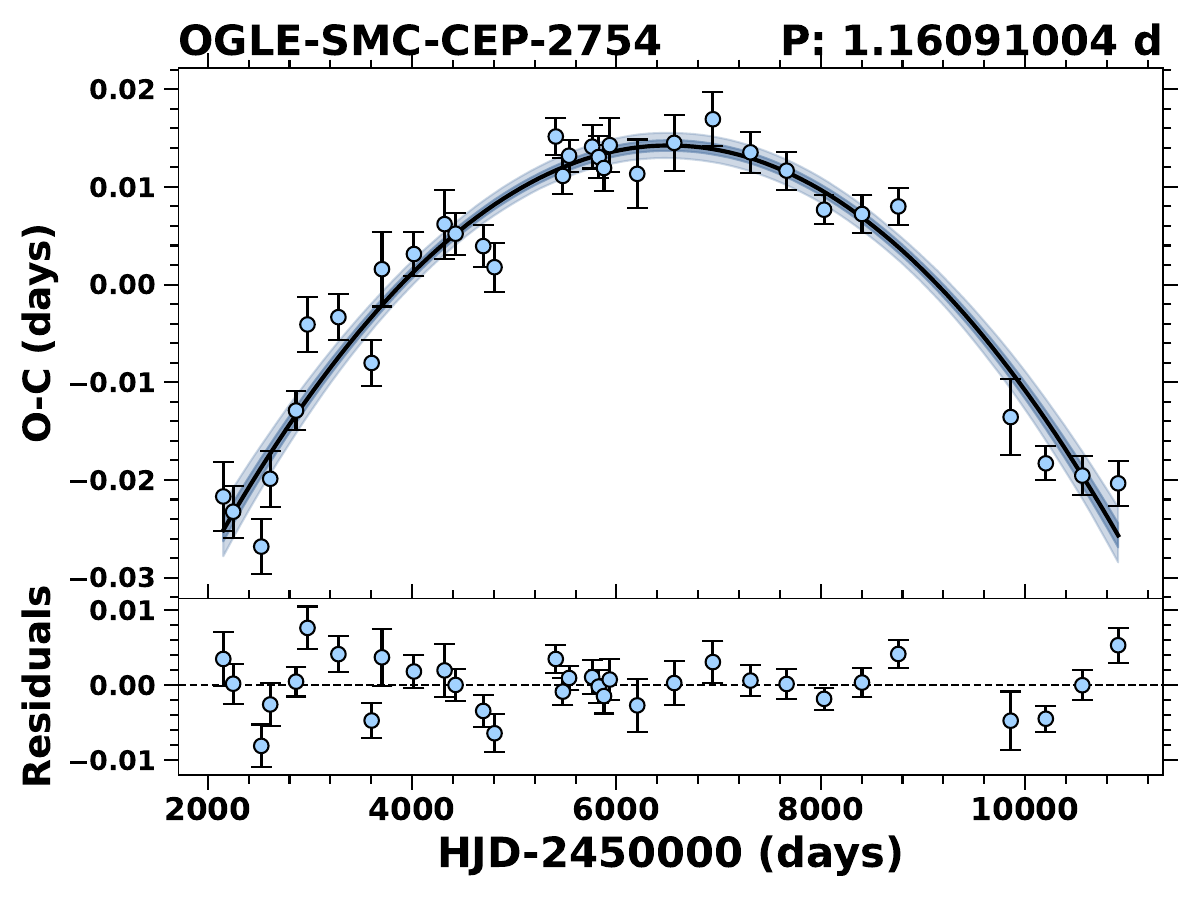}}
{\includegraphics[height=3.5cm,width=0.24\linewidth]{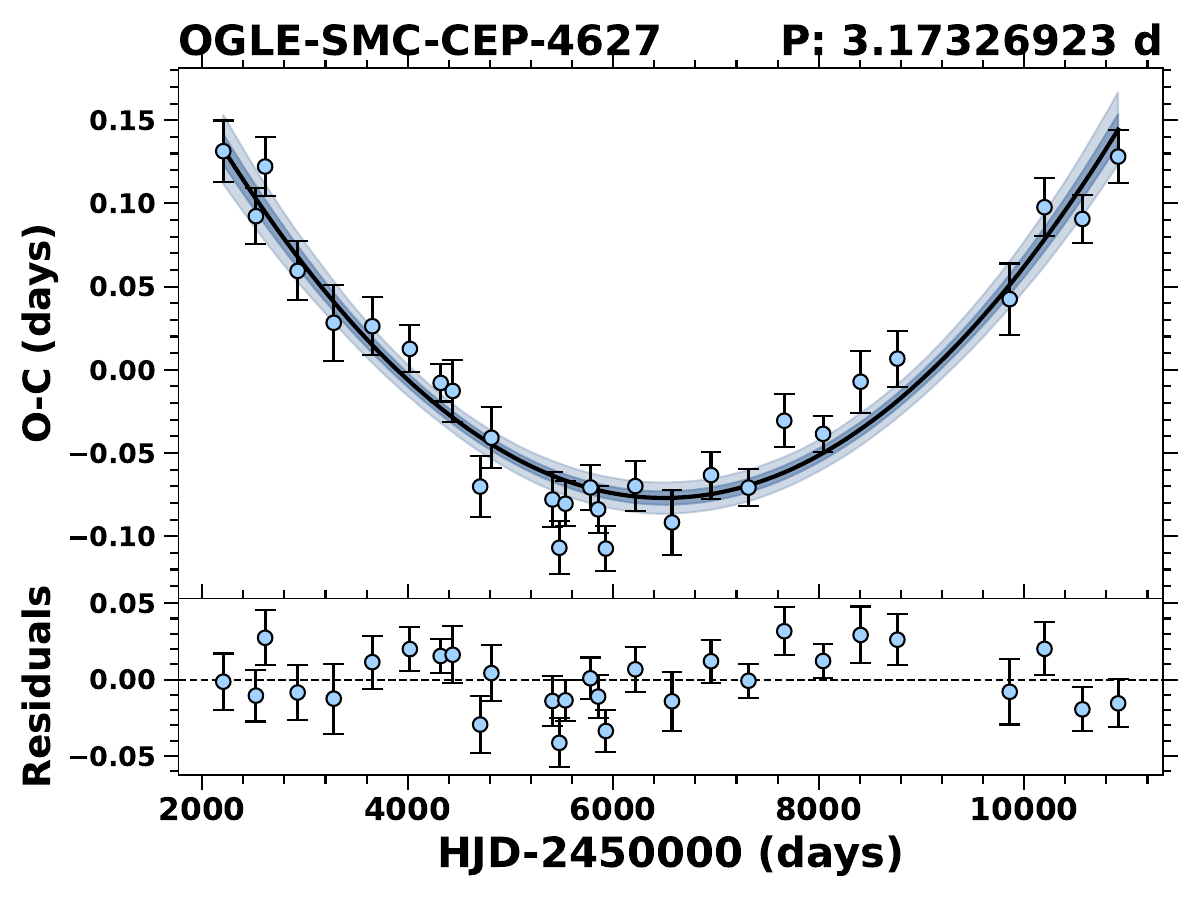}}

{\includegraphics[height=3.5cm,width=0.24\linewidth]{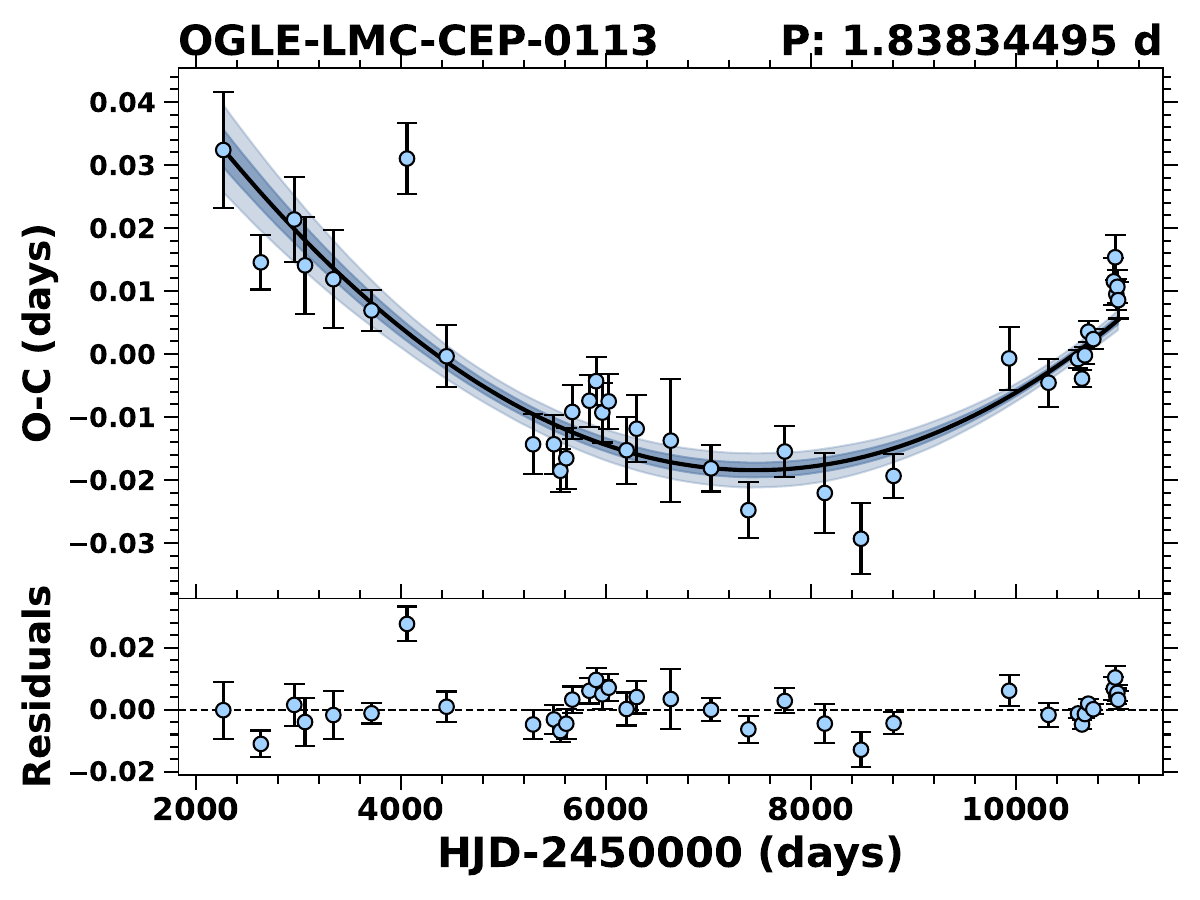}}
{\includegraphics[height=3.5cm,width=0.24\linewidth]{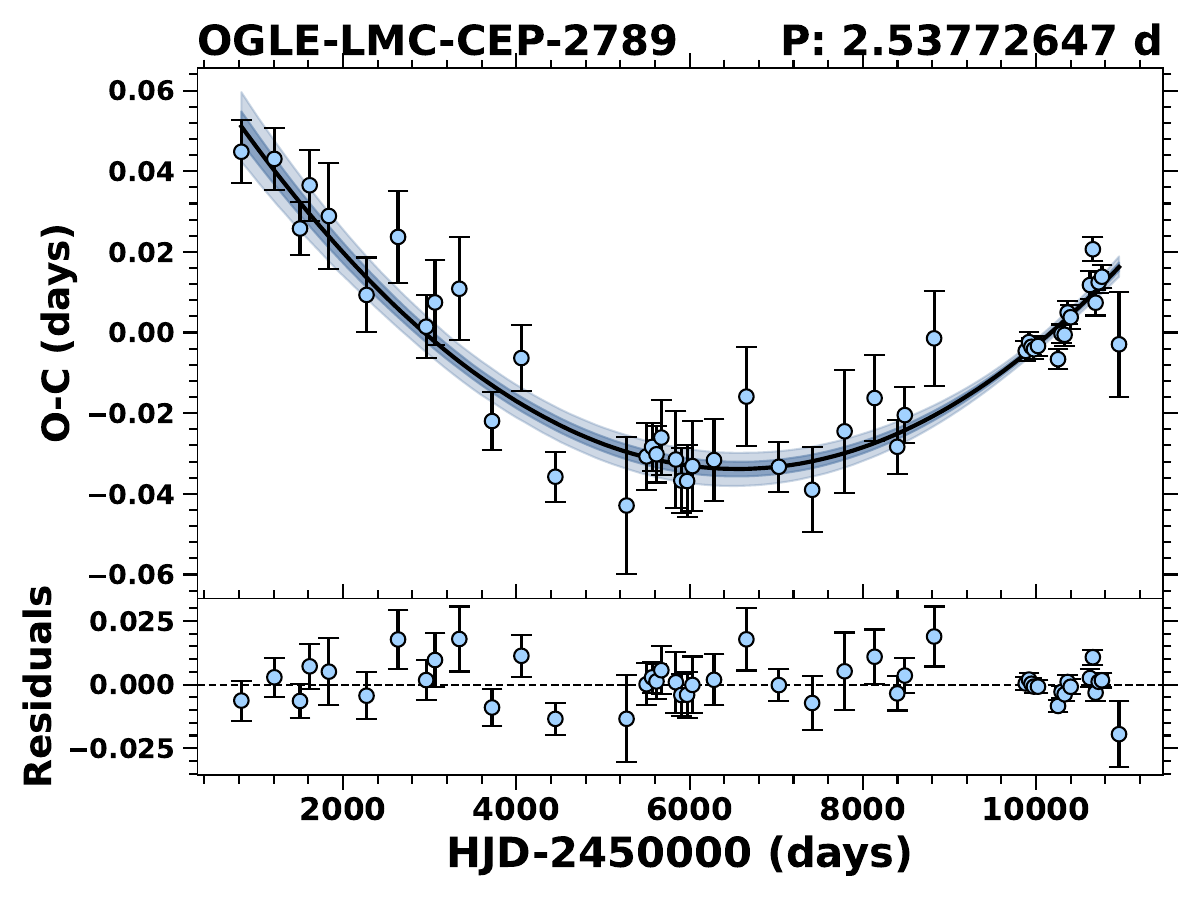}}
{\includegraphics[height=3.5cm,width=0.24\linewidth]{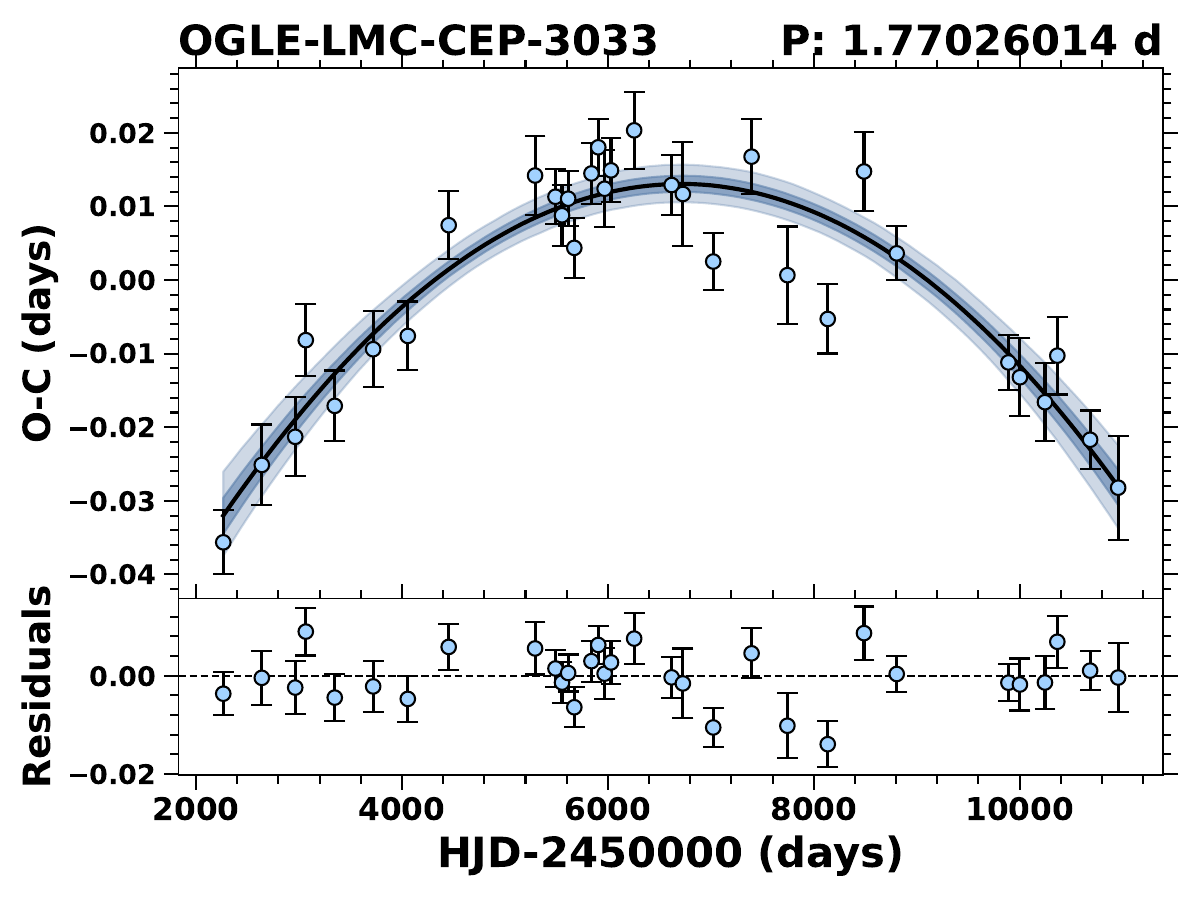}}
{\includegraphics[height=3.5cm,width=0.24\linewidth]{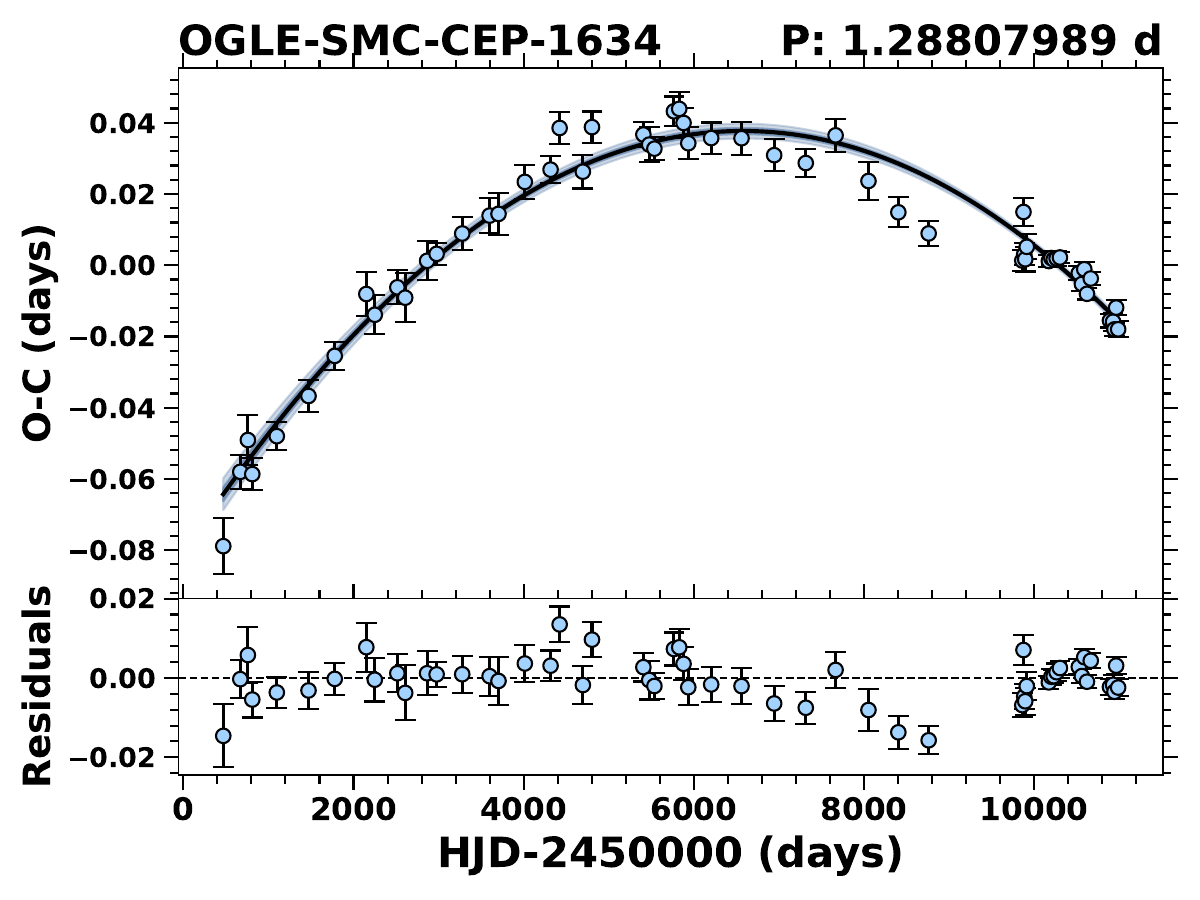}}

\caption{Representative \oc\ diagrams of \Cepa\ (top panel) and \Cepb\ (bottom panel) candidates belonging to class 2. The OGLE identifier and pulsation period are shown at the top of each panel.}

\label{fig:appendix_c2_ocplot_examples}
\end{center}
\end{figure*}

Table~\ref{tab:Class_2_list_cepa} lists the class 2 candidates along with their pulsation periods and estimated period change rates. Representative \oc\ diagrams are shown in Fig~\ref{fig:appendix_c2_ocplot_examples}.

\begin{table}[ht]
\caption{Class 2 Cepheid candidates.}
\label{tab:Class_2_list_cepa}
\centering\footnotesize
\setlength{\tabcolsep}{3pt}
\begin{tabular}{lrrr}
\hline
\hline
\textbf{OGLE ID} & \textbf{Pulsation period (d)} &  \textbf{d$P$/d$t$} &  \textbf{err d$P$/d$t$} \\
 &  &  \textbf{(d/Myr)} &  \textbf{(d/Myr)} \\
\hline
\hline
\textbf{LMC \Cepa\ sample} & & & \\
\hline
OGLE-LMC-CEP-0133 & 1.384010 & 19.675 & 0.196 \\
OGLE-LMC-CEP-0368 & 2.319331 & -43.108 & 0.192 \\
OGLE-LMC-CEP-0421 & 2.215381 & 7.721  & 0.212 \\
\ldots & \ldots  & \ldots  & \ldots   \\
\ldots & \ldots  & \ldots  & \ldots   \\

\hline
\textbf{SMC \Cepa\ sample} & & & \\
\hline
OGLE-SMC-CEP-2455 & 2.524637 & -6.465  & 0.675 \\
OGLE-SMC-CEP-2550 & 1.966623 & -1.120  & 0.085 \\
OGLE-SMC-CEP-2668 & 1.940816 & 1.816  & 0.082 \\
\ldots & \ldots  & \ldots  & \ldots   \\
\ldots & \ldots  & \ldots  & \ldots   \\

\hline
\textbf{LMC \Cepb\ sample} & & & \\
\hline
OGLE-LMC-CEP-0113 & 1.838345 & 2.549 & 0.153 \\
OGLE-LMC-CEP-1801 & 1.178179 & -0.963 & 0.064 \\
OGLE-LMC-CEP-2789 & 2.537726 & 4.792 & 0.228 \\
\ldots & \ldots  & \ldots  & \ldots   \\
\ldots & \ldots  & \ldots  & \ldots   \\

\hline
\textbf{SMC \Cepb\ sample} & & & \\
\hline
OGLE-SMC-CEP-1634 & 1.288080 & -2.580 & 0.057 \\
\hline

\end{tabular}
\tablefoot{The four columns give star's ID, pulsation period, period change rate and its error. Only first few entries are shown and the full table is available at Zenodo ( see Sect.~\ref{sec: Data availability})}
\end{table}

\section{Class 3 candidates}
Table~\ref{tab:Class_3_list_mainpaper} lists the class 3 candidates together with their corresponding characterisation parameters. Representative \oc\ diagrams are shown in Fig~\ref{fig:ocplot_sample_class3_examples}.

\begin{table*}[ht]
\caption{Sample of class 3 Cepheid candidates.}
\label{tab:Class_3_list_mainpaper}
\centering\footnotesize
\scalebox{0.82}{
\begin{tabular}{lccccrccccc}
\hline
\hline
\textbf{OGLE ID} & \textbf{Pulsation period (d)} & \textbf{$\log\epsilon$} & \textbf{Dominant} & \textbf{Dominant} & \textbf{$L$ index} & \textbf{$\log(\Delta P/P)$} & \textbf{$\log\sigma$} & \textbf{$\sigma_{\eta}$} & \textbf{$\sigma_{\xi}$} & \textbf{Best} \\
 & & & \textbf{Periodicity} & \textbf{Amplitude} & & & & ($\times10^{-5}$\,d) & ($\times10^{-5}$\,d) & \textbf{Model} \\
\hline
\hline
\multicolumn{11}{l}{\textbf{LMC \Cepa\ sample}} \\
\hline
OGLE-LMC-CEP-0031 & 4.33983633 & $-2.56$ & 4318.9 & 0.4943 & 0.398 & $-3.44$ & $-3.52$ & 259.819 & $-$ & M2 \\
OGLE-LMC-CEP-0077 & 2.96606336 & $-3.00$ & 4323.3 & 0.1844 & 0.390 & $-3.90$ & $-3.99$ & $-$ & 0.820 & M3 \\
OGLE-LMC-CEP-0080 & 3.42006805 & $-2.35$ & 2143.6 & 1.1041 & 1.420 & $-3.44$ & $-3.58$ & $-$ & 2.982 & M3 \\
\ldots & \ldots & \ldots & \ldots & \ldots & \ldots & \ldots & \ldots & \ldots & \ldots & \ldots \\
\ldots & \ldots & \ldots & \ldots & \ldots & \ldots & \ldots & \ldots & \ldots & \ldots & \ldots \\

\hline
\multicolumn{11}{l}{\textbf{SMC \Cepa\ sample}} \\
\hline

OGLE-SMC-CEP-0088 & 1.53847476 & $-3.11$ & 2873.4 & 0.0731 & 0.671 & $-3.93$ & $-4.37$ & $-$ & 0.288 & M3 \\
OGLE-SMC-CEP-0212 & 1.74103505 & $-3.08$ & 3430.4 & 0.1389 & 0.609 & $-3.93$ & $-4.26$ & $-$ & 0.318 & M3 \\
OGLE-SMC-CEP-0251 & 1.79680357 & $-3.34$ & 3186.9 & 0.0139 & 0.203 & $-3.75$ & $-4.22$ & 68.950 & $-$ & M2 \\
\ldots & \ldots & \ldots & \ldots & \ldots & \ldots & \ldots & \ldots & \ldots & \ldots & \ldots \\
\ldots & \ldots & \ldots & \ldots & \ldots & \ldots & \ldots & \ldots & \ldots & \ldots & \ldots \\

\hline
\multicolumn{11}{l}{\textbf{LMC \Cepb\ sample}} \\
\hline
OGLE-LMC-CEP-0676 & 2.54452114 & $-2.64$ & 4354.2 & 0.5027 & 1.827 & $-3.79$ & $-4.08$ & $-$ & 0.752 & M3 \\
OGLE-LMC-CEP-0817 & 1.87363459 & $-2.91$ & 3645.2 & 0.9685 & 1.082 & $-3.97$ & $-4.32$ & $-$ & 0.709 & M3 \\
OGLE-LMC-CEP-0925 & 2.04551965 & $-2.93$ & 2750.9 & 0.0619 & 0.949 & $-4.11$ & $-4.29$ & $-$ & 0.201 & M3 \\
\ldots & \ldots & \ldots & \ldots & \ldots & \ldots & \ldots & \ldots & \ldots & \ldots & \ldots \\
\ldots & \ldots & \ldots & \ldots & \ldots & \ldots & \ldots & \ldots & \ldots & \ldots & \ldots \\

\hline
\multicolumn{11}{l}{\textbf{SMC \Cepb\ sample}} \\
\hline
OGLE-SMC-CEP-0151 & 1.5329022 & $-2.99$ & 10362.2 & 0.0038 & 1.010 & $-4.11$ & $-4.41$ & $-$ & 0.152 & M3 \\
OGLE-SMC-CEP-1598 & 2.9940860 & $-2.92$ & 3346.3  & 0.0599 & 0.516 & $-3.78$ & $-3.91$ & $-$ & 0.545 & M3 \\

\hline
\end{tabular}
}
\tablefoot{The columns give star's ID, pulsation period, Eddington-Plakidis fluctuation parameter ($\log\epsilon$); dominant variability period and amplitude from the wavelet analysis; Stetson $L$ index; $\log_{10}(\Delta P/P)$ and $\log_{10}(\sigma)$ from the instantaneous period method; C2C jitter amplitude $\sigma_{\eta}$ and drift term $\sigma_{\xi}$ from the Koen C2C analysis and the corresponding best-fitting model. Only first few entries are shown and the full table is available at Zenodo ( see Sect.~\ref{sec: Data availability})}
\end{table*}

\section{Koen model distribution}
\label{appendix:Koen model Distribution}
To characterise the C2C period variability of the class~3 candidates, we apply the statistical framework of \citet{Koen2006MNRAS.365..489K}, as implemented for RR~Lyrae stars by \citet{Benko2025A&A...697A.154B}. Following \citet{Koen2006MNRAS.365..489K}, the instantaneous period of the $i$-th pulsation cycle is modelled as
\begin{equation}
    P_i = \langle P \rangle + \sum_{k=1}^{i} \xi_k + \eta_i,
\end{equation}
\noindent where $\langle P \rangle$ is the constant mean period, $\eta_i \sim \mathcal{N}(0,\sigma_\eta^2)$ is a zero-mean, uncorrelated cycle-to-cycle (C2C) perturbation of the cycle length, and the cumulative term $\sum_{k} \xi_k$, with $\xi_k \sim \mathcal{N}(0,\sigma_\xi^2)$, represents a smooth, persistent variation of the mean period. The measured epochs additionally carry an independent timing error $e_i \sim \mathcal{N}(0,\sigma_e^2)$. Propagating these contributions into the observed-minus-calculated residual after $n$ cycles gives
\begin{equation}
\label{eq:koen_oc}
    (\mathrm{O-C})_n = e_n + \sum_{i=1}^{n} \eta_i + \sum_{i=1}^{n}\sum_{k=1}^{i} \xi_k,
\end{equation}
\noindent so that the timing noise $\sigma_e$ enters per point, the C2C jitter $\sigma_\eta$ accumulates as a random walk, and the period drift $\sigma_\xi$ accumulates as an integrated random walk. The model therefore depends on three parameters, $\sigma_e$, $\sigma_\eta$, and $\sigma_\xi$.
\begin{figure*}[!t]
\centering
\includegraphics[height=6cm,width=0.47\textwidth]{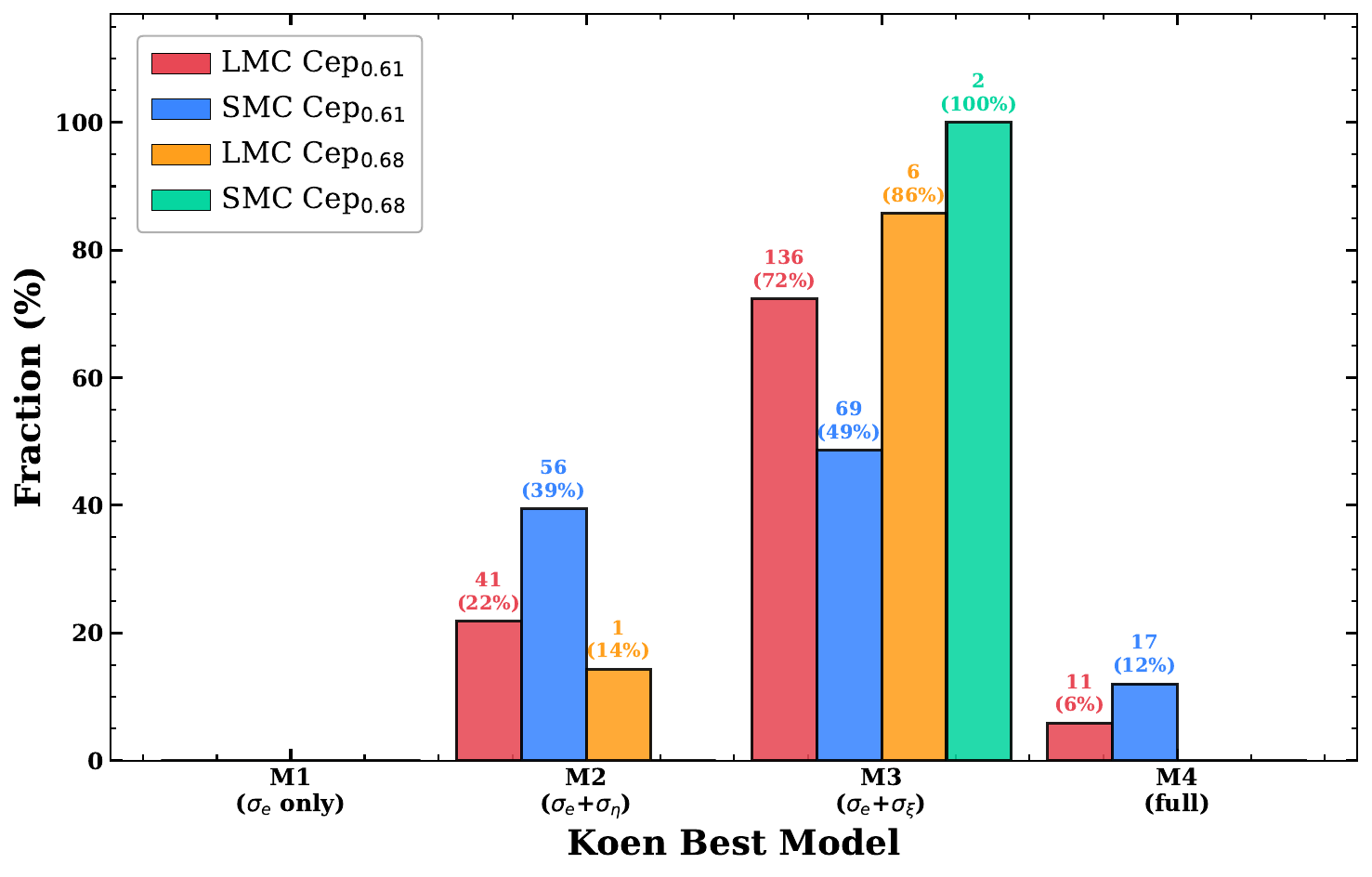}
\caption{Distribution of the preferred Koen models (M1--M4) across the four Cepheid subsamples.}
\label{fig:koen_mod}
\end{figure*}
Since the timing noise is unavoidable ($\sigma_e > 0$), four nested models are obtained by switching the C2C and drift terms on or off:
\begin{align}
    \mathrm{M1:} \quad & (\mathrm{O-C})_n = e_n, \\
    \mathrm{M2:} \quad & (\mathrm{O-C})_n = e_n + \textstyle\sum_{i=1}^{n} \eta_i, \\
    \mathrm{M3:} \quad & (\mathrm{O-C})_n = e_n + \textstyle\sum_{i=1}^{n}\sum_{k=1}^{i} \xi_k, \\
    \mathrm{M4:} \quad & (\mathrm{O-C})_n = e_n + \textstyle\sum_{i=1}^{n} \eta_i + \textstyle\sum_{i=1}^{n}\sum_{k=1}^{i} \xi_k,
\end{align}
\noindent corresponding to pure timing noise (M1), timing noise plus C2C jitter (M2), timing noise plus a smooth period drift (M3), and all three components (M4). Following \citet{Benko2025A&A...697A.154B}, each \oc\ series is first linearly transformed so that its first and last points are zero. The parameters of each model are then estimated by maximising the Gaussian likelihood computed from the full covariance matrices of the three components, evaluated at the cumulative cycle numbers of the \oc\ points \citep[see Appendix~A of][]{Koen2006MNRAS.365..489K}. The preferred model for each star is selected by minimising the AIC value.

Figure~\ref{fig:koen_mod} summarises the best-model distribution across the four subsamples. M1 is never preferred for any star. Pure timing noise therefore cannot reproduce the observed \oc\ scatter, and a genuine signal in the form of C2C jitter, a period drift, or both is always required. This is expected, since these are robustly classified class~3 candidates whose irregularity clearly exceeds the measurement noise. Model M3, consisting of timing noise plus a smooth period drift with no C2C term, is the most common, preferred by $72\%$ of LMC \Cepa, $49\%$ of SMC \Cepa, $86\%$ of LMC \Cepb, and all SMC \Cepb\ stars. For the majority of stars, therefore, the \oc\ variations are well described by measurement noise together with a smooth period drift, without invoking a dedicated C2C jitter term ($\sigma_\eta$). Only $37\%$ of stars assigned to M2 or M4 require $\sigma_\eta > 0$.

\end{appendix}

\end{document}